\documentclass[preprint,aps,nofootinbib,preprintnumbers,amsmath,amssymb,11pt,longbibliography]{revtex4-1}

\usepackage{amsmath}
\usepackage{graphicx}
\usepackage{dcolumn}
\usepackage{bbm}
\usepackage{bm}
\usepackage{subfigure}
\usepackage{amssymb}
\usepackage{hyperref}
\usepackage{xcolor}
\usepackage{subfigure}
\usepackage{makecell}
\usepackage{mathtools}
\usepackage{diagbox}
\usepackage[myheadings]{fullpage}

\usepackage{xparse}

\usepackage[a4paper,total={6in, 8in}]{geometry}

\usepackage{color}
\definecolor{darkblue}{rgb}{0.,0.,0.4}
\definecolor{darkred}{rgb}{0.5,0.,0.}
\definecolor{BlueViolet}{RGB}{138,43,226}
\definecolor{SkyBlue}{RGB}{30,144,255}
\definecolor{DarkGreen}{RGB}{0,100,0}

\stepcounter{secnumdepth}
\stepcounter{tocdepth}

\newcommand{\bfnc}{$\beta$-functions }

\newcommand{\bqa}{\begin{eqnarray}} 
\newcommand{\eqa}{\end{eqnarray}}
\newcommand{\nn}{\nonumber \\}
\newcommand{\eq}[1]{Eq. \eqref{#1}}
\newcommand{\fig}[1]{Fig. \ref{#1}}

\begin{document}

\title{Constraints on beta functions in field theories}

\author{\vspace*{1 cm}\large
Han Ma$^{\,\triangle}$, Sung-Sik Lee$^{\,\triangle, \dagger}$}
\affiliation{\vspace*{0.5 cm}$^\triangle$ Perimeter Institute for Theoretical Physics, Waterloo, Ontario N2L 2Y5, Canada,\\}
\affiliation{$^\dagger$ 
Department of Physics $\&$ Astronomy, McMaster University,
1280 Main St. W., Hamilton, Ontario L8S 4M1, Canada\\\\\\}

\begin{abstract}
The \bfnc   describe how couplings run under the renormalization group flow in field theories.
In general, all couplings  that respect the symmetry and locality are generated under the renormalization group flow,
and the exact renormalization group flow is characterized
by the \bfnc defined 
in the infinite dimensional space of couplings.
In this paper, we show that the renormalization group flow is highly constrained so that
the \bfnc defined in a measure zero subspace of couplings completely
determine the \bfnc in the entire space of couplings.
We provide  a quantum renormalization group-based algorithm
for reconstructing the full \bfnc from the \bfnc defined in the subspace.
As examples,
we derive the full $\beta$-functions for 
the $O(N)$ vector model and the $O_L(N) \times O_R(N)$ matrix model
entirely from the \bfnc defined in the subspace of single-trace couplings.

\end{abstract}

\maketitle

\tableofcontents

\section{Introduction}

One of the greatest advances in modern theoretical physics is the invention of the renormalization group (RG)
\cite{
PhysRev.95.1300,
PhysicsPhysiqueFizika.2.263,
PhysRevD.2.1541,
Symanzik:1970aa,
RevModPhys.47.773,
POLCHINSKI1984269,
1986JETPL..43..730Z,
WETTERICH199390,
1991NuPhB.363..486O,
CARDY1988749,
Komargodski:2011aa,
NAKAYAMA20151}. 
The idea is to organize a complicated  many-body system in terms of length scales of constituent degrees of freedom.
Thanks to locality that greatly limits the way short-distance modes influence long-distance modes,
one can understand coarse grained properties of the system in terms of effective field theories without considering all degrees of freedom in the system.
This opens the door for systematic understandings of many physical phenomena that are otherwise too difficult to study theoretically.

The central object in RG is the $\beta$-function.
It describes how an effective theory gradually changes as the length scale is increased. 
The renormalization of the couplings for long-wavelength modes is driven by fluctuations of short-wavelength modes,
which creates the RG flow in the space of theories.
While the $\beta$-function contains the full information on the fate of a system in the infrared limit, it is in practice hard to keep track of the exact RG flow.
Even if one starts with a relatively simple theory with a small number of couplings at a short distance scale, all couplings that respect symmetry and locality are eventually generated at larger distances.
In general, one has to deal with the RG flow in the infinite dimensional space of  couplings.

Therefore, it is desirable to take advantage of constraints that $\beta$-functions satisfy if there is any. 
It is easy to see that not all \bfnc are independent around free field theory fixed points. 
For example, 
the scaling dimension of $\phi^{2n}$ is $n$ times that of $\phi^2$ at the Gaussian fixed point.
Therefore, the $\beta$-function of the former is fixed by that of the latter to the linear order in the sources for the operators.  
It is then natural to ask whether such constraints exist for interacting theories and, if so, what the general rules are. There are proposals under special circumstances~\cite{damgaard1997constraints, poole2019constraints}.
In this paper, 
we show that \bfnc in all field theories are highly constrained : {\it the \bfnc defined in a measure zero subspace of  couplings completely fix the \bfnc in the entire space of couplings}.

Our result is beyond the well known constraint for beta functions present in continuum field theories. 
Consider a field theory defined non-perturbatively with a finite UV cutoff.
Examples include field theories  regularized on lattice. 
While infinitely many  couplings can be turned on 
in the UV theory,
at energy scales much smaller than the UV cutoff 
all couplings are fixed by a finite number of marginal and relevant couplings. 
As a result, the flow of most couplings is controlled by the marginal and relevant couplings in the low-energy limit.
However, the constraints discussed in our paper applies to  \bfnc at all energy scales including the scales that are comparable to the UV cutoff. 
At high energy scales close to the UV cutoff,
irrelevant couplings 
can be tuned independently,
and they are not fixed by the marginal and relevant couplings through the usual constraint that emerges only in the low-energy limit. 
In this paper, we are concerned about the general  kinematic constraints that  \bfnc obey at all energy scales.

To uncover the constraints that \bfnc satisfy in general field theories,  we use the quantum RG\cite{lee2012background,lee2014quantum}.
Quantum RG reformulates the Wilsonian RG by projecting the full RG flow onto a subspace of couplings.
The subspace is spanned by couplings for the so-called single-trace operators.
Single-trace operators are basic building blocks of general operators in that all operators that respect the symmetry can be written as composites of single-trace operators.
In large $N$ theories, the set of single-trace operators consists of the operators that involve one trace of flavor or color indices\cite{BECCHI2002250}. 
However, the notion of single-trace operators can be defined in any field theory once the fundamental degrees of freedom and the symmetry of the theory are fixed\cite{lee2014quantum}. 
Although quantum RG does not include composites of the single-trace operators (called multi-trace operators) directly,  it exactly takes into account their effects by promoting the single-trace couplings to dynamical variables. 
The pattern of fluctuations of the single-trace couplings precisely captures the multi-trace couplings.
As a result, the classical Wilsonian RG flow defined in the full space of couplings is replaced with a sum over RG paths defined in the subspace of single-trace couplings in the quantum RG.
The \bfnc of the Wilsonian RG is then replaced with an action for dynamical single-trace couplings that determines the weight of fluctuating  RG paths. 
The bulk theory constructed from quantum RG is free of UV divergence as far as the original theory is regularized.

For a $D$-dimensional field theory,  the theory for the dynamical single-trace couplings takes the form of a $(D+1)$-dimensional theory, 
where the dynamical couplings depend on the $D$-dimensional space and the RG scale.
The theory includes dynamical gravity because the coupling functions for the single-trace energy-momentum tensor is nothing but a  metric that is promoted to a dynamical variable in quantum RG\cite{lee2014quantum}.
For this reason, quantum RG  provides a natural framework for the AdS/CFT correspondence~\cite{maldacena1999large,gubser1998gauge,witten1998anti} in which the extra dimension in the bulk is interpreted as the RG scale
\cite{akhmedov1998remark,
de2000holographic,
Skenderis:2002wp,
Heemskerk:2009pn,
Heemskerk:2010hk,
2011JHEP...08..051F,
papadimitriou2016lectures,
doi:10.1002/prop.201400007}\footnote{
In order to construct a background independent gravitational theory for fluctuating couplings in quantum RG, one needs to use a coarse graining scheme\cite{Lee2020}
that does not introduce a fixed background
\cite{lee2012background,lee2016horizon}
and satisfy a consistency condition\cite{NAKAYAMA201437,PhysRevD.95.066003}.
For our purpose of demonstrating the existence of constraints of $\beta$-functions, however, the issue is not crucial.
The quantum RG is an exact reformulation of the Wilsonian RG in any coarse graining scheme irrespective of whether the scheme is background independent or not.}.

\begin{figure}[h]
\includegraphics[width=.6\textwidth]{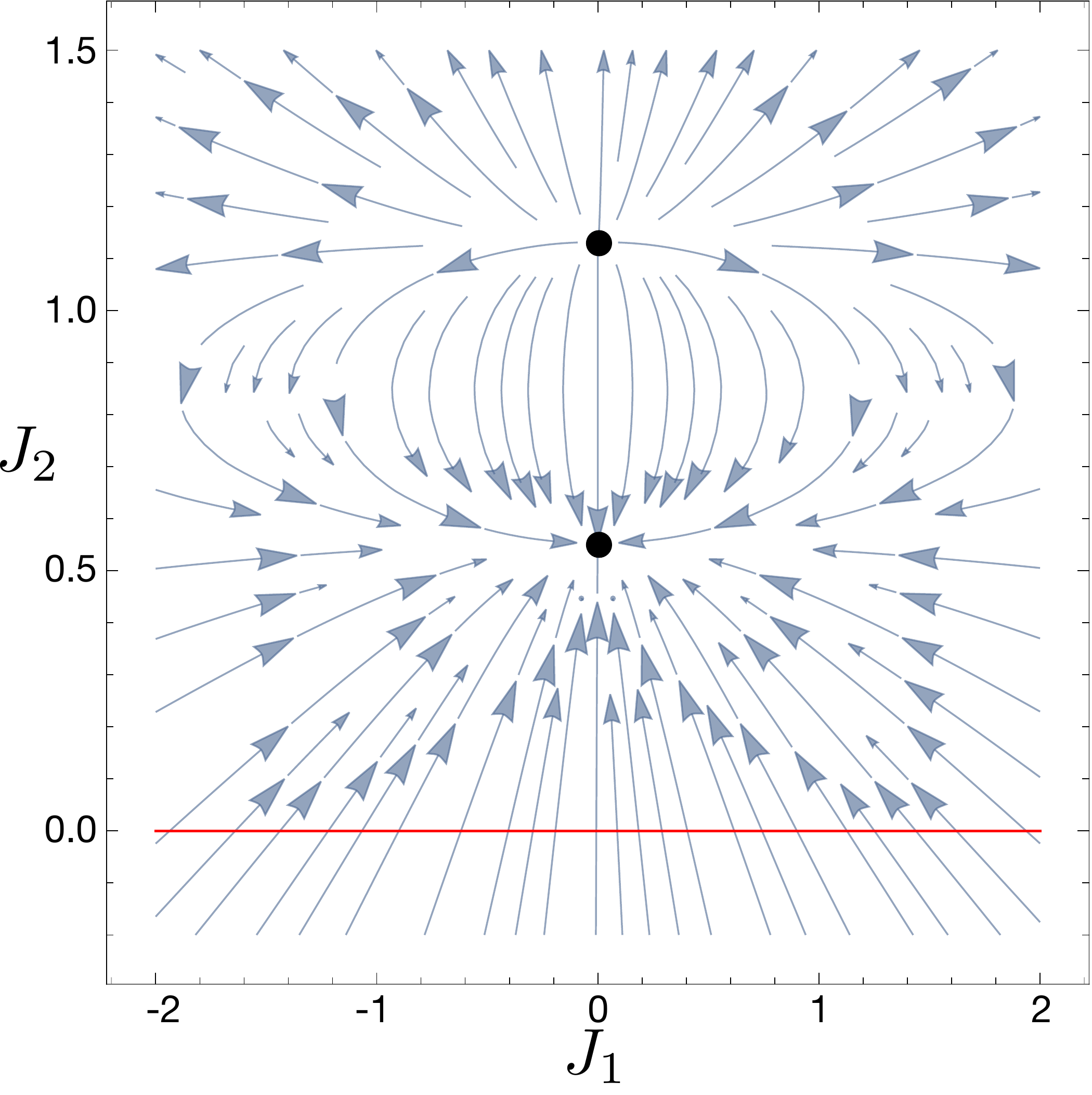}
\caption{
RG flow of a toy model considered in Sec.  \ref{sec:0d_example}.
$J_1$($J_2$) is the single-trace (double-trace) coupling.
The $\beta$-functions in the $J_2=0$  subspace, which is denoted as the red horizontal line, fix  the full $\beta$-functions in the space of $J_1$ and $J_2$.
The full $\beta$-functions exhibit rich structures that include one stable and one unstable fixed points away from the single-trace subspace, which are fully encoded in the $\beta$-functions within the subspace of  $J_2=0$. 
\label{fig:RG_flow}
} 
\end{figure}

In the Wilsonian RG, a field theory is represented as a point in the space of all couplings.
In quantum RG, a field theory is represented as a wavefunction  defined in the subspace of the single-trace couplings. 
The peak position of the wavefunction indicates the value of the single-trace coupling. 
Around the peak, the second and higher moments of the fluctuating single-trace coupling contain the information on the double-trace and higher-trace couplings.
The classical flow of couplings in the Wilsonian RG is replaced with a 
quantum evolution of the wavefunction in quantum RG.
The bulk theory that governs the quantum RG flow is entirely fixed by the \bfnc defined in the subspace of single-trace couplings\cite{lee2012background,lee2014quantum}.
Since the wavefunction at an RG scale encodes the full information on all couplings at the scale,
the bulk theory fully determines the \bfnc of all multi-trace couplings. 
This implies that the full \bfnc  is fixed by \bfnc defined in the subspace of single-trace couplings.
A simple example that illustrates the main result of this paper is shown in \fig{fig:RG_flow}.

In this work, we provide a general algorithm for extracting the full \bfnc from the \bfnc defined in the subspace of single-trace couplings.
The algorithm consists of the following steps. 
First, we construct the bulk theory for quantum RG from the \bfnc defined in the space of single-trace couplings. 
Second, we solve the (functional) Schrodinger equation that evolves an initial state fixed by the UV theory to IR.
Finally, we identify the ground state of the quantum RG Hamiltonian as the IR fixed point of the theory, and states with local excitations as the IR fixed point deformed with local operators. 
This allows us to extract the full \bfnc from the spectrum of the quantum RG Hamiltonian.
This dictionary in the algorithm is summarized  in Tab.~\ref{tab:correspondence}.

\begin{table}[h]
\caption{Dictionary of correspondence between boundary field theory and bulk theory. \label{tab:correspondence}}
\begin{tabular}{c|c} \hline
 {\bf boundary field theory } &{\bf bulk theory for quantum RG }  \\ \hline
 RG time (logarithmic length scale) 
 & extra bulk dimension  
 \\ \hline
 single-trace coupling (operator) & dynamic field (conjugate momentum) \\ \hline
 Boltzmann weight in the partition function & bulk state  \\ \hline
 \makecell{RG transformation 
 } & 
 \makecell{
radial quantum  evolution 
 }
 \\ \hline
 stable fixed point & ground state of the RG Hamiltonian \\  \hline
 local scaling operators & local excitations \\ \hline
 scaling dimensions & spectrum of the RG Hamiltonian \\ \hline
\end{tabular}
\end{table}

The rest of the paper is organized as follows. 
In Sec. \ref{sec:Wilsonian_RG},
we present the main result of our paper using two concrete models :
the $O(N)$ vector model and 
the  $O_L(N) \times O_R(N)$ matrix model regularized on a lattice.
Through quantum RG, the exact RG flow is mapped into quantum evolution of a wavefunction defined in the space of single-trace couplings.
We show that the resulting quantum theories
 from quantum RG flow 
of the regularized field theories are finite and well defined.
We explicitly compute the full \bfnc for these models
from the $\beta$-functions defined in the subspace of single-trace couplings.
\eq{eq:ONbetaHPsi}
and
\eq{eq:betarelationONmatrix}
are the main results.
In Secs.~\ref{sec:main}
and \ref{sec:QRG}, we generalize the results obtained through the  concrete examples.
In Sec.~\ref{sec:toymodels}, we consider toy models in which the bulk theory  is non-interacting, and the exact RG flow equation can be exactly solved through quantum RG.

\section{
Constraints on $\beta$ functions
\label{sec:Wilsonian_RG}}

In this section, 
we illustrate the main result of the paper 
using two examples.
The first example is the 
$O(N)$ vector model,
and the second one is the $O_L(N) \times O_R(N)$ matrix model.
To be concrete, we consider those theories regularized
on a  $D$-dimensional Euclidean lattice.
We first review 
how the exact Wilsonian RG defined in the space of all couplings can be  reformulated 
as a quantum evolution of wavefunction 
defined in the subspace of single-trace couplings\cite{lee2012background,lee2014quantum}.
From this, we constructively show that
the full beta functions are determined from the beta functions defined in the subspace of single-trace couplings.

\subsection{The O(N) vector model 
\label{sec:O(N)_vector}}

In describing RG flow of a theory,
it is convenient to choose a reference theory as the origin 
 of the space of theories.
A general theory is then viewed as a theory obtained by adding a deformation to the  reference theory. 
The RG flow then describes the change of the deformation as a function of length scale.
We write the $O(N)$ vector model as
\bqa
S [\phi]= S_0[\phi] + S_1[\phi],
\label{eq:Z_UV0}
\eqa
where $S_0$ is the reference action, 
\bqa
S_0[\phi] = \frac{1}{2}m^2 \sum_i 
(\phi_i \phi_i),
\eqa
and
$S_1$ is a deformation,
\begin{eqnarray}
S_1= 
 \sum_{ij} J^{(1)}_{ij} (\phi_{i}\phi_{j}) + \frac{J^{(2)}}{N}\sum_{i }  ( \phi_{i}^2)^2.
\label{eq:Z_UV0_S1}
\end{eqnarray}
$\phi^a_i$ is a real field  with flavour $a=1,2,..,N$ defined at site $i$,
and 
$(\phi_i \phi_j) \equiv \sum_a \phi^a_i \phi^a_j$.
$S_0$ represents the trivial gapped  fixed point.
$J^{(1)}_{ij}$ is the hopping amplitude between site $i$ and $j$, and
$J^{(2)}$ is the on-site quartic interaction.
Depending on the magnitudes of the deformations,
the theory may stay in 
the insulating phase,
or flows to a different fixed point associated with
the critical point
or the symmetry broken state.
Our goal is to understand the exact RG flow of the theory.
Since we choose the ultra-local gapped fixed point action as the reference action, 
we use the real space RG scheme
in which $S_0$ is invariant under the coarse graining transformation.
However, different RG schemes can be used
as is discussed in Appendix \ref{app:Gaussian_reference_action}. 
The choice of RG scheme 
does not affect the physics.

\subsubsection{Classical RG}

We first review the exact Wilsonian RG\footnote{
In this paper,
we use the terms Wilsonian RG and classical RG interchangeably.
}.
The exact Wilsonian RG flow is generated from the following steps\cite{POLCHINSKI1984269}.  
\begin{itemize}
\item 
Separation of $\phi$ into low-energy modes and high-energy modes :
    
This is needed before we integrate out the high-energy modes to obtain an effective action for the low-energy modes. 
In the real space, we usually consider a scheme where
a block of sites is merged to  generate a coarse-grained lattice\cite{KADANOFFRG}. 
However, this forces the RG steps to be discrete. 
To avoid this, we employ the scheme in which the field $\phi$ is partially integrated out without changing the number of sites.
For this,
we introduce an auxiliary field $\Phi$ with mass $\mu$.
The total action is written as
\begin{eqnarray}
S[\phi,\Phi] = S_0[\phi]+\tilde{S}_0[\Phi]+S_1 [\phi],
\end{eqnarray}
where
\begin{eqnarray}
\tilde{S}_0[\Phi] = \frac{1}{2}\mu^2 \sum_i  \Phi_i^2.
\end{eqnarray}
Now we rotate $\phi$ and $\Phi$ into a new pair of fields,
\begin{eqnarray}
\phi_i =\phi'_i+ \Phi'_i,\quad \Phi_i =A\phi'_i+B \Phi'_i,
\end{eqnarray}
where $A=\frac{m^2}{\tilde{\mu}\mu}$ and $B=-\frac{\tilde{\mu}}{\mu}$ with 
$\tilde{\mu}=\frac{m}{\sqrt{e^{2dz}-1}}\approx \frac{m}{\sqrt{2dz}}$. 
$dz$ is an infinitesimal parameter  that labels the continuous coarse graining steps.
The coefficients are chosen such that the original field $\phi$ is given by the sum of the low-energy field ($\phi'$) and the high-energy field ($\Phi'$),
and the low-energy field has mass $m e^{dz}$.
The action for $\phi'$ and $\Phi'$ becomes
\begin{eqnarray}
S[\phi', \Phi'] = \frac{1}{2} m^2 e^{2dz} \sum_i (\phi'_i)^2 +\frac{1}{2} \frac{m^2}{2dz} (\Phi'_i)^2 
+S_1 [\phi'+\Phi'].
\end{eqnarray}
The field $\phi'$ acquires the larger mass indicating that it has less fluctuation than $\phi$. 
The missing fluctuation is transferred to the higher energy field $\Phi'$.

\item
Coarse graining :

The high energy field $\Phi'$ is integrated out. 
This has the effect of partially including fluctuations of physical degrees of freedom without reducing the number of sites.
This gives rise to corrections that renormalize $S_1$ to $S_1+\delta S_1$, where 
\begin{eqnarray}
\delta S_1 = \frac{dz}{m^2} \left[\frac{\partial^2}{\partial (\phi'_i)^2} S_1 -\Big(\frac{\partial S_1}{\partial \phi'_i}\Big)^2 \right] 
\end{eqnarray}
up to the leading order in $dz$.

\item 
Rescaling of field;

To be able to perform this coarse graining procedure iteratively, we need to bring the mass from $me^{dz}$ back to $m$.
This can be done by rescaling the field as
\begin{eqnarray}
\phi'_i =e^{-dz}\phi''_i.
\end{eqnarray}
This restores the original reference action and 
generates an additional correction to $S_1$,
\begin{eqnarray}
\delta S_1' = 
-dz  \phi''_i \frac{\partial}{\partial \phi''_i}.
\end{eqnarray}

\end{itemize}

This completes one cycle of the coarse graining.
After this exact RG transformation,
the effective action becomes
$S(dz) = S + \delta S$,
where
\begin{eqnarray}
\delta S &=&
-N dz \Bigl[ 
 \sum_i \beta_i^{(0)} 
+
\frac{1}{N}\sum_{\{i_1,j_1\}}
\beta_{i_1,j_1}^{(1)} 
  (\phi_{i_1}\phi_{j_1})
+ 
\sum_{\{i_1,j_1, i_2, j_2\}}
\beta_{i_1,j_1,i_2,j_2}^{(2)} 
\frac{1}{N^2}  
(\phi_{i_1}\phi_{j_1})
(\phi_{i_2}\phi_{j_2}) \nn
&&
+ 
\sum_{\{i_1,j_1, i_2, j_2 i_3 j_3\}}
\beta_{i_1,j_1,i_2,j_2,i_3 j_3}^{(3)} 
\frac{1}{N^3}  
(\phi_{i_1}\phi_{j_1})
(\phi_{i_2}\phi_{j_2})
(\phi_{i_3}\phi_{j_3})
\Bigr].
\label{eq:classical_ON_vector_beta}
\end{eqnarray} 
Here the beta functions are 
$\beta^{(0)}_i
[J^{(1)},J^{(2)}] 
= 
- \frac{2}{m^2} J^{(1)}_{ii}$, ~
$\beta^{(1)}_{ij}
[J^{(1)},J^{(2)}] 
= 2 J^{(1)}_{ij}
+\frac{4}{m^2}\sum_{k}
J^{(1)}_{ki} J^{(1)}_{kj}
-\frac{4J^{(2)}}{m^2}(1+\frac{2}{N})\delta_{ij}$,~
$\beta^{(2)}_{i_1j_1i_2j_2}[J^{(1)},J^{(2)}] =
4 J^{(2)}  \delta_{i_1j_1}\delta_{j_1j_2}\delta_{i_1i_2}
+\frac{16}{ m^2}  J^{(2)} 
J^{(1)}_{i_1j_1}\delta_{i_1i_2}\delta_{i_2j_2}$,~
$\beta^{(3)}_{i_1j_1i_2j_2i_3j_3} [J^{(1)},J^{(2)}]$
$=$
$\frac{16 (J^{(2)})^2}{m^2}$
$\delta_{i_1j_1}$
$\delta_{i_2j_2} $
$\delta_{i_3j_3}$
$\delta_{i_1i_2}$
$\delta_{i_1i_3}$.
The exact RG transformation 
not only renormalizes the terms that are already present in the action but also generates new operators that are order of $\phi^6$.
In the subsequent RG steps, infinitely many other operators are generated.
The general effective action 
takes the form of
\begin{eqnarray}
S = S_0 +
N \sum_{K=1}^\infty 
~
\sum_{i_1,j_1,\dots, i_K,j_K} 
J^{(K)}_{i_1,j_1,\dots, i_K,j_K}
\mathcal{\bf O}_{i_1, j_1,\dots,i_K,j_K},
\label{eq:action_general_z}
\end{eqnarray}
where
\bqa
\mathcal{\bf O}_{i_1, j_1,\dots,i_K,j_K}
=
\frac{1}{N^K} \prod_{n=1}^K (\phi_{i_n}\phi_{j_n})
\label{eq:generalO}
\eqa
is the set of most general $O(N)$ invariant operators.
$\mathcal{\bf O}_{i_1 j_1; i_2 j_2; ..; i_K j_K} $
with $K=0$ is the identity operator.
$\mathcal{\bf O}_{i_1 j_1; i_2 j_2; ..; i_K j_K} $'s
with $K=1$ are referred to as  single-trace operators because they involve one summation of flavour indices.
Those with $K>1$ are multi-trace operators.
$J^{(K)}_{i_1,j_1,\dots, i_K,j_K}$
is the source for 
$\mathcal{\bf O}_{i_1, j_1,\dots,i_K,j_K}$.
In \eq{eq:action_general_z}
and \eq{eq:generalO},
factors of $N$ are chosen
so that 
$J^{(K)}_{i_1,j_1,\dots, i_K,j_K}$
and 
$\mathcal{\bf O}_{i_1, j_1,\dots,i_K,j_K}$
are $\mathcal{O}(1)$.
Even in the large $N$ limit,
multi-trace couplings are not negligible.
The exact RG flow is encoded in 
the beta functions,
\bqa
\frac{d J^{(K)}_{i_1,j_1,\dots, i_K,j_K}}{dz}
= -
\beta^{(K)}_{i_1,j_1,\dots, i_K,j_K} 
\left( 
J^{(1)},
J^{(2)},
\dots
\right)
\label{eq:ONfullbeta}
\eqa
that is defined in the infinite dimensional space of couplings,
$\left\{ J^{(K)}_{i_1,j_1,\dots, i_K,j_K} \right\}$.
Since each coupling in 
$\left\{ J^{(K)}_{i_1,j_1,\dots, i_K,j_K} \right\}$ 
can be added to the UV theory in \eq{eq:Z_UV0}
and tuned independently
\footnote{
For local theories, 
the multi-local couplings should decay exponentially  in space.
},
there is in general no particular relation among the couplings at high energy scales.
A universal relation among couplings emerge only in the long distance limit 
as all couplings are determined in terms of a few relevant and marginal couplings in the continuum limit.
Here our goal is to describe the entire RG flow that covers from the lattice scale to the long distance limit.
At short length scales,  couplings are not related to each other,
and the full beta functions 
at general values of couplings are needed
in \eq{eq:ONgeneralbeta}.
While the full $\beta$-functions define the vector field in the infinite dimensional space of all couplings,
we will show that the information of all $\beta$-functions is entirely encoded in the $\beta$-functions defined in a subspace of single-trace couplings.
We emphasize that this constraint among beta functions holds at all length scales even close to the lattice scale, 
and is not a consequence of the relations that emerge in the long distance limit.

\subsubsection{Quantum RG}

The exact RG flow of the effective action can be written as\cite{POLCHINSKI1984269}
\bqa
\frac{\partial S_1(z)}{\partial z}
&= &
 \frac{1}{m^2} \left[
 \frac{\partial}{\partial \phi_i} 
 \frac{\partial}{\partial \phi_i} 
 S_1(z) -\Big(\frac{\partial S_1(z)}{\partial \phi_i}\Big)^2  \right]
 -\phi_i \frac{\partial S_1(z)}{\partial \phi_i},
\label{eq:ONgeneralbeta}
\eqa
where the total effective action at scale $z$ is given by
$S(z) = S_0 + S_1(z)$.
This, in turn, can be written as a differential operator acting on $e^{-S}$ as
\bqa
e^{-S(z) } = e^{-\hat{H} z} e^{-S},
\label{eq:HeS}
\eqa
where 
\bqa
\hat H = 
e^{-\hat S_0}
\sum_{i,a} \left[i (\hat \phi_i \hat \pi_i) +\frac{1}{m^2} (\hat \pi_i \hat \pi_i) \right]
e^{\hat S_0},
\label{eq:PRGHamiltonian}
\eqa
and $\hat \pi_i^a = -i \frac{\delta}{\delta \phi_{i}^a}$ is
the conjugate momentum of $\hat \phi_i^a$. 
In \eq{eq:HeS},
$e^{-S}$ plays the role of a wavefunction,
and $\hat{H}$ acts as a quantum Hamiltonian for an imaginary time evolution.
Here, the imaginary time corresponds to the logarithmic length scale in RG.
For this reason, we call $\hat H$
 the {\it RG Hamiltonian}.

The observation that the RG flow can be generated from the quantum  RG Hamiltonian suggests that the space of theories can be viewed as a vector space.
In this picture, 
$e^{-S}$ is viewed as a wave function,
and the partition function becomes an overlap between two wavefunctions\cite{lee2016horizon},
\bqa
Z= \int \mathcal{D}\phi~ e^{-S} 
= \langle \mathbbm{1} | S \rangle,
\label{eq:Z1S}
\eqa
where
\bqa
| S \rangle = \int \mathcal{D} \phi ~ e^{-S[\phi]} | \phi \rangle
\label{eq:Sstate}
\eqa
is the state associated with the action $S$\cite{lee2016horizon},
and 
\bqa
| \mathbbm{1}  \rangle = \int \mathcal{D} \phi  ~ | \phi \rangle
\label{eq:1state}
\eqa
 is the state whose wavefunction is $1$.
$|\phi \rangle$ denotes the basis state whose inner product is given by
$\langle \phi' | \phi \rangle = \prod_{i,a} \delta \Bigl( \phi^{'a}_i - \phi^a_i  \Bigr)$.
Although $| \mathbbm{1} \rangle$ is not normalizable, the overlap in 
\eq{eq:Z1S} is well defined.
One can check that the RG Hamiltonian leaves \eq{eq:1state} invariant
when applied from the right :
$\langle \mathbbm{1} | H = 0$ or equivalently
$H^\dagger |  \mathbbm{1} \rangle  = 0$.
Therefore, the partition function is invariant under the insertion of the RG evolution operator in the overlap,
\bqa
Z=  
\langle \mathbbm{1} | e^{-z \hat H} | S \rangle,
\label{eq:Z1HS}
\eqa
where $dz$ is an infinitesimal change of the logarithmic length scale.
This reflects the fact that the partition function is unchanged under the RG transformation.
Only the form of the effective action  changes as a function of the length scale.
The flow of the effective action is encoded in the state evolution,
\bqa
 | S(z) \rangle 
 =
  e^{-z \hat H} | S \rangle,
\eqa
and the effective action at scale $z$ is given by
$S(z) = -\ln \langle \phi  | S(z) \rangle$.
Even if $S$ has only simple  interaction terms such as the ultra-local quartic interaction, 
all local operators that are invariant under the $O(N)$ symmetry are generated in $S(z)$ for $z>0$.
This makes it difficult to follow the exact RG flow in the space of actions.

The complication can be alleviated in quantum RG that takes advantage of 
the facts that the space of theories can be viewed as a Hilbert space, and 
that there exists a set of basis states that span the full Hilbert space.
The full Hilbert space associated with the $O(N)$ invariant action is spanned by a set of basis state whose wavefunctions include only the single-trace operators,
\bqa
{\cal O}_{ij} = \frac{ \phi_i \phi_j}{N}.
\label{eq:Oij}
\eqa
The basis states are written as
\bqa
| t \rangle =\int \mathcal{D}\phi~ e^{
-S_0
+ i N \sum_{ij}
t_{ij}
{\cal O}_{ij}
}|\phi\rangle.
\label{eq:t}
\eqa
Here  $S_0$ is the fixed reference action,
and the basis state is labeled by the bi-local field, $t_{ij}$.
Because general $O(N)$ invariant operators in  \eq{eq:generalO} can be written as polynomials of \eq{eq:Oij},
\eq{eq:t} forms a complete basis.
Suppose we start with the general $O(N)$ invariant action shown in \eq{eq:action_general_z}.
The state associated with \eq{eq:action_general_z}
can be written as a linear superposition of 
$| t \rangle$ as
\bqa
|S \rangle =\int 
\mathcal{D}t ~
\Psi_{J}[t]  |t\rangle,
\eqa
where
\begin{eqnarray}
\Psi_{J}[t] &=& \int \mathcal{D}p
~e^{  
-i N \sum_{ij}t_{ij} p_{ij}
- N \sum_{i_1,j_1,\dots,i_K,j_K}J^{(K)}_{i_1,j_1,\dots, i_K,j_K}
p_{i_1j_1} \dots
p_{i_Kj_K} 
}.
\label{eq:ONgeneralPsi}
\end{eqnarray}
Here the integrations 
over the dynamical sources $t_{ij}$ and
its conjugate field $p_{ij}$ are defined along the real axis as
$ \int \mathcal{D}t
\equiv 
\prod_{ij} \int_{-\infty}^\infty d t_{ij} $ 
and
$ \int \mathcal{D}p
\equiv 
\prod_{ij} \int_{-\infty}^\infty d p_{ij} $.
$\Psi_J[t]$ is the wavefunction defined in the space of the single-trace couplings.
Due to the linear superposition principle, 
the RG evolution of the general action can be  carried out solely in terms of how each basis state is evolved under the RG Hamiltonian.
After an infinitesimal RG evolution, we obtain
$|S(dz)\rangle  =e^{-dz \hat{H}}|S\rangle $,
where
\begin{eqnarray}
\label{eq:eHS_QRG}
|S(dz) \rangle 
&=& \int \mathcal{D}\phi \left[\int \mathcal{D}t
\Psi_{J}[t]
e^{ -S_0 + i N \sum_{ij}t_{ij} 
{\cal O}_{ij}
+ N \Big( \beta^{(0)}[-it,0,..] +\sum_{ij}\beta^{(1)}_{ij}[-it,0,..] 
{\cal O}_{ij}
\Big) dz}\right] 
|\phi\rangle.
\nonumber\\
\end{eqnarray}
The resulting state can be written as a linear superposition of the basis states as
$|S(dz)\rangle  =
\int \mathcal{D}t'~\Psi_{J}^{dz}[t'] | t' \rangle$,
where the new wavefunction is given by
\begin{eqnarray}
\Psi^{dz}_{J}[t'] =\int  \mathcal{D} p^{'} \mathcal{D}t~ 
e^{ 
dzN \left(
-i\sum_{ij}p^{'}_{ij} \frac{t^{'}_{ij}-t_{ij}}{dz}  + \beta^{(0)}[-it,0] + \sum_{ij}\beta^{(1)}_{ij}[-it,0]p^{'}_{ij} \right) }\Psi_{J}[t].
\label{eq:ONPsidz}
\end{eqnarray}
Now, \eq{eq:ONPsidz} can be viewed as an evolution of the wavefunction defined in the space of single-trace coupling $t$,
\bqa
\Psi^{dz}_{J}[t] =
e^{-N dz \hat{H}_{bulk} }
~ \Psi_{J}[t],
\label{eq:107}
\eqa
where $\hat{H}_{bulk}$
is the bulk Hamiltonian
given by
\begin{eqnarray}
\hat{H}_{bulk} &=& 
-\beta^{(0)}[-i\hat{t},0] 
- \sum_{ij}\hat{p}_{ij}\beta^{(1)}_{ij}[-i\hat{t},0] \nn
&=&\frac{2i}{m^2}\sum_k  \hat{t}_{kk}  
-2i \sum_{kl}\hat{p}_{kl} \hat{t}_{kl} 
+\frac{4}{m^2} \sum_{kji}
\hat{p}_{ij}
\hat{t}_{ik}
\hat{t}_{kj} 
.
\label{eq:Hbulk_O(N)}
\end{eqnarray}
Here,
$\hat{t}$ and $\hat{p}$ are conjugate operators $[\hat{t}_{ij},\hat{p}_{kl}]=
- \frac{i}{2N}
\left( \delta_{ik}\delta_{jl}+\delta_{il}\delta_{kj} \right)$.
The state at scale $z$ becomes
\bqa
\Psi^{z}_{J}[t] =
e^{-N z \hat{H}_{bulk}}
~ \Psi_{J}[t],
\label{eq:ONfinitez}
\eqa
and the exact renormalized effective action at scale $z$ is given by
$S(z) = - \log 
\int Dt ~
\Psi^z_J[t]
 \langle \phi | t \rangle$.
$\Psi^z_J[t]$
is not Gaussian,
and the non-Gaussianity of the wavefunction encodes the 
higher order operators that are generated in the effective action.
Through the standard mapping, 
\eq{eq:ONfinitez} 
can be written as a path integration as
\bqa
\Psi^{z}_J[t^z]
= 
\int 
\mathcal{\bf D} t 
\mathcal{\bf D} p 
~
e^{-N S_{bulk}} 
~\Psi^{0}_J[t^0],
\eqa
where 
$\int 
\mathcal{\bf D} t 
\mathcal{\bf D} p 
=
\int 
\prod_{0\leq z' < z} \mathcal{D} t^{z'}
\prod_{0 < z' \leq z} \mathcal{D} p^{z'}
$ 
sums over RG paths  for $t_{ij}(z)$ in the subspace of  the bi-local single-trace couplings (and the conjugate variables),
and $S_{bulk}$ is the bulk action that determines the weight for each RG path,
\bqa
S_{bulk} = 
\int_0^z dz'
\left[
i \sum_{ij} p_{ij}^{z'} \partial_{z'} t_{ij}^{z'}
+ H_{bulk}[ p^{z'}, t^{z'}]
\right].
\eqa
The bulk theory is fully regularized
and well defined because the bi-local variables are on the lattice on which the original field theory is defined.
For a system made of $L$ sites, there exist $L(L+1)/2$ independent bi-local fields,
and \eq{eq:ONfinitez} has no UV divergence.
The exact RG evolution originally defined in the space of all couplings is now replaced with a  path integration  of fluctuating RG paths in the subspace of single-trace couplings 
\cite{lee2012background}.
This mapping is exact at any $N$.
In the large $N$ limit,
one can use the semi-classical approximation to replace the bulk path integration with a saddle-point approximation\footnote{
For an explicit computation of the effective action in the large $N$ limit, 
see Ref. \cite{future}.
}.
For alternative approaches to the $O(N)$ vector model, 
see Refs. \cite{Das:2003vw,Douglas:2010rc,Leigh:2014tza}.

\subsubsection{Full \bfnc}

$\hat H_{bulk}$ 
in \eq{eq:Hbulk_O(N)}
is entirely fixed by
the \bfnc defined in the space of single-trace couplings
with $J^{(K)}=0$ for $K>1$.
For the $O(N)$ vector model, only 
$\beta^{(0)}$
and $\beta^{(1)}_{ij}$
are non-zero in the subspace,
and $\hat H_{bulk}$ depends only on
$\beta^{(0)}[-it,0,..]$ and $\beta^{(1)}[-it,0,..]$.
Since the full RG flow is controlled by $\hat H_{bulk}$,
the full beta functions can be recovered from these \bfnc.
To see this,
let us write 
$\delta S = S(dz)-S$
in terms of the beta functions as
\begin{eqnarray}
\label{eq:action_general_z+dz}
\delta S =
 - dz N \sum_{K}
\sum_{i_1,j_1,\dots,i_K,j_K} 
\beta^{(K)}_{i_1,j_1,\dots, i_K,j_K}
\prod_{n=1}^K
{\cal O}_{i_n j_n}.
\label{eq:deltaSdef}
\end{eqnarray}
In quantum RG, 
$\delta S$ can be written as
\begin{eqnarray}
\delta S 
&=&
-\ln \frac{ 
\int  \mathcal{D} t 
~ \left[ e^{-N dz 
\hat{H}_{bulk}} \Psi_J[t] \right] e^{i N \sum_{ij}t_{ij} {\cal O}_{ij} }
}
{ 
\int \mathcal{D} t ~   \Psi_J[t]
e^{iN \sum_{ij} t_{ij} {\cal O}_{ij} } 
}.
\label{eq:deltaSQRG}
\end{eqnarray}
Equating   \eq{eq:deltaSdef} and \eq{eq:deltaSQRG},
we have
\begin{eqnarray}
\label{eq:action_general_z+dz2}
 \sum_{K}
\sum_{i_1,j_1,\dots,i_K,j_K} 
\beta^{(K)}_{i_1,j_1,\dots, i_K,j_K}
\prod_{n=1}^K
{\cal O}_{i_n j_n}
&=&
- \frac{ 
\int  \mathcal{D} t 
~ \left[  
\hat{H}_{bulk} \Psi_J[t] \right] e^{i N \sum_{ij}t_{ij} {\cal O}_{ij} }
}
{ 
\int \mathcal{D} t ~   \Psi_J[t]
e^{iN \sum_{ij} t_{ij} {\cal O}_{ij} } 
}.
\end{eqnarray}
In the large $N$ limit,
all single-trace operators are independent,
and the general beta functions are obtained by equating the coefficient of monomials of single-trace operators in 
\eq{eq:action_general_z+dz2},
\begin{eqnarray}
&& \beta^{(K)}_{i_1,j_1,\dots,i_K,j_K}\left( 
J^{(1)},
J^{(2)},
\dots
\right) \nn
&&
= 
-
\frac{1}{K!}
\left.
\left( \frac{\partial}{ \partial {\cal O}_{i_1 j_1}} \right)
\dots
\left( \frac{\partial}{ \partial {\cal O}_{i_K j_K}} \right)
\frac{ 
\int  \mathcal{D} t 
~ \left[ \hat{H}_{bulk} \Psi_J[t] \right] 
e^{i N \sum_{ij}t_{ij} {\cal O}_{ij} }
}
{ 
\int \mathcal{D} t ~   \Psi_J[t]
e^{iN \sum_{ij} t_{ij} {\cal O}_{ij} } 
}
\right|_{{\cal O}=0}.
\label{eq:ONbetaHPsi}
\end{eqnarray}
For a finite $N$,
not all single-trace operators are independent.
For example, 
$O_{ij} O_{kl} = O_{ik} O_{jl}$ for $N=1$.
This leads to  multiple ways to express general operators in terms of  the single-trace operators.
This is analogous to a gauge freedom in which one physical state can be represented in multiple ways.
Namely, $\delta S$ is gauge invariant,
but there are multiple ways to express it as a polynomial of 
 $\mathcal{O}_{ij}$.
In this case,  one has to fix 
the gauge freedom 
to determine the $\beta$-functions unambiguously.
The natural choice is to 
treat $\mathcal{O}_{ij}$ as independent variables in
\eq{eq:action_general_z+dz2}.
This is possible because both 
\eq{eq:deltaSdef} and \eq{eq:deltaSQRG} are functions of $\mathcal{O}_{ij}$ only.
In this prescription, 
\eq{eq:ONbetaHPsi} 
holds for any $N$.
This is not the only prescription, but 
\eq{eq:ONbetaHPsi} 
certainly gives the exact RG flow of the effective action
for any $N$.
The right-hand side of \eq{eq:ONbetaHPsi} 
depends only on 
$J^{(K)}_{i_1,j_1,\dots, i_K,j_K}$,
$\beta^{(0)}[-it,0]$ and $\beta^{(1)}[-it,0]$
through \eq{eq:ONgeneralPsi}
and \eq{eq:Hbulk_O(N)}.
This shows that 
all beta functions in the presence of general couplings
$J^{(K)}_{i_1,j_1,\dots, i_K,j_K}$
are completely fixed by
the beta functions defined in the subspace of the single-trace couplings.

From \eq{eq:ONbetaHPsi}, we can find the general expression for $\beta^{(K)}[J^{(1)},J^{(2)},\dots]$ from 
$\beta^{(0),(1)}[-it,0]$. 
Starting from the general action in \eq{eq:action_general_z} associated with the wave function in \eq{eq:ONgeneralPsi}, 
we obtain 
\begin{eqnarray}
\hat{H}_{bulk}\Psi_J [t] &=& \Big(-\beta^{(0)}[-i \hat{t},0]-\sum_{ij}\hat{p}_{ij}\beta^{(1)}[-i\hat{t},0]\Big)\Psi_J [t] \\
&=&\int \mathcal{D}p 
e^{   - i N t_{ij} p_{ij} }
\Big(-\beta^{(0)}[-\frac{1}{N}\frac{\partial}{\partial p},0]-\sum_{ij}{p}_{ij}\beta^{(1)}[-\frac{1}{N}\frac{\partial}{\partial p},0]\Big)
e^{
- N J^{(K)}_{i_1,j_1,\dots, i_K,j_K}
p_{i_1j_1} \dots
p_{i_Kj_K} 
}.
\nonumber
\end{eqnarray}
This gives 
\begin{eqnarray}
\int \mathcal{D }t \left[ \hat{H}_{bulk} \Psi_J[t] \right] 
e^{i N \sum_{ij}t_{ij} {\cal O}_{ij} }
&=&\Big(-\beta^{(0)}[-\frac{1}{N}\frac{\partial}{\partial \mathcal{O}},0]-\sum_{ij}\mathcal{O}_{ij}\beta^{(1)}[-\frac{1}{N}\frac{\partial}{\partial \mathcal{O}},0]\Big) \nonumber\\
&\times&~\exp \Big\{
- N J^{(K)}_{i_1,j_1,\dots, i_K,j_K}
\mathcal{O}_{i_1j_1} \dots
\mathcal{O}_{i_Kj_K} 
\Big\}.
\label{eq:ONHPsit}
\end{eqnarray}
In the O(N) vector model, $\beta^{(0)}$ and $\beta^{(1)}$ take the form of 
\begin{eqnarray}
\beta^{(0)} [J^{(1)},0] = \sum_{kl} {\beta}^{(0),1}_{kl} J^{(1)}_{kl} ,\quad \beta^{(1)}_{ij}[J^{(1)},0] =  {\beta}^{(1),1}_{ij} J^{(1)}_{ij} + \sum_{kl}  {\beta}^{(1),2}_{ijkl} 
J^{(1)}_{ik}
J^{(1)}_{jl},
\end{eqnarray}
where
\bqa
 {\beta}^{(0),1}_{kl} = -\frac{2}{m^2}\delta_{kl}, ~~~
 {\beta}^{(1),1}_{ij} = 2, ~~~
{\beta}^{(1),2}_{ijkl} = \frac{4}{m^2}\delta_{kl}.
\label{eq:betacomponentsON}
\eqa
Through \eq{eq:ONHPsit}
the full $\beta$-functions can be written solely in terms of
$ {\beta}^{(0),1}_{kl}$
$ {\beta}^{(1),1}_{ij}$ 
and 
$ {\beta}^{(1),2}_{ijkl}$
as
\begin{eqnarray}
\beta^{(K)}_{i_1,j_1,\dots,i_K,j_K} &=& 
 (K+1)\sum_{kl} {\beta}^{(0),1}_{kl} J^{(K+1)}_{k,l,i_1,j_1,\dots,i_{K},j_{K}}
+ K  {\beta}^{(1),1}_{i_1,j_1} J^{(K)}_{i_1,j_1,\dots,i_{K},j_{K}} \nonumber\\
&-& \frac{(K+1)K}{N}\sum_{kl}  {\beta}^{(1),2}_{i_1,j_1,k,l}
 J^{(K+1)}_{i_1,k,j_1,l,i_2,j_2\dots,i_{K},j_{K}}\nonumber\\
&+& \sum_{M=1}^KM(K+1-M) \sum_{kl}  {\beta}^{(1),2}_{i_1,j_1,k,l} J^{(M)}_{i_1,k,i_2,j_2,\dots,i_{M},j_{M}}J^{(K+1-M)}_{j_1,l,i_{M+1},j_{M+1},\dots,i_{K},j_{K}}
\label{eq:beta_from_single_trace}
\end{eqnarray}
It is noted that 
the full $\beta$-functions at general couplings are completely characterized by
 the data in \eq{eq:betacomponentsON}
that defines
the $\beta$-functions in the subspace of single-trace coupling\footnote{
As a special case, one can check that the $\beta$-functions in the presence of  $J^{(1)}_{ij}$ and $J^{(2)}_{iiii}$  are reproduced.}.
This shows that \bfnc away from the subspace of single-trace couplings is fixed by \bfnc in the subspace.

The $O(N)$ model is rather special in that
$\hat H_{bulk}$ is linear in $p_{ij}$\cite{doi:10.1142/S0217751X95001273},
and the single-trace coupling $t_{ij}$ is non-dynamical in the bulk.
However, the constraints among \bfnc
hold for general theories in which
 $\hat H_{bulk}$ is not linear in the conjugate momenta.
To see this, we consider a matrix model as our next example\cite{lee2012background}.

\subsection{A matrix model 
\label{sec:O(N)_matrix}}

As a next example,  
we consider 
a matrix model
defined  on a D-dimensional Euclidean lattice. 
The fundamental field is 
 a real $N \times N$ matrix field, $\phi_i^{aa'}$,
where $i$ is the lattice site 
and $a,a'=1,2,..,N$ are the flavour indices.
Under the global $O_L(N) \times O_R(N)$ symmetry, 
the matrix field transforms
as $\phi_i \rightarrow A \phi_i B$, where 
$A \in O_L(N)$
and 
$B \in O_R(N)$.
Single-trace operators 
that are invariant under 
$O_L(N) \times O_R(N)$ symmetry
are denoted as 
\bqa
\mathcal{ O}_{I}
=
\frac{1}{N}\textrm{tr}(
\phi_{i_1}\phi_{i_2}^T
\phi_{i_3}\phi_{i_4}^T
\dots 
\phi_{i_{2m-1}}
\phi_{i_{2m}}^T
),
\label{eq:Omatrixsingletrace}
\eqa
where the trace sums over the flavour indices,
and
$I= (i_1,i_2,\dots,i_{2m})$
is a short-hand notation for a series of sites that form a loop 
through a trace over flavour indices.
Because the trace is invariant under cyclic permutation and transpose,
$ (i_1,i_2,i_3,i_4,\dots,i_{2m-1},i_{2m}) = (i_3,i_4,\dots,i_{2m-1},i_{2m},i_1,i_2) $
and
$ (i_1,i_2,\dots,i_{2m}) = (i_{2m},i_{2m-1},\dots,i_2,i_1) $.
Henceforth, we refer to $I$ as a loop.
The set of single-trace operators
plays the special role because general $O(N)$ invariant operators can be written as polynomials of the single-trace operators as
\bqa
\mathcal{\bf O}_{
I_{1}, 
I_{2}, 
\dots,
I_{K} 
}
=
\prod_{n=1}^K 
\mathcal{O}_{I_n}.
\label{eq:generalOmatrix}
\eqa

\subsubsection{Classical RG}

The general action that is invariant under the symmetry  can be written as
\begin{eqnarray}
S(z^\ast) = S_0+
N^2
\sum_{K=1}^\infty
\sum_{\{I_1,\dots,I_K\}} 
J^{(K)}_{I_1,\dots,I_K}
\mathcal{O}_{I_1}\dots \mathcal{O}_{I_K},
\label{eq:generalONactionmatrix}
\end{eqnarray}
where
$S_0= \frac{N}{2} m^2
\textrm{tr}(\phi_i \phi_i^T)$ is the reference action,
and 
$J^{(K)}_{I_1,I_2,\dots, I_K}$ is the coupling for the $K$-trace operators.
Under the exact renormalization group flow,
the scale dependent couplings obey
\bqa
\frac{d J^{(K)}_{I_1,I_2,\dots, I_K}}{dz}
= -
\beta^{(K)}_{I_1,I_2,\dots, I_K} 
\left( 
J^{(1)},
J^{(2)},
\dots
\right),
\label{eq:ONmatrixfullbeta}
\eqa
where 
$\beta^{(K)}_{I_1,I_2,\dots, I_K}$
is the beta functions.
Even if one starts with a simple UV action, all multi-trace operators are generated under the RG flow.
Each coupling can be tuned independently at UV,
and the exact RG flow is encoded in the beta functions of all couplings defined in the presence of  general couplings.
Below, we show that the general $\beta$-functions are completely fixed by the \bfnc defined in the subspace 
with $J^{(K)}=0$ for $K>1$
as is the case for the $O(N)$ vector model.

\subsubsection{Quantum RG}

As is discussed in the previous section, the space of theories is identified as a Hilbert space.
The Hilbert space can be spanned by a set of basis states whose wavefunctions include only single-trace operators.
For the matrix model,
the basis states are chosen to be
\begin{eqnarray}
|t \rangle =\int \mathcal{D}\phi~
e^{-S_0
+ i N^2 \sum_{I} t_I \mathcal{O}_I }
~ |\phi\rangle,
\label{eq:tmatrix}
\end{eqnarray}
where $t_I$ is the source for the single-trace operator $\mathcal{O}_I$.
The quantum state associated with the general $O(N)$ invariant action in
\eq{eq:generalONactionmatrix} can be written as 
a linear superposition of the basis states as
\begin{eqnarray}
|S\rangle &=& \int \mathcal{D} t  ~\Psi^0_{J}[t]|t \rangle,
\end{eqnarray}
where $\Psi_{J}^0[t]$ is
the wave function 
defined in the space of  $t_I$,
\begin{eqnarray}
\Psi_{J}^0[t] &=&\int 
\mathcal{D}p
~\exp\Big\{
- iN^2 \sum_{I}t_{I}p_{I}
-
N^2
\sum_{K=1}^\infty
\sum_{\{I_1,\dots,I_K\}} 
J^{(K)}_{I_1,\dots,I_K}
p_{I_1}\dots 
p_{I_K}
\Big\}.
\label{eq:Psi_UV_matrix}
\end{eqnarray}
Here 
$\mathcal{D}t
= \prod_I \int_{-\infty}^\infty
dt_I$
represents the integration over each single-trace coupling along the real axis.
$p_I$ represents the field that is conjugate to $t_I$,
and
$\mathcal{D}p
= \prod_I \int_{-\infty}^\infty
dp_I$.
It is straightforward to check that the integration over $t_I$ and $p_I$ reproduce the original action : 
$| S \rangle = \int 
 \mathcal{D}\phi~e^{-S}
~ |\phi\rangle
$.

As is shown in the previous section, 
the exact real space RG flow  is generated by the RG Hamiltonian,
\begin{eqnarray}
\hat{H}=e^{-\hat S_0}\sum_i\left[i \textrm{tr}(\hat \phi_i\hat \pi_i)
+\frac{1}{N m^2}\textrm{tr}(\hat \pi_i \hat \pi_i^T)\right]e^{\hat S_0},
\end{eqnarray}
where 
$\hat \pi_i^{ba}$
is the  conjugate momentum of 
$\hat \phi_i^{ab}$.
In the $| \phi \rangle$ basis, $\hat \pi_i^{ba}
=-i\frac{\partial}{\partial \phi_i^{ab}}$. 
The renormalized action at scale $z$ is obtained by
$S(z) = - \log \langle \phi | S(z) \rangle$,
where
$| S(z) \rangle =
e^{-z \hat H} 
| S \rangle$.
Because the set of basis states $\{ | t \rangle \}$ is complete,
it is enough to know how each basis state is evolved under $e^{-z \hat H}$.
Furthermore,
$| S(z) \rangle$
can be also written as 
$|S(z)\rangle
= \int \mathcal{D} t 
~\Psi_{J}^{z}[t]
|t \rangle
$,
where
$\Psi_{J}^{z}[t]$
is the wave function at scale $z$.
Following the steps used in the previous section, it is straightforward to show that 
$\Psi_{J}^{z}[t]$
is related to 
$\Psi_{J}^{0}[t]$
through the evolution given by
\bqa
\Psi_{J}^{z}[t] = e^{-N^2 z \hat{H}_{bulk}} 
\Psi_{J}^0[t]
\label{eq:matrixevolution}
\eqa
where $\hat H_{bulk}$ is the induced RG Hamiltonian for the dynamical single-trace couplings $t_I$
and their conjugate momenta $p_I$, 
\begin{eqnarray}
\hat{H}_{bulk}=
-\beta^{(0)}[-i\hat{t},0]
-\sum_{I}\hat{p}_{I}\beta^{(1)}_{I}[-i\hat{t},0] 
-\sum_{I_1 I_2}\hat{p}_{I_1}
\hat{p}_{I_2}\beta^{(2)}_{I_1 I_2 }[-i\hat{t},0] 
.
\label{eq:ONmatrixH}
\end{eqnarray}
$\beta^{(0)}[J^{(1)},0]$, 
$\beta^{(1)}[J^{(1)},0]$,
and
$\beta^{(2)}[J^{(1)},0]$ 
are the $\beta$-functions
for the identity operator,
the single-trace operators
and the double-trace operators,
respectively,
in the presence of single-trace couplings $J^{(1)}$ only,
\bqa
\beta^{(0)}[J^{(1)},0]
& = & 
- \sum_i \frac{2}{m^2} J^{(1)}_{(ii)}, \nn
\beta^{(1)}_I[J^{(1)},0]
& = & 
 n_I J^{(1)}_I 
- \frac{2}{m^2}
\sum_{I' \in 
L_{I+2}
}
J^{(1)}_{I'} 
- \frac{2}{N m^2}
\sum_{I' \in 
L_{I+2}'
}
J^{(1)}_{I'} 
+ \frac{1}{m^2} \sum_{(I',I'') \in
L_I^+
}
J^{(1)}_{I'} J^{(1)}_{I''},
\nn
\beta^{(2)}_{I_1 I_2}[J^{(1)},0]
& = & 
- \frac{2}{m^2}
\sum_{I'\in 
L_{I_1,I_2}^-
} J^{(1)}_{I'}.
\label{eq:ONmatrixbetasingle}
\eqa
$\hat t_I$ and $\hat p_{I'}$ obeys the commutation relation,
$[ \hat t_I, \hat p_{I'} ]=
-\frac{i}{N^2} 
\delta_{I,I'}
$,
where
$\delta_{I,I'}$ denotes the Kronecker delta function defined in the space of loops.
$J^{(1)}_{(ii)}$ 
denotes the source of
$\textrm{tr}(\phi_i
\phi_i^T)$.
$n_I$ denotes the number of $\phi$ fields in $\mathcal{O}_I$.
$L_{I+2}$  denotes 
the set of loops that can be made by adding two identical sites  to $I$ consecutively.
$L_{I+2}'$  denotes 
the set of loops that can be made by adding two identical sites 
at two even positions 
or at two odd positions 
of loop $I$.
 For 
 $I= (i_1,i_2,i_3,i_4)$,
 \bqa
 && L_{I+2} = 
 \Bigl\{  
 (j,j,i_1,i_2,i_3,i_4),
 (i_1,j,j,i_2,i_3,i_4),
 (i_1,i_2,j,j,i_3,i_4),
 (i_1,i_2,i_3,j,j,i_4)
 ~\Big|~
 j \in {\bf R}
 \Bigr\}, \nn
 && L_{I+2}' = 
 \Bigl\{  
 (j,i_1,j,i_2,i_3,i_4),
 (i_1,j,i_2,j,i_3,i_4),
 (i_1,i_2,j,i_3,j,i_4),
 (i_1,i_2,i_3,j,i_4,j)
 ~\Big|~
 j \in {\bf R}
 \Bigr\}, \nn
 \eqa 
 where ${\bf R}$
 represents  all possible sites.
$L_I^+$ is the set of pairs of loops that can be merged into loop $I$ by removing one common site from each of the two loops.
For example,
for 
$I= (i_1,i_2,i_3,i_4)$,
\bqa
&& L_I^+
=
\Biggl\{
\Bigl(
(i_1,i_2,i_3,j),
(j,i_4)
\Bigr),
\Bigl(
(i_4,i_1,i_2,j),
(j,i_3)
\Bigr),
\Bigl(
(i_3,i_4,i_1,j),
(j,i_2)
\Bigr), 
\Bigl(
(i_2,i_3,i_4,j),
(j,i_1)
\Bigr), 
\nn
&&
\Bigl(
(j,i_4),
(i_1,i_2,i_3,j)
\Bigr),
\Bigl(
(j,i_3),
(i_4,i_1,i_2,j)
\Bigr),
\Bigl(
(j,i_2),
(i_3,i_4,i_1,j)
\Bigr), 
\Bigl(
(j,i_1),
(i_2,i_3,i_4,j)
\Bigr)
\Big|
 j \in {\bf R}
 \Biggr\}. \nn
\eqa
Finally,
$L_{I_1,I_2}^-$
denotes the set of loops that can be split into two loops $I_1$ and $I_2$ by removing two identical sites, 
one at an even position
and the other at an odd position.
For
$I_1= (i_1,i_2)$
and
$I_2= (i_3,i_4)$,
$L_{I_1,I_2}^-
=
\Bigl\{
(i_1,i_2,j,i_3,i_4,j)
|
 j \in {\bf R}
 \Bigr\}$.
 In the path integral representation,
\eq{eq:matrixevolution} can be written as\cite{lee2012background,LEE2012781}
\bqa
\Psi^{z}_J[t^z]
= 
\int 
\mathcal{\bf D} t 
\mathcal{\bf D} p 
~
e^{- N^2 S_{bulk}} 
~\Psi^{0}_J[t^0],
\eqa
where 
$\int 
\mathcal{\bf D} t 
\mathcal{\bf D} p 
=
\int 
\prod_{0\leq z' < z} \mathcal{D} t^{z'}
\prod_{0 < z' \leq z} \mathcal{D} p^{z'}
$ represents the sum over RG paths in the space of single-trace couplings
and 
\bqa
S_{bulk} = 
\int_0^z dz'
\left[
i \sum_I p_I^{z'} \partial_{z'} t_I^{z'}
+ H_{bulk}[ p_I^{z'}, t_I^{z'}]
\right]
\eqa
is the bulk action.
The bulk theory, which is fully regularized
and well defined, 
describes dynamics of loop variables in the bulk.
Unlike the vector model, 
the bulk action is quadratic in $p_I$
and the loop variables are genuinely dynamical in the bulk\footnote{
There are no higher order terms in $p_I$ because double-trace operators are generated,
but triple or higher trace operators are not generated
in the subspace of single-trace couplings.}.
While the mapping itself is exact at any $N$, the bulk theory becomes weakly interacting only in the large $N$ limit.
For general $N$, 
one has to solve the quantum theory of strongly interacting loops, which is a hard problem.
However, one can extract the general constraints that \bfnc obey
without solving the full quantum problem for general $N$.

\subsubsection{Full \bfnc}

The bulk Hamiltonian is given by the beta functions in the presence of the single-trace couplings only\cite{lee2012background,lee2014quantum}.
Because the evolution generated by
\eq{eq:ONmatrixH}
contains the full information of the exact RG flow,
all beta functions can be extracted from
$\beta^{(0)}[-i\hat{t},0]$,
$\beta^{(1)}[-i\hat{t},0]$ and
$\beta^{(2)}[-i\hat{t},0]$.
The change of the effective action under an infinitesimal RG transformation, 
\begin{eqnarray}
\delta S =
- N^2 dz \sum_{K} 
\sum_{\{I_1,\dots,I_K\}} 
\beta^{(K)}_{I_1,\dots,I_K}[J^{(1)},
J^{(2)},
\dots]
\mathcal{O}_{I_1}
\mathcal{O}_{I_2}
..
\mathcal{O}_{I_K}
,
\label{eq:deltaSmatrixON}
\end{eqnarray}
can be also written as
\begin{eqnarray}
\delta S 
&=&
-\ln \frac{ 
\int  \mathcal{D} t 
~ \left[ e^{-N^2 dz \hat{H}_{bulk}} \Psi_J[t] \right] e^{i N^2 \sum_{I}t_{I} {\cal O}_{I} }
}
{ 
\int \mathcal{D} t ~   \Psi_J[t]
e^{iN^2 \sum_{I} t_{I} {\cal O}_{I} } 
}
\label{eq:deltaSQRGmatrix}
\end{eqnarray}
in quantum RG.
Equating 
\eq{eq:deltaSmatrixON}
and
\eq{eq:deltaSQRGmatrix},
the general beta functions can be extracted from $\hat H_{bulk}$ as
\bqa
&&
\beta^{(K)}_{I_1,I_2,\dots,I_K}
\left( 
J^{(1)},
J^{(2)},
\dots
\right)  \nn
&&
= 
-
\frac{1}{K!}
\left.
\left( \frac{\partial}{ \partial {\cal O}_{I_1}} \right)
\dots
\left( \frac{\partial}{ \partial {\cal O}_{I_K }} \right)
\frac{ 
\int  \mathcal{D} t 
~ \left[ \hat{H}_{bulk} \Psi_J[t] \right] e^{i N^2 \sum_{I}t_{I} {\cal O}_{I} }
}
{ 
\int \mathcal{D} t ~   \Psi_J[t]
e^{iN^2 \sum_{I} t_{I} {\cal O}_{I} } 
}
\right|_{{\cal O}=0}.
\label{eq:betarelationONmatrix}
\eqa
From
\begin{eqnarray}
&& \int \mathcal{D}t \left[\hat{H}_{bulk}\Psi_J[t] \right]e^{iN^2\sum_I t_I \mathcal{O}_I}
= \Bigg(-\beta^{(0)}\left[-\frac{1}{N^2}\frac{\partial}{\partial\mathcal{O}},0\right]-\sum_I \mathcal{O}_I\beta_I^{(1)}\left[-\frac{1}{N^2}\frac{\partial}{\partial\mathcal{O}},0\right] \nn
&& 
\hspace{1cm}
-\sum_{I_1,I_2}\mathcal{O}_{I_1}\mathcal{O}_{I_2}\beta^{(2)}_{I_1,I_2}\left[-\frac{1}{N^2}\frac{\partial}{\partial\mathcal{O}},0\right]\Bigg)  
~e^{- N^2\sum_K \sum_{I_1,\dots,I_K}J^{(K)}_{I_1,\dots, I_K}\mathcal{O}_{I_1}\dots \mathcal{O}_{I_K}},
\end{eqnarray}
the general $\beta$-functions
are obtained to be
\begin{eqnarray}
\beta^{(0)} & = & -  \frac{2}{m^2}\sum_i    J_{(ii)}^{(1)}, \nn
\beta^{(K>0)}_{I_1,\dots,I_K}
&=& 
-  (K+1) \frac{2}{m^2}\sum_i   J_{(ii),I_1,\dots,I_{K}}^{(K+1)}
+    
\left( \sum_{t=1}^K n_{I_t}  \right)
J^{(K)}_{I_1,I_2,\dots,I_{K}} 
-K \frac{2}{m^2}   \sum_{I'\in L_{I_1+2}} J^{(K)}_{I',I_2,\dots,I_{K}}  \nn
&-&
 K \frac{2}{N m^2}   \sum_{I'\in L_{I_1+2}'} J^{(K)}_{I',I_2,\dots,I_{K}}  
 -K(K+1) \frac{1}{m^2N^2} \sum_{I',I''\in L^+_{I_K}}  J^{(K+1)}_{I',I'',I_1,\dots,I_{K-1}}  \nonumber\\
&+& \frac{1}{m^2}  \sum_{I',I''\in L^+_{I_K}}\sum_{M=1}^{K}(K-M+1)M J^{(K-M+1)}_{I'',I_1,\dots,I_{K-M}}J^{(M)}_{I',I_{K-M+1},\dots,I_{K-1}}
\nonumber\\
&-&(K-1) \frac{2}{m^2}\sum_{I'\in L^-_{I_1I_2}}
 J^{(K-1)}_{I',I_3,\dots,I_{K}}.
\label{eq:generalbetaONmatrix}
\end{eqnarray}
It is noted that the full $\beta$-functions in
\eq{eq:generalbetaONmatrix}
are entirely determined from
$\beta^{(0)}[-i\hat{t},0]$,
$\beta^{(1)}[-i\hat{t},0]$ and
$\beta^{(2)}[-i\hat{t},0]$.

We close this section with a few remarks.
First,
the \bfnc in
\eq{eq:beta_from_single_trace}
and \eq{eq:generalbetaONmatrix},
which have been
derived from the \bfnc defined on the subspace of single-trace couplings,
are exact for any $N$.
The validity of the mapping from the exact Wilsonian RG to quantum RG
and the constraints that  are derived from it 
do not require that the bulk theory is in the semi-classical limit.
Second, the constraints can be extended to general theories because the notion of single-trace operators can be defined in any theory.
This follows from the fact that the space of theories can be in general viewed as a Hilbert space, 
where an action $S[\phi]$ for fundamental field $\phi$ defines a wavefunction $e^{-S[\phi]}$ in the Hilbert space.
Furthermore,
there exists a set of basis states that span the Hilbert space.
In general, there exist multiple choices of basis states.
The basis states do not need to be orthogonal, 
and an over-complete set is an acceptable choice.
All we need for quantum RG is one choice of complete basis states
\footnote{
It may well be the case that one choice of basis states gives a simpler bulk theory than others.
}.
For the vector model and the matrix model, the complete set of basis states are given in
\eq{eq:t}
and 
\eq{eq:tmatrix}, respectively.
Once a complete set of  basis is chosen, the wavefunctions 
of the basis states define a set of actions.
The operators that are needed to construct the wavefunctions of the basis states 
define the set of single-trace operators 
 in general theories.
They are given in 
\eq{eq:Oij}
and 
\eq{eq:Omatrixsingletrace}
for the vector model 
and the matrix model, respectively.
Because the Hilbert space structure and the basis states can be defined in any theory,
 there exist constraints among 
$\beta$-functions in general theories.
The generalization is discussed in the next section.
Finally, 
the full \bfnc could have been computed directly from the exact Polchinski RG equation in \eq{eq:ONgeneralbeta}.
The salient point of our paper is not the exact \bfnc itself but the fact that the entire \bfnc are fully characterized by a small set of data defined in the subspace of single-trace couplings. 
As a result,
it is impossible
to change a theory or RG scheme
such that the $\beta$-functions away from the subspace of single-trace couplings are modified without modifying \bfnc in the subspace. 
Because multi-trace operators are composites of the single-trace operators,
the RG flow in the presence of  general multi-trace operators are completely fixed by the \bfnc defined in the subspace of single-trace couplings.
This constraint holds even when multi-trace operators have large anomalous dimensions.

\section{Generalization\label{sec:main}
}

In this section,
we generalize the results  obtained for the two concrete models in the previous section.
Let us consider a field theory in the $D$-dimensional Euclidean space with the partition function,
\begin{equation}
Z  =\int \mathcal{D} \phi ~ e^{-S \left[ \phi \right]},
\label{eq:partition_function_general0}
\end{equation}
where $\phi(x)$ represents a set of fundamental fields and 
$S$ is an action that is invariant under a symmetry group $G$.
The RG flow is defined in the space of local theories in a given symmetry sector.
To describe the RG flow, 
one first needs to coordinatize the space of theories.
For this, we divide $S$ into a reference action $S_0$ and a deformation,
\begin{equation}
S =  S_0 \left[ \phi \right] +
\sum_M \int d^D x  ~ J_{M}(x) \mathcal{\bf O}_M (x).
\label{eq:partition_function_general1}
\end{equation}
Here the reference action $S_0$  sets the origin in the space of theories.
$\{ \mathcal{\bf O}_M(x) \}$ represents the complete set of local operators that are invariant under the symmetry $G$.
We call these operators `symmetry-allowed operators'.
$J_M(x)$ is the coupling function that deforms the reference theory.
One infinitesimal step of coarse graining consists of 
integrating out fast modes of the fundamental fields,
and rescaling the space and fields.
The one cycle of coarse graining puts the theory into the same form as before except for renormalized sources, 
\begin{equation}
Z =  \int \mathcal{D} \phi ~ e^{- S_0 \left[ \phi \right] -
\sum_M \int d^D x \Big( 
J_{M}(x) 
- \beta_{M}[J;x] dz 
\Big)
\mathcal{\bf O}_M (x) },
\label{eq:partition_function_general2}
\end{equation}
where $dz$ is an infinitesimal parameter
and $ \beta_{M}[J;x]$ is the beta function
for the local operator $\mathcal{\bf O}_M (x)$.
$ \beta_{M}[J;x]$ is a function of $x$ and 
a functional of coupling functions $J_N(x)$.
In weakly coupled field theories,
one may ignore operators whose couplings remain small
in the perturbative series.
In general, one has to keep all couplings. 
Successive applications of the coarse graining
give rise to the exact Wilsonian renormalization group (RG) flow  
in the infinite dimensional space of couplings\cite{POLCHINSKI1984269}.

Alternatively, the RG flow can be projected to a subspace of couplings 
at the price of promoting the deterministic RG flow 
to a path integration over RG paths (quantum RG) within the subspace\cite{lee2012background,lee2014quantum}.
To see this, we define a quantum state from the action
by promoting the Boltzmann weight to a wave function
as in \eq{eq:Sstate} \cite{lee2016horizon}.
This correspondence between a $D$-dimensional action
and a $D$-dimensional quantum state is not the same as the correspondence between 
a $D$-dimensional action and the ground state defined 
on a $(D-1)$-dimensional slice with a fixed imaginary time.
As we pointed out in Eq.~(\ref{eq:Z1S}), the partition function in  \eq{eq:partition_function_general0} can be written as
an overlap between two states,
$
Z = \langle \mathbbm{1} |  S \rangle
$,
where
$| \mathbbm{1} \rangle $ is defined in \eq{eq:1state},
represents the trivial fixed point with zero correlation length. 
In this picture, one infinitesimal step of coarse graining is generated
by a quantum operator inserted between the overlap of $\langle \mathbbm{1}|$ and  $|S\rangle$. Here, we rewrite the Eq.~(\ref{eq:Z1HS}) for convenience,
\bqa
Z = \langle \mathbbm{1} | e^{-dz H } |  S \rangle, 
\label{eq:0RS}
\eqa
where $H$ is the RG Hamiltonian 
that generates the coarse graining transformation
that satisfies $\langle \mathbbm{1} | H = 0$\cite{lee2016horizon}.
A concrete example of RG Hamiltonian is discussed in  Sec.~\ref{sec:Wilsonian_RG}.
Since $| \mathbbm{1} \rangle$ is invariant under the evolution generated by $H$,
the partition function remains the same under the insertion of the operator.
Nonetheless, $H$ generates a non-trivial evolution of $| S \rangle$ 
once it is applied to the right in   \eq{eq:0RS}
\footnote{
In more generality, 
one may choose 
$| S_* \rangle 
= \int \mathcal{D}\phi ~ e^{-S_*} |\phi \rangle$ 
with a different fixed point action $S_*$
instead of  
$| \mathbbm{1} \rangle $\cite{lee2016horizon}. 
In this case, the partition function is written as
$Z = \langle S_* |  S_1 \rangle$,
where $S_1 = S-S_*$ is the deformation measured with respect to $S_*$.
In this case, the coarse graining Hamiltonian that satisfies 
$\langle S_* | H' = 0$ is 
related to $H$ through a similarity transformation,
$H' = e^{S_*} H e^{-S_*}$. 
}.

The one-to-one correspondence between states and actions implies
that the resulting state corresponds to a renormalized action, 
\bqa
e^{-dz H } |  S \rangle =  |  S + \delta S \rangle,
\label{eq:eHSdS}
\eqa
where 
\bqa
\delta S = 
- dz \sum_M \int dx \beta_{M}[J;x]  \mathcal{\bf O}_M (x).
\eqa
Successive applications of the coarse graining transformations
give rise to a scale dependent quantum state 
which corresponds to the scale dependent Wilsonian action,
\bqa
e^{-z H } |  S \rangle = 
  \int \mathcal{D} \phi ~ e^{- S_0 \left[ \phi \right] -
\sum_M \int d^D x
J_{M}(x,z) 
\mathcal{\bf O}_M (x) }
| \phi \rangle,
\label{eq:zRS}
\eqa  
where $J_M(x,z)$ is the renormalized coupling function that satisfies
\bqa
\frac{\partial J_{M}(x,z)}{\partial z} = - \beta_{M}[J;x]
\label{eq:Wilsonbeta}
\eqa
with the initial condition $J_{M}(x,0) = J_{M}(x)$.
In \eq{eq:zRS}, $J_{M}(x,z)$'s are classical parameters 
that keep track of the exact Wilsonian RG flow.
However,  \eq{eq:Wilsonbeta} is a rather inefficient way 
of keeping track of the evolution of quantum state in that 
the number of classical parameters one needs to keep
is far greater than the number of linearly independent quantum states. 

Once we realize that the space of theories can be viewed as a vector space,
a more natural description of the RG flow is to take advantage of the 
linear superposition principle.
Instead of labeling a quantum state in terms of classical parameters,
a quantum state is expressed as a linear superposition of basis states,
which as a set is much smaller than  $\{ J_M(x) \}$.
A complete set of basis states can be chosen to be
\bqa
| t \rangle = 
  \int \mathcal{D} \phi ~ e^{- S_0 \left[ \phi \right] + i 
\sum_{m} \int d^D x ~
t_{m}(x) 
\mathcal{O}_m (x) }
| \phi \rangle,
\label{eq:basis}
\eqa 
where $\{ \mathcal{O}_m(x) \}$ is a subset of symmetry-allowed local operators
from which all symmetry-allowed local operators can be written as composites,
\bqa
\mathcal{\bf O}_M(x) = 
\mathcal{O}_{m_1}(x) 
\left[ \partial_{\mu_1}..\partial_{\mu_{l_1}} \mathcal{O}_{m_2}(x) \right]
\left[ \partial_{\mu_1}..\partial_{\mu_{l_2}} \mathcal{O}_{m_3}(x) \right]
..
\left[ \partial_{\mu_1}..\partial_{\mu_{l_{k-1}}} \mathcal{O}_{m_k}(x) \right].
\label{eq:multiO}
\eqa
We call this subset of operators single-trace operators
because they are the set of operators that involve
one trace in  large $N$ matrix models.

In the example of the scalar field theory with the $Z_2$ symmetry, the  single-trace operators are given by 
 the set of quadratic operators, $\mathcal{O}_{\mu_1,\mu_2,..,\mu_n}(x) =
\phi(x) 
\partial_{\mu_1}
\partial_{\mu_2}
..
\partial_{\mu_n}
\phi(x)$.
Other $Z_2$ invariant operators that are quartic or higher order in $\phi$ can be written as composites of the single-trace operators, and they are called multi-trace operators.
It is noted that the distinction between single-trace and multi-trace operators depends not only on the field content of the theory but also on the symmetry.
In the absence of the $Z_2$ symmetry, the fundamental field $\phi$ is the only single-trace operator, and everything else is regarded as multi-trace operator.
We emphasize that operators can be divided into single-trace operators and multi-trace operators in any theory.

It is straightforward to see that \eq{eq:basis} forms a complete basis,
and \eq{eq:Sstate} with \eq{eq:partition_function_general1} can be written as \cite{lee2012background,lee2014quantum}
\bqa
| S \rangle = \int \mathcal{D} t ~ \Psi_J[t] | t \rangle, 
\label{eq:Sinj}
\eqa
where $\Psi_J[t]$ is a wavefunction defined in the space of the single-trace sources. 
In the next section, 
we will provide a prescription 
to find $\Psi_J[t]$
from a general action.
The measure and the integration path of the sources in \eq{eq:Sinj} will be carefully defined
when we discuss examples in the following sections.
The state for a general theory with multi-trace operators can be written as 
a linear superposition of those whose wavefunctions only include single-trace operators. 
The multi-trace operators that are not explicitly included in the description
endows the single-trace couplings with quantum fluctuations. 

\begin{figure}[h]
    \centering
    \includegraphics[width=.8\textwidth]{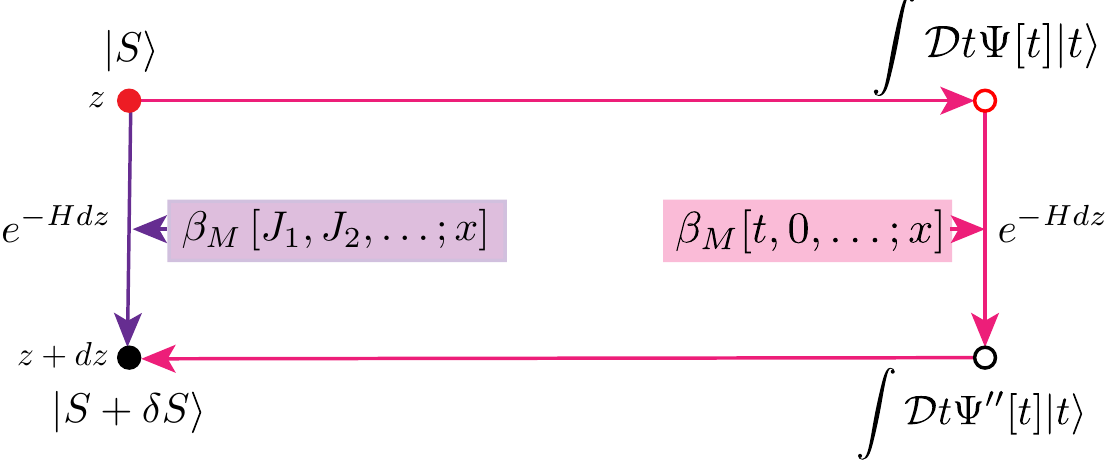}
    \caption{
    A field theory $S$ at scale $z$ is represented by a quantum state $| S \rangle$ on the upper left corner.
    The vertical arrows represent
    an infinitesimal step of coarse graining generated by $H$.
    The coarse graining directly applied to $| S \rangle$, denoted as the vertical arrow on the left, maps 
    $ | S \rangle$ to 
    $ | S + \delta S \rangle$.
    $  S + \delta S $ is 
    the renormalized action at scale $z+dz$, and
    $\delta S$ encodes the information on the full $\beta$-functions.
   Alternatively, $| S \rangle$ 
   is first written as a linear superposition of basis states whose action only includes single-trace deformations through the horizontal arrow that points to the right.
   The result of the coarse graining applied to that state, denoted as the vertical arrow on the right, only depends on 
the $\beta$ functions defined in the space of single-trace couplings. 
The resulting coarse grained state  on the lower right corner is finally mapped back to a renormalized action on the lower left corner.
    The commutativity of the  diagram implies the full $\beta$-functions are fixed by the $\beta$-functions defined in the subspace of single-trace couplings.
    }
    \label{fig:equivalent_RG}
\end{figure}

Because  $H$ is a linear operator,
the RG flow 
in \eq{eq:eHSdS}
is entirely fixed by how 
the coarse graining operator acts on the basis states,
\bqa
e^{-dz H }  |S \rangle & = &
\int \mathcal{D}t ~\Psi_J\left[t\right]  e^{-dz H}  |t\rangle.
\label{eq:RG_wave_function}
\eqa
To figure out the resulting state,
we simply use the expression in \eq{eq:zRS}
by turning off all multi-trace sources.
 From \eq{eq:basis}, we obtain
\bqa
 e^{-dz H } | t \rangle  
&=&
  \int \mathcal{D} \phi ~ 
  e^{- S_0 \left[ \phi \right] +i
\sum_{n} \int d^D x~
t_{n}(x) 
\mathcal{O}_m (x)
- dz \sum_M \int d^Dx 
 \beta_{M}[-it,0,..;x] 
\mathcal{\bf O}_M (x)
 }
| \phi \rangle.
\label{eq:edzRj}
\eqa
Here
the beta functions are expressed as
 $\beta_M[J;x] = \beta_{M}[-it,-it^{[2]},..;x] $,
 where $-it$ is the single-trace coupling
 and $-it^{[k]}$'s with $k \geq 2$ are the couplings for $k$-trace 
 operators which are composites of $k$ single-trace operators. 
$ \beta_{M}[-it,0,..;x] $ is the beta function 
defined in the subspace of single-trace sources.
Because $\{ | t \rangle \}$ is complete,
\eq{eq:edzRj} can in turn be expressed as a linear superposition of $|t \rangle$ as
\bqa
e^{-dz H } | t \rangle = \int \mathcal{D}t'~
\Phi_{dz}\left[t; t'\right] |t'\rangle,
\label{eq:jPhi}
\eqa
where 
$\Phi_{dz}\left[t; t'\right]$
is a propagator of the RG transformation that is determined from $\beta_{M}[-it,0,..;x] $.
This allows us to write the resulting state after a coarse graining transformation as 
\bqa
e^{-dz H }  |S \rangle & = &
\int \mathcal{D}t'_{}\Psi'' \left[t'_{}\right] |t'\rangle,
\label{eq:RG_wave_function2}
\eqa
where
\begin{equation}
\Psi''\left[t'\right]= \int \mathcal{D}t ~ \Psi_J\left[t\right] \Phi_{dz}\left[t; t'\right].
\end{equation}
In the end, the RG transformation leads to the evolution of wave function,  
$\Psi''  = e^{- \mathcal{ H} dz}  \Psi $,
where 
$\mathcal{ H}$  is an induced coarse graining operator defined by
$\mathcal{ H} \Psi [t']
=  -\frac{1}{dz}
\Bigl(
\int \mathcal{D}t \Psi\left[t\right] 
\Phi_{dz}\left[t; t'\right]
- \Psi[t']
\Bigr)
$.
By  equating  
\eq{eq:RG_wave_function} 
with \eq{eq:eHSdS}, we obtain
\bqa
e^{-dz H}  |S \rangle & = &
\int \mathcal{D} \phi ~  e^{- S_0}
\left[ \int \mathcal{D}t'(x)
 \Psi'' \left[t' \right] 
  e^{ i
\sum_{n} \int d^D x~
t'_{n}(x) 
\mathcal{O}_m (x) }
\right]
| \phi \rangle  \nn
&=&
\int \mathcal{D} \phi 
  e^{- S_0 -
   \sum_M \int dx 
   \Big(
   J_{M}(x) - dz   \beta_{M}[J;x] \Big)  \mathcal{\bf O}_M (x) }
| \phi \rangle.
\eqa
This shows that 
\bqa
\int \mathcal{D}t'
\mathcal{D}t \Psi_J\left[t\right] 
\Phi_{dz}\left[t; t'\right]
  e^{ i
\sum_{n} \int d^D x~
t'_{n}(x) 
\mathcal{O}_m (x) }
=
  e^{ -
   \sum_M \int dx 
   \Big(
   J_{M}(x) - dz   \beta_{M}[J;x] \Big)  \mathcal{\bf O}_M (x) }.
 \label{eq:mainresult}
\eqa
\eq{eq:mainresult} is the main result of the paper.
On the left hand side of \eq{eq:mainresult},
$\Psi_J[t]$ is fixed by the theory at scale $z$ through \eq{eq:Sinj},
and $\Phi_{dz}\left[t; t'\right]$ is entirely determined from  $ \beta_{M}[-it,0,..;x]$ 
through Eqs. (\ref{eq:edzRj}) and (\ref{eq:jPhi}).
This, in turn, fixes the full beta functions $ \beta_{M}[J;x] $ through \eq{eq:mainresult}.
{\it Therefore, the beta functions  defined in the subspace of the single-trace operators
completely fix the full beta functions away from the subspace.}
This is illustrated in \fig{fig:equivalent_RG}.

\section{
Quantum Renormalization group
\label{sec:QRG}}

In this section, 
we lay out an algorithm for extracting general scaling operators, scaling dimensions and operator product expansion coefficients from \bfnc defined in the subspace of single-trace operators.

\subsection{Action-state correspondence
\label{sec:correspondence_QFT}}

To be concrete, we consider a partition function given by
\begin{equation}
Z \left[ J_{1},J_{2},\dots \right] = \langle \mathbbm{1}|S_{J_1,J_2,\dots}\rangle,
\label{eq:partition_function_general}
\end{equation}
where 
\begin{equation}
|S_{J_{1},J_{2},\dots} \rangle =\int \mathcal{D}\phi ~  
e^{ -S_0 -\int d^D x \Big( \sum_{n } J_{n}(x)\mathcal{O}^n  \left[\phi (x)\right]+\dots\Big)}  
|\phi \rangle.
\label{eq:SJJ}
\end{equation}
Here $S_0$ is the reference action.
$\mathcal{O}(x)$ is a real and local single-trace operator.
$\mathcal{O}^n (x)$ with $n>1$ represents local multi-trace operators.
The ellipsis includes multi-trace operators with derivatives as is shown \eq{eq:multiO}.
We assume that the deformation is bounded  from below, and 
the highest multi-trace operator is an even power of $\mathcal{O}$ with a positive coupling. 
We can remove the multi-trace operator  in the action by using an identity,
\bqa
f( \mathcal{O}(y) ) 
&=&\int_R \mathcal{D}t' \mathcal{D} p' ~ e^{-i\int d^D x ~ t'(x) \left[ p'(x)-\mathcal{O}(x)\right]} 
  f( p'(y) ) 
 \nn
&=&\int_I \mathcal{D}t \mathcal{D} p ~ e^{-i\int d^D x ~ t(x) \left[ p(x)-i\mathcal{O}(x)\right]} f( -i p(y) ),
\eqa
for any $f(x)$.
The integration of $t'(x)$ and $p'(x)$ are defined along the real axes. 
In the second line,
we define
$t(x) = -i t'(x)$ and $p(x)= i p'(x)$
so that the integration for $t(x)$ and $p(x)$ are defined along the imaginary axes\footnote{
This Wick's rotation has the advantage that the source for  $\mathcal{O}$ is simply represented by $t(x)$.
}.
The path of the integration variables is denoted by the subscripts $R$ (real) and $I$ (imaginary).
Then, \eq{eq:SJJ} can be rewritten as
\begin{equation}
\begin{aligned}
|S_{J_{1},J_{2},\dots} \rangle 
&=\int \mathcal{D} \phi  ~
e^{-S_0}
 \int_I  \mathcal{D}t  \mathcal{D} p~  e^{-i\int d^D x  t(x) \left[ p(x)-i\mathcal{O}(x)\right]} e^{-  \int d^D x \Big( \sum_{n }  J_{n}(x) \left[-ip(x)\right]^n  + ...\Big) } |\phi \rangle\\
&=\int_I   \mathcal{D}t  \mathcal{D} p ~ e^{ -\int d^D x \Big(i  t(x) p(x)+\sum_{n}  (-i)^n J_{n}(x) p^n (x)  + ...\Big)  } \left[ \int \mathcal{D} \phi ~ 
e^{-S_0 -\int d^Dx~ t(x)  \mathcal{O}(x) }|\phi\rangle\right].
\label{eq:correspondence_partition_function_wave_function}
\end{aligned}
\end{equation}
This implies that the state can be written as a linear superposition of the basis states as
\begin{equation}
|S_{J_{1},J_{2},\dots} \rangle = \int_I \mathcal{D} t ~  \Psi_{J_{1},J_{2},\dots} \left[t\right] |t \rangle,
\label{eq:basis_jx}
\end{equation}
where
\begin{equation}
|t \rangle 
=\int \mathcal{D}\phi ~   
|\phi \rangle T_{\phi,t}
\end{equation}
is the complete basis state whose wavefunction is made of the reference action and the single-trace deformation only,
\bqa
T_{\phi,t}=e^{-S_0-\int d^D x ~ t (x) \mathcal{O} (x)},
\eqa
and
\begin{equation}
\Psi_{J_{1},J_{2},\dots } \left[ t \right] = \int_I \mathcal{D} p~ 
e^{ -\int d^D x \Big(i  t(x) p(x)+ \sum_{n}  (-i)^n J_{n}(x) p^n(x) + ... \Big)  }
 \label{eq:wave_function_general}
\end{equation}
is the wavefunction of the dynamical single-trace source. 
The integration over $p$ in \eq{eq:wave_function_general}
 is convergent for deformations that are bounded from below.
This allows us to write the original partition function
in terms of partition functions that involve only single-trace deformations,
$Z \left[ J_{1},J_{2},\dots  \right]=\int_I   \mathcal{D}t ~ \Psi_{J_{1},J_{2},\dots} \left[ t \right] Z \left[ t,0, \dots \right]$ with $\langle \mathbbm{1} | t \rangle = Z\left[t,0,\dots\right]$.

Let us first consider a simple example with an ultra-local deformation (no derivative terms) with $J_{n>2}  =0$.
In this case, the wave function is Gaussian,
\begin{equation}
\begin{aligned}
\Psi_{J_{1},J_{2}} \left[ t \right] &= \int_I \mathcal{D} p~   e^{- \int d^D x \Big(i  t p-iJ_{1} p -J_{2} p^2 \Big)  }
=\left[\prod_x \sqrt{\frac{\pi}{J_{2}(x)}}\right] e^{\int d^D x \frac{\left[t(x)- J_{1}(x) \right]^2}{4J_{2}(x)}}. 
\label{eq:wavefunction_gaussian_general}
\end{aligned}
\end{equation}
We note that $t$ in \eq{eq:wavefunction_gaussian_general}
is to be integrated along the imaginary axis,
and the wavefunction is normalizable if $J_{2}>0$. 
This shows how both single-trace and multi-trace couplings are encoded in the wavefunction : 
\begin{equation}
\begin{aligned}
\langle \Psi_{J_{1},J_{2}} | t(x) |\Psi_{J_{1},J_{2}} \rangle &=
J_{1}(x), \\
\langle \Psi_{J_{1},J_{2}}| t^2(x) | \Psi_{J_{1},J_{2}}\rangle-\langle \Psi_{J_{1},J_{2}} | t(x) |\Psi_{J_{1},J_{2}} \rangle^2  &= -2J_{2}(x).
\end{aligned}
\end{equation}
The expectation value of $t(x)$ gives the single-trace coupling, and the second cumulant gives the double-trace coupling.
The second cumulant is negative because $t(x)$ fluctuates along the imaginary axis.

If double-trace operators have a support over a finite region,
the action becomes
\begin{equation}
S' =S_0+ \int d^D x \Big( J_{1}(x) \mathcal{O}(x) + \int d^D x' J_{2}(x-x') \mathcal{O}(x) \mathcal{O}(x') \Big),
\end{equation}
where $J_2(x-x')$ is the source for the bi-local double-trace operator.
The corresponding wave function is written as
\begin{equation}
\begin{aligned}
\Psi_{J_{1},J_{2}}\left[ t \right] &=\int_I  \mathcal{D} p(x)  e^{- \int d^D x \Big(i  t(x) p(x)-iJ_{1}(x) p(x) -\int d^D x' J_{2}(x-x') p(x) p(x') \Big)  } \\
&=\sqrt{\textrm{det} \left[\pi J_{2}^{-1}(x-x')\right]} 
e^{\frac{1}{4} \int d^D x \int d^D x' \left[t(x)- J_{1}(x) \right] J_{2}^{-1}(x-x') \left[t (x')- J_{1}(x') \right] }. \end{aligned}
\end{equation}
The non-zero correlation length for $j$ gives
\bqa
&& \langle t(x) t(x') \rangle - \langle t(x) \rangle \langle t(x') \rangle= - 2 J_{2}(x-x'). 
\eqa
This shows that the correlation between fluctuating single-trace sources
encodes the information on how the source for the bi-local double-trace operator
decays in space.  
In general, all multi-trace couplings can be extracted from higher order cumulants. 
It is noted that the single-trace coupling has non-trivial quantum fluctuations only in the presence of multi-trace couplings\footnote{
In the context of the holographic renormalization group, this amounts to the fact that integrating out bulk degrees of freedom generates multi-trace operators at a new UV boundary
\cite{
heemskerk2011holographic,2011JHEP...08..051F}.
}
Moreover, \eq{eq:SJJ} can be written as
\begin{equation}
|S_{J_{1},J_{2},\dots} \rangle 
= \int \mathcal{D}\phi ~  \mathcal{W}_{\phi} \left[J_{1},J_{2},\dots\right] |\phi \rangle,
\label{eq:SJJ2}
\end{equation}
where
\begin{equation}
\mathcal{W}_{\phi}\left[J_{1},J_{2},\dots\right] = 
\int_I \mathcal{D}t ~T_{\phi,t} \Psi_{J_{1},J_{2},\dots} \left[t\right].  
\label{eq:matrix_multiplication}
\end{equation}
We denote the vector spaces formed by $\{ \mathcal{W}_\phi \}$ and $\{ \Psi_J[t] \}$ as $W$ and $V$, respectively.
$W$ is the space of Boltzmann weights within a given symmetry sector 
with inner product, 
$( \mathcal{W}, \mathcal{W}' ) = 
\int \mathcal{D} \phi~
\mathcal{W}^*_\phi  \mathcal{W}'_\phi$.
$V$ is the space of wavefunctions of the single-trace sources with inner product,
$( \Psi, \Psi' )= 
\int \mathcal{D} t \mathcal{D} t'   ~
 \Psi^*[t] \langle t | t' \rangle \Psi'[t']$.
\eq{eq:matrix_multiplication} provides a  bijective map,
 $\mathcal{T}: V \rightarrow W$.
Accordingly, for every linear operator  $\hat{\mathcal{A}}_\phi$ that acts on $W$,
there exists a linear operator  $\hat{\mathcal{A}}_j$ that acts on $V$
such that
\begin{equation}
(\hat{\mathcal{A}}_\phi\mathcal{W})_{\phi} \left[J_{1},J_{2},\dots\right] = 
\int_I  \mathcal{D}t ~ \hat{T}_{\phi,t} (\hat{\mathcal{A}}_t \Psi)_{J_{1},J_{2},\dots}  \left[t\right].  \label{eq:A_on_matrix_multiplication}
\end{equation}
In order to find the correspondence  between $\hat{\mathcal{A}}_\phi$ and $\hat{\mathcal{A}}_t$, we consider the detailed form of $\Psi$ in \eq{eq:wave_function_general}. $\hat{\mathcal{A}}_t$, 
generated by $\frac{\delta}{\delta t}$ and $t$, gives the relations
\begin{equation}
\begin{aligned}
\frac{\delta}{\delta t(x)} \Psi_{J_{1},\dots }  \left[ t \right] &=  \int_I \mathcal{D} p ~  (-i)p(x) e^{ -\int d^D x \Big(i  t(x) p(x)+ \sum_{n}  (-i)^n J_{n}(x) p^n (x) \Big)  }, \\
 t(x) \Psi_{J_{1},\dots }  \left[ t \right] &= \int_I \mathcal{D} p ~ \Big(- i \frac{\delta}{\delta p(x)}  e^{ -\int d^D x  \sum_{n}  (-i)^n J_{n}(x) p^n (x)  } \Big) e^{ -i\int d^D x   ~t(x) p(x)  }.
\end{aligned}
\end{equation}
Since $p(x)$ is already identified as $i\mathcal{O}(x)$ in \eq{eq:correspondence_partition_function_wave_function}, 
$t(x)$ and $\frac{\delta}{\delta t(x)}$ acting on $V$
correspond to following operator acting on $W$
\footnote{We can explicitly derive this correspondence as 
\begin{eqnarray}
&& \int_I  \mathcal{D}t \Big(  \frac{\delta}{\delta t}  \Psi_{J_{1},J_2,\dots }   \left[ t \right] \Big) | t\rangle  
=- \int_I   \mathcal{D}t ~   \Psi_{J_{1},J_2,\dots }   \left[ t \right] \frac{\delta}{\delta t} | t\rangle \nonumber\\
&=&- \int_I   \mathcal{D}t ~   \Psi_{J_{1},J_2,\dots }   \left[ t \right]  \int \mathcal{D}\phi
\frac{\delta}{\delta t} e^{-S_0 -\int d^D x~t(x) \mathcal{O}(x)}| \phi \rangle 
=   \int \mathcal{D}\phi (x) \Big( \mathcal{O} \mathcal{W}_{\phi}\Big) |\phi \rangle, \nonumber \\
&& \int_I   \mathcal{D}t ~ \Big( t(x) \Psi_{J_{1},J_2,\dots }   \left[ t \right] \Big) | t\rangle \nonumber 
= \int_I \mathcal{D}t
\int    \mathcal{D} \phi ~
 \Psi_{J_{1},J_2,\dots }[t]
~  t(x) e^{ -S_0 -\int d^D x   ~t(x) \mathcal{O}(x)  }  | \phi \rangle  \nonumber \\
&=&\int \mathcal{D} \phi ~ 
\Big(
-e^{-S_0} \frac{\delta}{\delta \mathcal{O}} 
e^{S_0}   
\mathcal{W}_\phi \Big) \left[J_1,J_2,\dots\right]
| \phi\rangle. \nonumber
\end{eqnarray}
}.
\begin{equation}
\begin{aligned}
 \frac{\delta}{\delta t(x)}   &
 \Leftrightarrow   \mathcal{O}(x),  \\
 t(x)  &\Leftrightarrow 
 -  e^{-S_0} 
 \frac{\delta}{\delta  \mathcal{O}(x)}
 e^{S_0}. 
\end{aligned}
\end{equation}
Therefore, we obtain 
$\hat{\mathcal{A}}_{t}\left[t, \frac{\delta}{\delta t} \right]\Leftrightarrow 
\hat{\mathcal{A}}_{\phi}\left[  - e^{-S_0} \frac{\delta}{\delta  \mathcal{O}} e^{S_0},   \mathcal{O} \right]$.

\subsection{RG flow as quantum evolution \label{sec:QRG_Hamiltonian}}

Identifying the space of theories as a vector space 
naturally leads to the quantum RG\cite{lee2014quantum}. 
We first consider the  field theory in \eq{eq:partition_function_general}
 defined in a finite box with the linear system size $1/\lambda$,
where $ \lambda$ corresponds to an IR cutoff energy scale.
From \eq{eq:correspondence_partition_function_wave_function}, 
the  partition function is equivalent to 
\begin{equation}
\begin{aligned}
Z_{\lambda} \left[ J_{1}, J_{2},\dots\right] 
&=\int_I  \mathcal{D} t^{(0)} ~
\Psi_{J_{1},J_{2},\dots } \left[ t^{(0)} \right] ~
Z_{\lambda} \left[ t^{(0)}, 0,\dots \right],
\end{aligned}
\end{equation}
where
$\Psi_{J_1,J_2,...}[ t^{(0)} ]$ is 
the  wavefunction defined in  \eq{eq:wave_function_general},
and 
$Z_{\lambda} \left[ t^{(0)}, 0,\dots \right]=\langle \mathbbm{1} | t^{(0)} \rangle=\int \mathcal{D}\phi ~  e^{- S_0 - \int^{1/\lambda} d^D x ~t^{(0)}(x) \mathcal{O}(x)  }$. 
The subscript $\lambda$ of $Z$ keeps track of the IR cutoff. 
Next we perform a coarse graining transformation on $Z_{\lambda} \left[ t^{(0)}, 0,\dots \right]$
 as is discussed in the previous section  : 
integrating out high-energy modes of $\phi$ 
which reduces the UV cutoff from $\Lambda$ to $\Lambda'=b\Lambda$ with $b=e^{-dz} <1$. 
In general, not only the single-trace source is renormalized but also multi-trace operators are generated.
Under rescaling of the field and the space, 
$\phi'(x') = b^{-\Delta_\phi} \phi(x)$ with $x'=b x$, 
 $\Lambda'$ is brought back to $ \Lambda$,
but the system size decreases as
$1/ \lambda' = b/ \lambda$.
The resulting partition function becomes\footnote{Here we omit an additional subscript $\lambda/b$ in $S_0$ to avoid clutter in notation.}
\begin{equation}
\begin{aligned}
& Z_{\lambda} \left[ t^{(0)}, 0,\dots \right]   
= \int \mathcal{D} \phi ~  e^{- S_0\left[ \phi \right]} e^{ - \int^{b/\lambda}  d^D x' 
 \Big\{  
  t^{(0)}(x')  \mathcal{O}(x') 
 -dz  \sum_{n \geq 0} \beta_n  \left[t^{(0)},0,\dots\right]  \mathcal{O}^n (x') \Big\} }.
\label{eq:Zlambdaj0}
\end{aligned}
\end{equation}
The change of the coupling  for $\mathcal{O}^n(x)$
with increasing length scale is given by
$ - \beta_n dz$.
$- \beta_0 dz$ corresponds to the free energy contributed from the high-energy modes that are integrated out in 
the infinitesimal coarse graining step.
In general, the $\beta$-functions depend on all the couplings $J_{n}$.
However, what enters in
\eq{eq:Zlambdaj0}
 are the beta functions in the subspace of single-trace couplings only.
From Eq.~(\ref{eq:correspondence_partition_function_wave_function}), 
the partition function 
in \eq{eq:Zlambdaj0} can be written as
\begin{equation}
\begin{aligned}
Z_{\lambda} &\left[ t^{(0)}, 0, ... \right]  \\
&=\int_I  \mathcal{D} ~t^{(1)}(x) \mathcal{D} p^{(1)}(x)  e^{- \int^{b/\lambda} d^D x 
 \left[ i p^{(1)}  
\Big(t^{(1)}  -t^{(0)} \Big) - dz \sum_{n} \beta_{n} dz (-ip^{(1)})^n \right]}
 Z_{b^{-1}\lambda} \left[  t^{(1)}, 0,\dots\right].
\end{aligned}
\end{equation}
After $M$ steps of coarse graining, 
we take the $dz \rightarrow 0$ limit, keeping $M dz=z^*$ fixed.
This results in
\begin{equation}
\begin{aligned}
Z_\lambda \left[ J_1,J_2,\dots \right] 
&= \int_I  \mathcal{D} t(x,z) \mathcal{D} p(x,z) 
\Psi_{J_1,J_2,...}[ t(0) ]
e^{- \int_0^{z^*} dz L\left[ t, p, z \right]}
Z_{e^{z^\ast}\lambda}\left[ t(z^*), 0 ,\dots \right],
\label{eq:partition_function_evolve}
\end{aligned}
\end{equation}
where 
$z$ is the extra direction in the bulk that labels the logarithmic length scale. 
The bulk Lagrangian is given by
\bqa
L \left[ t, p, z \right] 
&=&\int^{e^{-z}/\lambda} d^D x \Big( ip(x,z) \partial_z t(x,z) 
- \sum_{n\geq 0} \beta_{n}[t(z),0,..;x]  \left(-i {p}(x,z) \right)^n \Bigr).
\label{eq:Lagrangian_evolution}
\eqa
In \eq{eq:partition_function_evolve},
the RG flow of the $D$-dimensional field theory
is replaced with a $(D+1)$-dimensional path integration
of the dynamical single-trace sources\cite{lee2012background,lee2014quantum}.
The fluctuations of the RG paths encode the information of the multi-trace operators.
The sum over all possible RG paths within the subspace of the single-trace sources
is weighted with the $(D+1)$-dimensional bulk action.
Equivalently,  the RG flow is described by the quantum evolution of the wavefunction of the single-trace source.
In the Hamiltonian picture, 
$t(x)$ and $p(x)$ are canonical conjugate operators
that satisfy the commutation relation
$\left[ t(x),p(x') \right]= - i \delta(x-x')$.
The RG evolution can be viewed as an imaginary time evolution, 
$\mathcal{R}(z+dz,z) =e^{- \mathcal{H}^\lambda[t,p,z] dz}$ generated by the bulk  RG Hamiltonian, 
\begin{equation}
\begin{aligned}
\mathcal{H}^\lambda \left[ t, p,z \right]&=  - \sum_{n \geq 0} \int^{e^{-z}/\lambda} d^D x ~
\beta_{n}[t,0,..;x]   \left( -i p(x) \right)^n,
\label{eq:Hamiltonian_evolve}
\end{aligned}
\end{equation}
where the superscript $\lambda$ represents the IR cutoff scale associated with the finite system size.
$\mathcal{H}^\lambda$ depends on $z$ explicitly through $\lambda$
as the system size decreases with increasing $z$.
In the thermodynamic limit ($\lambda = 0$),
$\mathcal{H}^0$ is independent of $z$.
Being an operator that acts on wavefunctions defined in the space of single-trace sources, 
$\mathcal{H}^\lambda$ generates the quantum evolution of the state associated with the RG flow.
The RG Hamiltonian is fixed by the $\beta$-functions within the subspace of the single-trace sources only.

\subsection{Reconstruction of the Wilsonian RG from the quantum RG \label{sec:reconstruction}}

In this section, we explain how the full $\beta$-functions of a field theory
can be reconstructed from the quantum evolution with the RG Hamiltonian,
although the latter is fixed by
the $\beta$-functions within the subspace of the single-trace couplings. 
Suppose that there exists a unique IR fixed point in the thermodynamic limit.
The fixed point action is invariant under the RG transformation,
and the corresponding quantum state should be an eigenstate of $\mathcal{H}^\lambda$ at $\lambda = 0$.
Furthermore, the stable IR fixed point must correspond to the ground state of $\mathcal{H}^0$ 
because generic initial state is always projected onto it in the large $z$ limit.
More generally, one can consider excited states of $\mathcal{H}^0$. 
States of particular interest are eigenstates that support excitations local in space.
Those states correspond to the IR fixed point perturbed with local operators with definite scaling dimensions. 
These scaling dimensions are given by the energy differences between the excited states and the ground state.
In the rest of this section, we establish the correspondence 
between the ground state (excited states) 
and the stable fixed point (the stable fixed point with  operator insertions). 

\subsubsection{Stable IR fixed point as the bulk ground state}

We begin with the discussion of the ground state. 
The ground state of $\mathcal{H}^0$ satisfies
\bqa
\mathcal{H}^0 \psi_0 \left[t \right] 
=
 \mathcal{E}_{0}
\psi_0 \left[t \right] 
\eqa
with the lowest eigenvalue.
The partition function for the the IR fixed point  is given by 
\bqa
Z^\ast \left[ J_{1}^\ast,J_{2}^\ast, \dots  \right] 
& =
\int_I \mathcal{D} t ~  \psi_0 \left[t \right] \int \mathcal{D}\phi ~ e^{-S_0 - \int d^D x ~t(x) \mathcal{O}(x)},
\label{eq:partition_function_evolution}
\eqa
where $\mathcal{O}$ is the single-trace operator. 
To extract the multi-trace couplings at the fixed point, we use the cumulant expansion $\langle e^{-\Omega} \rangle = e^{-\langle \Omega \rangle + \frac{1}{2}(\langle \Omega^2 \rangle-\langle \Omega \rangle^2) +\dots}$ to rewrite \eq{eq:partition_function_evolution} as
\begin{equation}
\begin{aligned}
Z^\ast \left[ J_{1}^\ast,J_{2}^\ast, \dots  \right] &= \int \mathcal{D}\phi ~ e^{-S_0}   \langle  e^{ - \int d^D x ~t (x) \mathcal{O}(x)} \rangle_{\psi_0} \\
&= 
\int \mathcal{D}\phi ~ e^{-S_0} e^{- \int d^D x \langle t (x) \rangle_{\psi_0} \mathcal{O}(x) 
+\frac{1}{2}\int d^D x d^D y (\langle t (x) t(y) \rangle_{\psi_0}- \langle t(x) \rangle_{\psi_0}  \langle t(y) \rangle_{\psi_0}) 
\mathcal{O} (x) 
\mathcal{O} (y) 
+\dots},
\end{aligned}
\end{equation}
where $
\langle F[t] \rangle_{\psi_0} 
\equiv 
 \int_I \mathcal{D} t ~  \psi_0 \left[t \right] F[t] 
$.
Identifying this 
as the fixed point action,
\bqa
Z^\ast \left[ J_{1}^\ast,J_{2}^\ast, \dots  \right] & =
\int \mathcal{D}\phi ~ e^{
-S_0 - \int d^D x  J_{1}^\ast(x) \mathcal{O}(x) - \int d^Dx d^D y J_{2}^\ast(x-y) \mathcal{O}(x) \mathcal{O}(y) - ..},
\label{eq:partition_function_evolution0}
\eqa
 we obtain the couplings at the fixed point,
\begin{equation}
\begin{aligned}
J_{1}^\ast (x)&= \langle t(x) \rangle_{\psi_0}  \\
J_{2}^\ast (x-y )&= -\frac{1}{2}
\Big( 
\langle t(x) t(y)  \rangle_{\psi_0}  - \langle t(x) \rangle_{\psi_0}  \langle t(y) \rangle_{\psi_0}  
\Big). 
\label{eq:coupling_gs}
\end{aligned}
\end{equation}
Sources for higher-trace operators at the fixed point are given by the higher cumulants.

\subsubsection{Scaling operators as local excitations of the bulk theory}

Next, let us study excited states of $\mathcal{H}^0$.
We start by considering states with local excitations. 
The $n$-th excited state that supports a local excitation at $x$ satisfies
\bqa
e^{- \mathcal{H}^0 z} \psi_{n,x} = e^{-\mathcal{E}_{n} z} \psi_{n,e^{-z}x},
\eqa
where $\mathcal{E}_{n}$ is the $n$-th eigenvalue.
For general $x$, this is not a genuine eigenvalue equation 
because the dilatation generator in the RG Hamiltonian
preserves only one point in space
which is chosen to be $x=0$ here.
A local operator inserted at $x$ is transported to $e^{-z} x$ under the RG flow.
Only the states that support local excitations at $x=0$ 
can remain invariant under the RG evolution.
$\{ \psi_{n,0} \}$ forms the complete basis of states that support local excitations at $x=0$.
Generic excited states can be obtained by applying local operators, denoted as $\hat{\mathcal{A}}_{n,x,t}$, 
to the ground state as
\begin{equation}
\psi_{n,x} \left[ t\right]  = (\hat{\mathcal{A}}_{n,x,t} \psi_0 )\left[ t\right].
\end{equation}
$\hat{\mathcal{A}}_{n,x,t}$,
which consists of $t(x)$ and $\frac{\delta}{\delta t(x)}$, 
creates local excitations at $x$,
where
$\hat{\mathcal{A}}_{0,x,t}=\mathbbm{1}$.
From the correspondence in \eq{eq:A_on_matrix_multiplication},
 $\hat{\mathcal{A}}_{n,x,t}$ is dual to an operator 
 that acts on the Boltzmann weight,  
\begin{equation}
\begin{aligned}
\int_I \mathcal{D} t ~(\hat{\mathcal{A}}_{n,x,t} \psi_0 )\left[ t\right]\int \mathcal{D}\phi ~ e^{ -S_0- \int d^D x ~t(x) \mathcal{O}(x)}= \int \mathcal{D} \phi ~ 
(\hat{\mathcal{A}}_{n,x,\phi} \mathcal{W} )_\phi\left[ J_1^\ast,J_2^\ast,\dots\right], 
\label{eq:j_phi_A}
\end{aligned}
\end{equation}
where $\hat{\mathcal{A}}_{n,x,\phi}$ consists of 
$\mathcal{O}(x)$ and 
$\frac{\delta}{\delta \mathcal{O}(x)}$. 
$\hat{\mathcal{A}}_{n,x,t}$ ($\hat{\mathcal{A}}_{n,x,\phi}$) 
is the representation of the operator in space $V$ ($W$). 
Henceforth, we use  $\hat{\mathcal{A}}_{n,x}$  to denote the operator itself when there is no need to specify its representation.

We now verify that  excited states indeed correspond to the fixed point perturbed
with local operators that have definite scaling dimensions.
For this, we consider the IR fixed point theory with a small perturbation added at the origin,
\begin{equation}
 e^{ \epsilon_{n,0}  \hat{\mathcal{A}}_{n,0}} |S_{J_1^\ast,J_2^\ast,\dots}\rangle 
=\int \mathcal{D} \phi ~ (e^{ \epsilon_{n,0}  \hat{\mathcal{A}}_{n,0,\phi}} \mathcal{W} )_\phi\left[ J_1^\ast,J_2^\ast,\dots\right] |\phi \rangle,
\end{equation} 
where $\epsilon_{n,0}$ is an infinitesimal parameter.
The RG Hamiltonian generates an evolution of this state as
\begin{equation}
\begin{aligned}
 e^{-H dz} 
  \left[e^{ \epsilon_{n,0}  \hat{\mathcal{A}}_{n,0}} |S_{J_1^\ast,J_2^\ast,\dots}\rangle\right] 
&= \int_I \mathcal{D} t ~e^{-\mathcal{H}^0 dz}   \Big( \psi_0[t] + \epsilon_{n,0} \psi_{n,0} \left[t \right] \Big) \int \mathcal{D}\phi ~ e^{-S_0 - \int d^D x t(x) \mathcal{O}(x)}  | \phi \rangle
\\&= e^{-\mathcal{E}_{0}dz} 
e^{ \epsilon_{n,0} e^{-( \mathcal{E}_{n} - \mathcal{E}_{0}) dz}  \hat{\mathcal{A}}_{n,0} }
| S_{J_1^\ast,J_2^\ast,\dots}\rangle
\end{aligned}
\end{equation}
to the linear order in $\epsilon_{n,0}$.
This implies that the infinitesimal source evolves as
\begin{equation}
\frac{d \epsilon_{n,0}}{dz} = -(\mathcal{E}_{n}-\mathcal{E}_{0}) \epsilon_{n,0}
\end{equation}
under the RG flow, and
$\hat{\mathcal{A}}_{n,0,\phi}$ is a scaling operators with the scaling dimensions
\bqa
\Delta_{n} = \mathcal{E}_{n}-\mathcal{E}_0.
\label{eq:scalingDofA}
\eqa
The  spectrum of $\mathcal{H}^0$ encodes the information of all scaling operators and their scaling dimensions.
$\hat{\mathcal{A}}_{n,x}$'s create  local excitations with definite scaling dimensions, and 
are called scaling operators.
On the other hand, 
$\mathcal{\bf O}_M(x)$'s
represent general multi-trace operators in terms of which the UV action is written,
and are called UV operators.
In general, the scaling operators can be written as linear superpositions of UV operators :
\begin{equation}
\hat{\mathcal{A}}_{n,x,\phi} =\sum_M  g_{n,M} \mathcal{\bf O}_M(x).
\label{eq:O_poly}
\end{equation} 
The inverse of  \eq{eq:O_poly} can be written as 
$\mathcal{\bf O}_M (x)= \sum_{n}  (g^{-1})^{M,n} \hat{\mathcal{A}}_{n,x,\phi}$.

From the local scaling operators, 
one can generate translationally invariant eigenstates
by turning on the sources uniformly in space.
For example, 
\begin{equation}
\psi_{n} \left[ t\right]  = \int d^D x ~  \psi_{n,x}\left[ t\right]
\end{equation}
is an eigenstate with energy $\Delta_n-D$. 
The shift in the energy by $-D$  is from the spatial dilatation included in $\mathcal{H}^0$,
\bqa
e^{-\mathcal{H}^0 dz} 
\psi_{n} \left[ t\right]  
=
e^{- \Delta_n dz} 
 \int d^D x ~ 
  \psi_{n,e^{-dz} x}\left[ t\right]
=
e^{- (\Delta_n-D) dz} 
 \int d^D \tilde x ~ 
  \psi_{n,\tilde  x}\left[ t\right],
\eqa
where $x = e^{dz} \tilde x$ is used. 
If  $\Delta_n > D$ for all $n$,
the fixed point is stable as all deformations with spatially uniform sources are irrelevant.
Local excitations with energy gap less than (equal to) $D$ correspond
to relevant (marginal) operators.
Throughout this paper, we assume that local excitations of the RG Hamiltonian  have a non-zero gap
$(\Delta_n > 0)$.

\subsubsection{Mixing matrix}

According to \eq{eq:O_poly}, 
$g_{n,M}$ encodes the relation between scaling operators, 
$\hat{\mathcal{A}}_{n,x,\phi}$ 
and UV operators, $\mathcal{\bf O}_M(x)$. 
By writing 
\begin{equation}
\sum_{n,x}  \epsilon_{n,x} \hat{\mathcal{A}}_{n,x,\phi} 
 =\sum_{M,x}  \epsilon^{UV}_{M}(x) \mathcal{\bf O}_M(x)
\end{equation}
and using  \eq{eq:O_poly},
we obtain the relation between the sources for the scaling operators
and the UV operators,
\begin{equation}
\epsilon^{UV}_{M}(x) = \sum_{n} \epsilon_{n,x}  g_{n,M} , 
\quad 
\epsilon_{n,x} = \sum_{M}  \epsilon^{UV}_{M}(x)  (g^{-1})^{M,n}. 
\label{eq:linearTR}
\end{equation}
To the linear order in $\epsilon$, 
the $\beta$-functions of these UV couplings are
\begin{equation}
\begin{aligned}
\frac{d \epsilon_{M}^{UV}(x)}{dz} &= - \sum_{n} (\mathcal{E}_{n}-\mathcal{E}_0) \epsilon_{n,x}  g_{n,M}\\
&= - \sum_{n,M'} 
\epsilon^{UV}_{M'}(x) 
 (g^{-1})^{M' n} 
g_{n,M} (\mathcal{E}_{n}-\mathcal{E}_0).  
\end{aligned}
\end{equation}
This gives the mixing matrix 
\begin{equation}
{\mathcal{M}^{M'}}_{M } = - \sum_{n} 
 (g^{-1})^{M' n} 
g_{n,M} (\mathcal{E}_{n}-\mathcal{E}_0).  
\label{eq:ggE}
\end{equation}
It is noted that what appears on the right hand side of \eq{eq:ggE} is determined 
from the spectrum of $\mathcal{H}^0$ which is fixed by the beta functions defined in the subspace of the single-trace couplings.
\eq{eq:ggE} shows that this small set of data completely fixes the full mixing matrix
that involves all multi-trace operators.
This shows that the $\beta$-functions for general multi-trace operators
are fixed by those for the single-trace operator to the linear order in the deformation.
In the following two subsections, we show that this holds 
beyond the linear order.

\subsubsection{Operator product expansion \label{sec:ope}}

\begin{figure}[h]
    \centering
    \includegraphics[width=.6\textwidth]{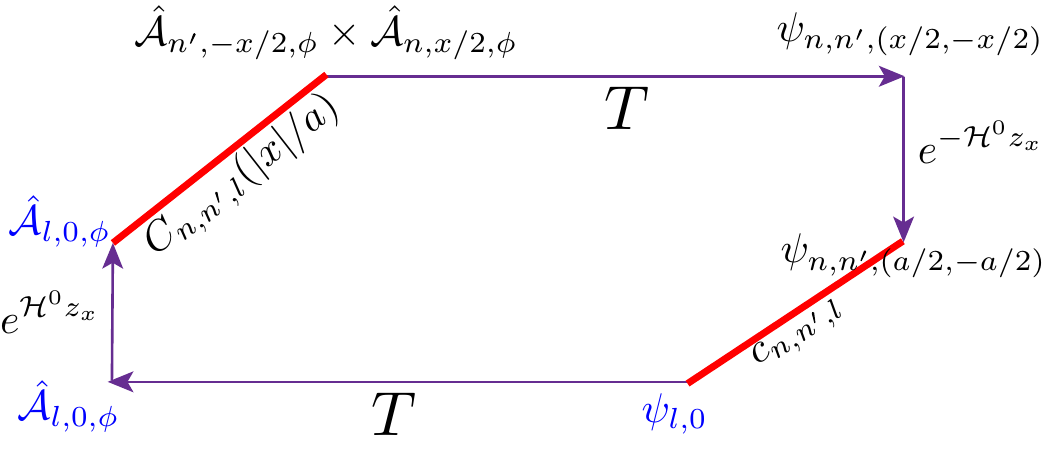}
    \caption{
The procedure of extracting OPE between two local operators.
Two local operators  
$\hat{\mathcal{A}}_{n',-x/2,\phi}$ and $\hat{\mathcal{A}}_{n',x/2,\phi}$ 
shown in the upper left corner undergo a series of transformations 
following the arrows in the clockwise direction :
1) 
the operators inserted to the IR fixed point corresponds to the ground state with two local excitations through the action-state correspondence
(right arrow $T$);
2) 
the state is evolved with the RG Hamiltonian for 
$ z_x = \ln \frac{|x|}{a}$
(down arrow $e^{-\mathcal{H}^0 z}$ );
3) 
the resulting state supports local excitation near the origin which is then expressed as a linear superposition of the eigenstates $\psi_{l,0}$ with local excitations at the origin weighted with $c_{n,n',l}$;
4) the resulting state corresponds to the fixed point with local operator insertions at the origin (left arrow $T$);
5) the theory is evolved backward in RG time by  $- z_x$
(up arrow $e^{\mathcal{H}^0 z}$ ).
The identification of the final operator with the initial product of two operators gives the desired operator product expansion.
    }
    \label{fig:ope}
\end{figure}

The operator product expansion (OPE) between general multi-trace operators 
is also fully encoded in the spectrum of $\mathcal{H}^0$.
Suppose we insert two local scaling operators 
$\hat{\mathcal{A}}_{n,x/2}$ and $\hat{\mathcal{A}}_{n',-x/2}$ 
at $x/2$ and $-x/2$, respectively.
The wavefunction for the resulting theory is
$\psi_{n,n',(x/2,-x/2)}\left[t\right] =(\hat{\mathcal{A}}_{n,x/2,t}\hat{\mathcal{A}}_{n',-x/2,t} \psi_{0} )\left[t\right]$. 
Under the  evolution generated by $\mathcal{H}^0$,  
the separation between the two operators decreases exponentially in $z$.
For $ z \ll \ln \frac{|x|}{a}$, where $a = \Lambda^{-1}$ is the short distance cutoff length scale,
the two local excitations remain  well-separated in space, and evolve independently,
\bqa
e^{-\mathcal{H}^0 z} (\hat{\mathcal{A}}_{n,x/2,t}\hat{\mathcal{A}}_{n',-x/2,t} \psi_{0} )\left[t\right] 
\approx
 e^{  -(\Delta_{n} +\Delta_{n'}+\mathcal{E}_0)z} 
 (
 \hat{\mathcal{A}}_{n,xe^{-z}/2,t}
 \hat{\mathcal{A}}_{n',-x e^{-z}/2,t} 
 \psi_{0} )\left[t\right].
 \label{eq:premerging}
\eqa
This follows from the facts that
1) $\hat{\mathcal{A}}_{n',-x/2}$ and $\hat{\mathcal{A}}_{n,x/2}$
create local excitations with energies  
$\Delta_{n}$ and $\Delta_{n'}$ above the ground state; 
2) $\mathcal{H}^0$ is local at length scales larger than $a$,
and two operators evolve independently 
with the total energy given by 
 $\mathcal{E}_{n,n'} = \mathcal{E}_0 +  ( \Delta_n + \Delta_{n'} )$
in the limit that their separation is large.
As two excitations approach,
they interact, and the state evolves into a more complicated state.
Nonetheless, the state obtained after evolving the initial state for 
$ z_x = \ln \frac{|x|}{a}$ can be written as a linear superposition 
of eigenstates of the RG Hamiltonian with local excitations located at the origin,
\bqa
&& e^{-\mathcal{H}^0 z_x} (\hat{\mathcal{A}}_{n,x/2,t}\hat{\mathcal{A}}_{n',-x/2,t} \psi_{0} )\left[t\right] 
=  
 e^{   -(\Delta_{n} +\Delta_{n'}+\mathcal{E}_0)z_x} 
 \sum_{l} F_{n,n',l}(x)  ( \hat{\mathcal{A}}_{l,0,t}  \psi_{0} )\left[t\right],
 \label{eq:merging}
\eqa
where $F_{n,n',l}(x)$ is a function 
that captures the effect of interaction. 
It is a regularization dependent function which can be computed from the RG Hamiltonian.
There is no interaction between two operators when the separation is much larger than $a$.
This follows from the fact that  one only integrates out modes whose wavelengths are order of $a$ in each coarse graining step.
As a result,  $F_{n,n',l}(x)$  exponentially approaches a constant in the large $|x|$ limit,
$\lim_{|x| \rightarrow \infty}  | F_{n,n',l}(x) - c_{n,n',l}| \sim e^{-|x|/a}$, where
$ c_{n,n',l} \equiv F_{n,n',l}(\infty)$.
Now the state in  \eq{eq:merging} is evolved backward in RG time $z_x$,
which results in
\bqa
(\hat{\mathcal{A}}_{n,x/2,t}\hat{\mathcal{A}}_{n',-x/2,t} \psi_{0} )\left[t\right] 
= 
\sum_l F_{n,n',l}(x) ~
 e^{ (\Delta_l - \Delta_{n} -\Delta_{n'})z_x} 
 ( \hat{\mathcal{A}}_{l,0,t}  \psi_{0} )\left[t\right].
 \label{eq:merging2}
\eqa
From this,  we obtain the OPE of scaling operators :
\begin{equation}
\hat{\mathcal{A}}_{n,x/2,\phi} \times \hat{\mathcal{A}}_{n',-x/2,\phi} = 
\sum_l C_{n,n',l} \left( x \right) \hat{\mathcal{A}}_{l,0,\phi},
\label{eq:ope}
\end{equation}
where
\bqa
&& C_{n,n',l} ( x ) =   \frac{ F_{n,n',l}(x)  }{|x/a|^{\Delta_{n} + \Delta_{n'} - \Delta_{l}}}.
\eqa
In the large $|x|$ limit, this reduces to the standard form,
$C_{n,n',l} (x) \sim  \frac{ c_{n,n',l}  }{|x|^{\Delta_{n} + \Delta_{n'} - \Delta_{l} } }$.
The procedure used to extract the OPE is summarized in the \fig{fig:ope}.

\subsubsection{$\beta$-functionals \label{sec:beta_func}}

Now we are ready to derive the full beta functions.
For generality, we consider the case in which the sources are position dependent,
and derive the beta functionals of the Wilsonian RG from the quantum RG. 
We consider a UV theory with general deformations added to the fixed point theory $S^\ast$ parametrized by $J_{n,x}^\ast$,
\bqa
S = S^\ast + \int d^D x ~ \epsilon_{n,x}   \hat{\mathcal{A}}_{n,x,\phi},
\eqa 
where $\epsilon_{n,x} = J_{n,x} - J_{n,x}^*$,
and repeated indices are summed over.

The state that corresponds to this deformed theory is given by
\bqa
| \Psi  \rangle =  | \psi_0 \rangle - 
\int d^D x\left( \epsilon_{n,x} 
- \frac{1}{2}\int d^D x'   \epsilon_{l,x+\frac{x'}{2}} \epsilon_{m,x-\frac{x'}{2}} C_{lmn}(x')
\right)
 | \psi_{n,x} \rangle + O(\epsilon^3)
 \label{eq:deformedstate}
\eqa 
to the second order in $\epsilon_n $,
where we use
$\hat{\mathcal{A}}_{l,x+x'/2,\phi} \times \hat{\mathcal{A}}_{m,x-x'/2,\phi} 
=\sum_n C_{lmn}(x') \hat{\mathcal{A}}_{n,x,\phi}$ obtained in \eq{eq:ope}. 
Under the evolution generated by the  RG Hamiltonian,
the state evolves to
\begin{equation}
\begin{aligned}
&| \Psi(z)  \rangle
=e^{-\mathcal{E}_0 dz} 
\left[ | \psi_0 \rangle - 
e^{-  \Delta_n dz}  \int d^D x\left( \epsilon_{n,x} 
- \frac{1}{2}\int d^D x'   \epsilon_{l,x+\frac{x'}{2}} \epsilon_{m,x-\frac{x'}{2}} C_{lmn}(x')
\right)  | \psi_{n,x e^{-dz} } \rangle \right] + O(\epsilon^3) \\
&=e^{-\mathcal{E}_0 dz} 
\left[ | \psi_0 \rangle - 
e^{-   \Delta_n dz}  \int d^D \tilde x e^{D dz}  \left( \epsilon_{n,e^{dz} \tilde x } 
- \frac{1}{2}\int d^D \tilde x'   e^{D dz}  
 \epsilon_{l,e^{dz} ( \tilde x + \frac{\tilde x'}{2})} 
\epsilon_{m,e^{dz} (\tilde x - \frac{\tilde x'}{2})} 
C_{lmn}(e^{dz} \tilde x')
\right)  | \psi_{n,\tilde x } \rangle \right] \\
&+ O(\epsilon^3),
\end{aligned}
\end{equation}
where the renormalized spatial coordinate is defined as $ x =\tilde{x} e^{dz}$ in the second line.
The final state can be written in the form of \eq{eq:deformedstate}
provided $\epsilon_{n,x}$ is replaced with 
renormalized source,
\bqa
&& \epsilon'_{n,\tilde x} = 
e^{-\tilde \Delta_n dz}  \epsilon_{n,e^{dz} \tilde x}  \nn
&&
+\frac{1}{2}\int d^D \tilde x'~
 \epsilon_{l,e^{dz} ( \tilde x + \frac{\tilde x'}{2})} 
\epsilon_{m,e^{dz} (\tilde x - \frac{\tilde x'}{2})} 
\left[
 e^{-(\tilde \Delta_l + \tilde \Delta_m) dz} C_{lmn} ( \tilde x')
 -
  e^{(D-\tilde \Delta_n) dz }C_{lmn} (e^{dz} \tilde x')\right] 
+ O(\epsilon^3),\nn
\label{eq:epsilonprime}
\eqa
where $ \tilde \Delta_n = \Delta_n - D$.
$| \Psi(z)  \rangle$ corresponds to the 
action,
\bqa
\begin{aligned}
S(z) &= S^* + \int d^D x~  \epsilon'_{n,x}
\hat{\mathcal{A}}_{n,x,\phi},
\end{aligned}
\eqa 
where the renormalized source is given by 
\bqa 
\epsilon'_{n, x} &=& 
\epsilon_{n, x} + dz
\Big\{
-\tilde \Delta_n   \epsilon_{n,  x}  
+  x \frac{\partial}{\partial  x}  \epsilon_{n,  x}
+ \frac{1}{2}\int d^D  x'~ \epsilon_{l, x + \frac{ x'}{2}} 
\epsilon_{m,x - \frac{ x'}{2}} G_{lmn}( x' ) 
\Big\}
+  O(\epsilon^3)
\label{eq:fusion73}
\eqa
to the linear order in $dz$ with
\bqa
&& G_{lmn}(y )  = 
-C_{lmn}(y) y \frac{\partial}{\partial y}  \ln F_{lmn}(y). 
\eqa
The term quadratic in $\epsilon$ 
in \eq{eq:fusion73}
describes the fusion of two operators into one.
Since 
$G_{lmn}(y ) $ decays exponentially in  the large $y$ limit,
operators whose separation is smaller than $a$ mainly contribute to the fusion process.
This shows that the $\beta$-functionals of $J_{n,x}$ are local,
\begin{equation}
\begin{aligned}
\frac{d}{dz} J_{n,x}  &= - ( \Delta_n - D) 
(J_{n,x}-J_{n,x}^\ast) 
+  x \frac{\partial}{\partial  x}(J_{n,x}-J_{n,x}^\ast)   
\\
&+\frac{1}{2}\int d^D  x'~ (J_{l,x+\frac{x'}{2}}-J_{l,x+\frac{x'}{2}}^\ast) 
(J_{m,x-\frac{x'}{2}}-J_{m,x+\frac{x'}{2}}^\ast)  G_{lmn}(x') 
 + O \Bigl( (J-J^\ast)^3 \Bigr). 
\label{eq:generallinearbeta}
\end{aligned}
\end{equation}
This also confirms that 
couplings are irrelevant (relevant)
if $ \Delta_n > D$ ($\Delta_n < D$).
It is noted that these are $\beta$ functions for sources of the scaling operators.
The $\beta$ functions for the sources of the UV operators are obtained 
from \eq{eq:generallinearbeta}
through  a linear transformation in  \eq{eq:linearTR}.

\section{Toy models}
\label{sec:toymodels}

In the previous two sections,
it is shown
that there exist general constraints that $\beta$-functions satisfy due to the fact that the full Wilsonian RG can be replaced with a quantum RG defined in the subspace of single-trace couplings.
While the quantum theory is well-defined, 
it is in general no easier than the original problem.
Only in the large $N$ limit, the quantum problem can be solved with a semi-classical approximation.
See Refs. \cite{future}
for recent development.
In this section, 
we consider toy models
in which the bulk theory is non-interacting
and can be solved explicitly.

\subsection{$0$-dimensional solvable Example 
\label{sec:0d_example}}

In this section, we consider a toy model of a $0$-dimensional  theory.
For simplicity,  
we assume that there is only one single-trace operator,
and that only single-trace and double-trace operators are generated
under the coarse graining
when the reference theory $S_0$ is deformed by the single-trace operator. 
We assume that the reference theory is invariant under a $\mathbbm{Z}_2$ symmetry, and the single-trace operator is odd under the symmetry.
The symmetry constrains the form of the $\beta$-functions.
We assume that the $\beta$-functions in the subspace of the single-trace deformation take the following form
\bqa
\beta_0(t, 0, ..) &=& f- w t^2, \nn
 \beta_1(t,0,..) &=& at, \nn
 \beta_2 (t,0,..) &=& -b, \nn
 \beta_{n > 2} (t,0,..) &=& 0.
 \label{eq:0Dbeta}
 \eqa 
Here $t$ stands for the source for the single-trace operator.
$\beta_0$  describes the  flow of the identity operator.
$\beta_1$ ($\beta_2$) is the beta function for the single (double)-trace coupling. 
$a$, $b$, $w$ and $f$ are non-zero real parameters. 
Under the $\mathbbm{Z}_2$ transformation,
the single-trace coupling transforms as $t \rightarrow -t$.
This guarantees that $\beta_0$ and $\beta_2$ are even in $t$,
and  $\beta_1$ is odd in $t$.
\eq{eq:0Dbeta} describes how couplings are renormalized 
when $S_0$ is deformed by the single-trace operator.
In particular, $b \neq 0$ implies that 
the double-trace operator is generated 
and the RG flow leaves the subspace of single-trace deformation
even if the UV theory has only single-trace deformation.
From \eq{eq:0Dbeta}, it is unclear where the double-trace coupling $t_2$ eventually flows 
under the RG flow.
Depending on the sign of $\beta_2(t, t_2)$ at large $t_2$,
the theory may or may not flow to a scale invariant fixed point in the IR.
Remarkably,  $\beta_2(t, t_2)$ at general values of $t_2$ 
is already encoded in \eq{eq:0Dbeta},
which determines the fate of the RG flow in the space of all couplings.

Following the formalism in Sec.~\ref{sec:QRG_Hamiltonian}, 
we obtain the bulk RG Hamiltonian in \eq{eq:Hamiltonian_evolve}.
It is convenient to shift the RG Hamiltonian to remove a constant piece,
\begin{equation}
\begin{aligned}
\tilde{\mathcal{H}}=\mathcal{H}+
\left( f -\frac{a}{2} \right)
&= b\left[ \tilde{p}+\frac{ia}{2b}\tilde{t} \right]^2 + \left[  \frac{a^2-4bw}{4b}\right] \tilde{t}^2
\label{eq:ham_0d},
\end{aligned}
\end{equation}
where 
$\tilde{t}=it$ and $\tilde{p}=-ip$ fluctuate along the real axis
\footnote{ The additional shift $a/2$ is generated from the normal ordering.}.
They satisfy commutation relation $\left[ \tilde p, \tilde t\right]=i$. 
This is a TP-symmetric non-Hermitian quadratic Hamiltonian\cite{BenderPT}.
As is shown in Appendix \ref{app:non_Hermitian}, the spectrum of this RG Hamiltonian is given by
\begin{equation}
E_{n}  = (n+\frac{1}{2})\sqrt{\eta},
\label{eq:EnH}
\end{equation}
where $\eta=a^2-4bw$. 
Unlike the Hermitian cases, the left and right eigenstates take different forms,
\begin{equation}
\begin{aligned}
\psi^{R}_{n}(\tilde{t}) &=
\frac{1}{\sqrt{2^n n!}} (\frac{\varepsilon}{2\pi })^{1/4} 
e^{-\xi_R \tilde{t}^2}H_n \left[\sqrt{\frac{\varepsilon}{2}} \tilde{t} \right], \\
\psi^{L}_{n}(\tilde{t}) &=
\frac{1}{\sqrt{2^n n!}} (\frac{\varepsilon}{2\pi })^{1/4} 
e^{-\xi_L \tilde{t}^2}H_n \left[\sqrt{\frac{\varepsilon}{2}} \tilde{t}\right], \\
\end{aligned}
\end{equation}
where 
$\xi_{R,L}=\frac{1}{4b}(\sqrt{\eta}  \pm a)$, 
$\varepsilon=\frac{1}{b}\sqrt{\eta}$ and
$H_n(x)$ is the Hermite polynomial : $H_0(x)=1$, $H_1(x) =2x$, $H_2(x) =4x^2-2$, 
$\dots$.

The spectrum of the Hamiltonian is determined by the parameters in the $\beta$-functions, $a$, $b$ and $w$.
First, the eigenvalues of the Hamiltonian are real  for  $\eta > 0$. 
Second, the eigenstates are square-integrable for $\xi_{R,L} > 0$.
These conditions are satisfied for
$b>0$ and $w<0$ for any real $a$.
In the following, we first focus on this parameter region that supports a real spectrum with normalizable eigenstates.
At the end of this section, we will see that violation of these conditions is associated with a loss of stable fixed point in the IR.

\subsubsection{Fixed point \label{sec:gs_0d_example}}

A generic initial state evolves to the right ground state
of $\tilde{\mathcal{H}}$ 
in the large $z$ limit,
\begin{equation}
\begin{aligned}
\psi^{R}_{0}(\tilde{t}) &=
 (\frac{\varepsilon}{2\pi })^{1/4} 
e^{-\xi_R \tilde{t}^2}.
\end{aligned}
\end{equation}
As  discussed in Sec.~\ref{sec:reconstruction}, 
the right ground state 
corresponds to the stable fixed point of the theory. 
To extract the fixed point action,
we write the right ground state as
\begin{equation}
\begin{aligned}
| 0 \rangle
&= \int \mathcal{D} \phi \int_R \mathcal{D} \tilde{t} ~ \psi^{R}_{0}(\tilde{t})  e^{- S_0+i \tilde{t} \mathcal{O}}  | \phi \rangle \\
&=  (\frac{\varepsilon}{2\pi })^{1/4}  \sqrt{\frac{\pi}{\xi_R}}  \int \mathcal{D} \phi ~  e^{- S_0- \frac{1}{4\xi_R}\mathcal{O}^2 }  | \phi \rangle.
\end{aligned}
\end{equation}
The logarithm of its wavefunction  in the $\phi$ basis gives the fixed point action,
\bqa
S^* & = &
 S_0 + J_1^\ast \mathcal{O} + J_2^\ast \mathcal{O}^2 
 \label{eq:0Dfp}
\eqa
with
$J_1^\ast=0$ and $J_2^\ast=\frac{1}{4\xi_R}$. 
We emphasize that the stable fixed point exists away from the subspace of the single-trace coupling,
yet the position of the fixed point is fully determined from the beta functions defined in the subspace.

\subsubsection{Scaling operators and their OPEs\label{sec:0d_example_excitation}}

Now we turn our attention to excited states of $\tilde{\mathcal{H}}$. 
Each right eigenstate corresponds to the fixed point theory with an operator insertion. 
The $n$-th excited state is 
\begin{equation}
\begin{aligned}
| n \rangle &= \int \mathcal{D} \phi   \int_R \mathcal{D} \tilde{t} ~ \psi^{R}_{n}(\tilde{t})  
 e^{- S_0+i \tilde{t} \mathcal{O}} | \phi \rangle \\
&=  \int \mathcal{D} \phi  \frac{1}{\sqrt{2^n n!}} (\frac{\varepsilon}{2\pi })^{1/4} \sqrt{\frac{\pi}{\xi_R}}    e^{-S_0} H_n \left[-i\sqrt{\frac{\varepsilon}{2}}  \frac{\partial}{\partial \mathcal{O}}\right]  e^{-\frac{1}{4\xi_R}  \mathcal{O}^2}  | \phi \rangle.
\end{aligned}
\end{equation}
The excited states can be reached by
applying `raising' operators to the ground state,
\begin{equation}
\begin{aligned}
| n \rangle & =   \hat{\mathcal{A}}_n | 0 \rangle, 
\end{aligned}
\end{equation}
 where $\hat{\mathcal{A}}_n$ is the operator that maps the ground state
 to the $n$-th excited states that has scaling dimension $\sqrt{\eta}n$.
In general, the $n$-th scaling operator is given by a linear superposition of all $k$-trace operators with $k=n, n-2, n-4, ..$.
For $n=0,1,2,3,4$, we obtain
\bqa
\hat{\mathcal{A}}_0  &=&  \mathbbm{1}, \nn 
\hat{\mathcal{A}}_1  &=&  i \frac{\sqrt{\varepsilon} }{2\xi_R}  \mathcal{O}, \nn
\hat{\mathcal{A}}_2  &=&  \frac{1}{\sqrt{2}} (\frac{\varepsilon}{2\xi_R}-1)  \mathbbm{1} 
- \frac{1}{\sqrt{2}}\frac{\varepsilon}{4\xi_R^2}  \mathcal{O}^2, \nn
\hat{\mathcal{A}}_3 &=& - \frac{i \sqrt{\varepsilon^3}}{8\sqrt{6}\xi_R^3} \mathcal{O}^3 + \frac{i \sqrt{3\varepsilon}}{2\sqrt{2}\xi_R}\Big(\frac{\varepsilon}{2\xi_R}-1\Big)\mathcal{O}, \nn
\hat{\mathcal{A}}_4 &=& \frac{\varepsilon^2}{32\sqrt{6}\xi_R^4}\mathcal{O}^4 +\frac{\sqrt{3}\varepsilon}{4\sqrt{2}\xi_R^2}\Big(1-\frac{\varepsilon}{2\xi_R}\Big)\mathcal{O}^2+\frac{\sqrt{3}}{2\sqrt{2}} (1-\frac{\varepsilon}{2\xi_R})^2.
\label{eq:0draisingOp}
\eqa
It is straightforward to identify all scaling operators in this way.

The OPE coefficient can be computed accordingly. 
For instance, two $\hat{\mathcal{A}}_1$ fuse to 
\begin{equation}
\hat{\mathcal{A}}_1 \times \hat{\mathcal{A}}_1 =   (1-\frac{\varepsilon}{2\xi_R})  \hat{\mathcal{A}}_0 + \sqrt{2}\hat{\mathcal{A}}_2, 
\end{equation}
and the associated OPE coefficients are given by 
$C_{110}=   (1-\frac{\varepsilon}{2\xi_R}) $, 
$C_{112}=\sqrt{2}$. 
Similarly, all OPE coefficients can be extracted from the eigenstates of the  RG Hamiltonian. 
In Tab.~\ref{tab:ope}, we list all OPE for $\hat{\mathcal{A}}_n \times \hat{\mathcal{A}}_m$ up to $n,m=2$.

\begin{table}[h]
    \centering
    \caption{
    Operator product expansion of 
    $\hat{\mathcal{A}}_{n} \times 
    \hat{\mathcal{A}}_{m}$
     for $n,m=0,1,2$.
}
    \begin{tabular}{|c|c|c|c|}
    \hline
           \backslashbox{$m$}{$n$}        &  $0$ & $1$ & $2$ \\ \hline 
     $0$&  $\hat{\mathcal{A}}_0$ & $\hat{\mathcal{A}}_1$ & $\hat{\mathcal{A}}_2$ \\ \hline
     $1$&  $\hat{\mathcal{A}}_1$ & $(1-\frac{\varepsilon}{2\xi_R})  \hat{\mathcal{A}}_0 + \sqrt{2}\hat{\mathcal{A}}_2$ & $\sqrt{2}(1-\frac{\varepsilon}{2\xi_R})\hat{\mathcal{A}}_1 +\sqrt{3}\hat{\mathcal{A}}_3$ \\ \hline
     $2$  & $\hat{\mathcal{A}}_2$  & $\sqrt{2}(1-\frac{\varepsilon}{2\xi_R})\hat{\mathcal{A}}_1 + \sqrt{3}\hat{\mathcal{A}}_3$ & $(1-\frac{\varepsilon}{2\xi_R})^2\hat{\mathcal{A}}_0 + 2\sqrt{2}(1-\frac{\varepsilon}{2\xi_R}) \hat{\mathcal{A}}_2 +\sqrt{6}\hat{\mathcal{A}}_4$ \\ \hline
    \end{tabular}
    \label{tab:ope}
\end{table}

\subsubsection{Full $\beta$-functions}

Based on Sec.~\ref{sec:beta_func}, we can immediately write down the full $\beta$-functions of the theory.
In $0$ dimension, there is no $x$ dependence of the couplings and OPE coefficients.
By setting $D=0$ and $C_{lmn}(x)=c_{lmn}$ in \eq{eq:epsilonprime}, 
we readily obtain  the $\beta$-function,
\begin{equation}
\begin{aligned}
\frac{d}{dz} J_{n}  &= - \Delta_n (J_{n}-J_{n}^\ast) \\
&+ \frac{1}{2} ( \Delta_n - \Delta_l - \Delta_m) c_{lmn}  (J_{l}-J_{l}^\ast)(J_{m}-J_{m}^\ast)  + O\Big((J-J^\ast)^3 \Big),
\label{eq:0Dlinearbeta}
\end{aligned}
\end{equation}
From \eq{eq:EnH} that implies $\Delta_n = \sqrt{\eta} n$  and Table \ref{tab:ope},
we obtain the beta functions for the couplings of $\hat{\mathcal{A}}_1$ and $\hat{\mathcal{A}}_2$, 
\bqa
\frac{d}{dz} J_{1}  &=& - \sqrt{\eta}\left[1+ \frac{\varepsilon}{2\sqrt{2}\xi_R^2} -\frac{1}{\sqrt{2}\xi_R} \right] J_{1} 
-2 \sqrt{2\eta}  (1-\frac{\varepsilon}{2\xi_R}) J_{1} J_{2}  + O\Big((J-J^\ast)^3 \Big), 
\label{eq:0DJ1} \\
\frac{d}{dz} J_{2}  & = &
 \frac{\sqrt{\eta}}{2\xi_R} + \frac{\varepsilon \sqrt{ \eta}}{8\sqrt{2}\xi_R^3}-\frac{\sqrt{ \eta}}{4\sqrt{2}\xi_R^2}  \nn
&& - 2\sqrt{\eta}\left[ 1 +\frac{\varepsilon}{2\sqrt{2}\xi_R^2}-\frac{1}{\sqrt{2}\xi_R} \right]J_{2} 
-2 \sqrt{2 \eta}(1-\frac{\varepsilon}{2\xi_R}) J_{2}^2 + O\Big((J-J^\ast)^3 \Big),
 \label{eq:0DJ2}
\eqa
where $\frac{d}{dz} J_{n} = -\beta_n$.
Although $\beta_{n>2}=0$ in the subspace of single-trace coupling, they are in general non-zero away from the subspace.
It is straightforward to compute $\beta_n$ for any $n$ order by order in $(J-J^*)$.
This shows that the full beta functions are indeed encoded in the  RG Hamiltonian  
that is fixed by the beta functions defined in the space of single-trace couplings.

\begin{figure}
    \centering
    \includegraphics[width=.98\textwidth]{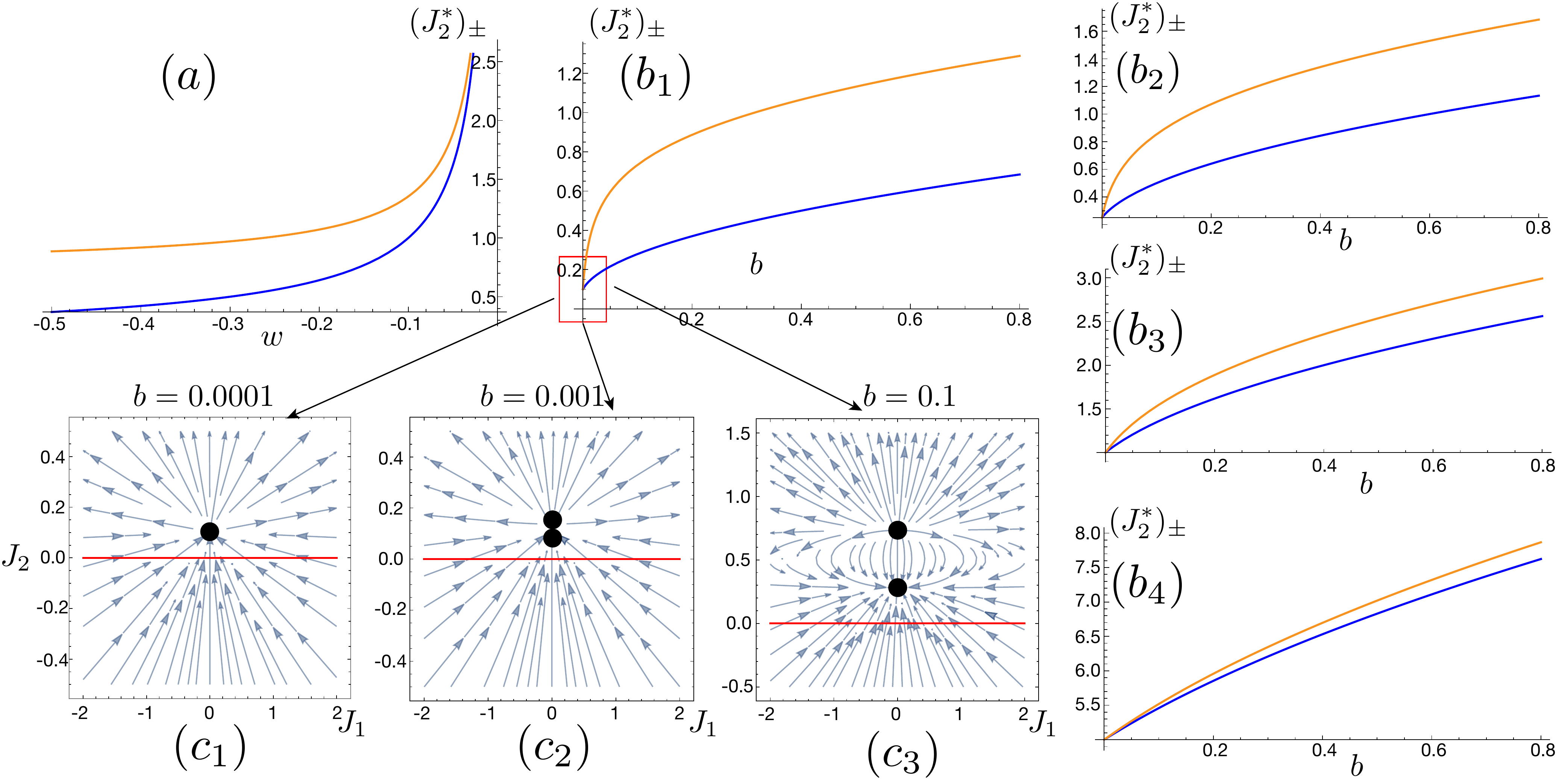}
    \caption{
   The double-trace coupling $J_2$ at the stable (blue line) and unstable (orange line) fixed point 
   as a function of $w$ at $b=0.2$ (a); 
   as a function of $b$ at $w=-0.5$ ($\rm b_1$), $w=-0.2$ ($\rm b_2$), $w=-0.05$ ($\rm b_3$), $w=-0.01$ ($\rm b_4$). 
   Three RG flow diagrams at $w=-0.5$ and $b=0.0001$ ($\rm c_1$), $b=0.001$ ($\rm c_2$), $b=0.1$ ($\rm c_3$), $b=0.5$ (\fig{fig:RG_flow}) are also presented. 
   $a=-0.1$ is used for all plots.
  }
    \label{fig:fp_move}
\end{figure}

The beta functions for multi-trace operators allow us to explore the RG flow away from the subspace of the single-trace coupling.
Eqs. (\ref{eq:0DJ1}) and (\ref{eq:0DJ2}) 
computed to the quadratic order in $\delta J = (J-J^*)$ can be trusted near $J=J^*$.
To describe the RG flow far away from the stable fixed point, one needs to take into account terms that are higher order in $\delta J$ and higher trace couplings. 
Here we focus on the flow in the space of $J_1$ and $J_2$ near $J=J^*$.
To the quadratic order in $\delta J$,
$J_n$ with $n>2$ are not generated,
and we can trust 
Eqs. (\ref{eq:0DJ1}) and (\ref{eq:0DJ2}) 
near $J=J^*$.
The RG flow in the space of $J_1$ and $J_2$ is shown in  \fig{fig:RG_flow} for $a=-0.1$, $b=0.5$ and $w=-0.5$. 
We find two fixed points at
\bqa
(J_1^\ast, J_{2}^\ast)_- &=& \left(0, 
\frac{1 }{4\xi_R } 
\right), \label{eq:stablefp} \\
(J_1^\ast, J_{2}^\ast)_+ &=& \left(0,  
\frac{1 }{4\xi_R } +
\frac{ \sqrt{2} \xi_R }{ (\varepsilon  -2\xi_R) } 
\right), \label{eq:unstablefp}
\eqa
where $0<(J_{2}^\ast)_- <  (J_{2}^\ast)_+  $ 
because  $\xi_R>0$ and $\varepsilon>2\xi_R$. 
In either the small $\xi_R$ or large $\varepsilon $ limit, 
 the second fixed point is close to the stable fixed point,
and the terms that are higher order in $\delta J$ in  Eqs. (\ref{eq:0DJ1}) and (\ref{eq:0DJ2}) are negligible near these two fixed points.
The first fixed point in  \eq{eq:stablefp} is the stable fixed point  
identified  from the ground state of the RG Hamiltonian in \eq{eq:0Dfp}.
Both $\delta J_1$ and $\delta J_2$ are irrelevant 
whose scaling dimensions are $- \sqrt{\eta}$ and $-2 \sqrt{\eta}$, respectively.
The second fixed point in  \eq{eq:unstablefp} is an unstable fixed point.
Both $\delta J_1$ and $\delta J_2$ are relevant 
with scaling dimensions, $\sqrt{\eta}$ and $2 \sqrt{\eta}$, respectively. 
In  \fig{fig:fp_move}(a) and \fig{fig:fp_move}($\rm b_{1\sim  4}$), we plot the value of $J_2$ at the two fixed points as $b$ and $w$ are varied, respectively.
For $b>0$ and $w<0$, the spectrum of the RG Hamiltonian is real and the eigenstates are normalizable.
In this case, 
$\xi_R $ and $\varepsilon-2\xi_R $ are both positive and finite such that
the two fixed points remain separated,
as shown in \fig{fig:fp_move}($\rm c_{2\sim 3}$).
The RG flow changes qualitatively if  $b$ or $w$ approaches $0$.
In  Appendix \ref{app:limitingcases}, 
we examine the RG flow in the 
$w \rightarrow 0^-$ 
and 
$b \rightarrow 0^+$ 
limits in more details.

\subsection{$D$-dimensional solvable example \label{sec:dd_example}}

In this section, we extend the discussion in the previous section to a $D$-dimensional field theory.
For simplicity, 
we continue to assume that there is only one single-trace operator,
and that multi-trace operators higher than double-trace operator
are not generated when the reference action is deformed 
only by the single-trace operator.
We also assume that the reference theory is invariant under the spatial translation, the rotation and the inversion symmetry,
and has an internal $\mathbbm{Z}_2$ symmetry
under which the single-trace operator is odd.
The symmetry largely fixes the form of the $\beta$-functionals in the subspace of  the single-trace couplings order by order in the coupling.
To be concrete, we consider the following $\beta$-functionals in the subspace of single-trace couplings,
\bqa
\beta_0 \left[t,0,..; x\right] &=& f - g  \left[\partial_x t(x)\right]^2  - w t^2(x), \nn \beta_1\left[t,0,..; x\right] &=& a t(x)
- x \partial_x t(x), \nn
\beta_2 \left[t,0,..; x\right]&=& -b, \nn
\beta_{k \geq 3} \left[t,0,..; x\right]&=& 0. 
\label{eq:DDbeta}
\eqa
Here  
$\beta_k[t^{(1)},t^{(2)},t^{(3)},..;x]$ represents the
$\beta$-functional for the $k$-trace operator at 
$\left( t^{(1)},t^{(2)},t^{(3)},..\right)$, 
where $t^{(m)}$ is the $m$-trace coupling.
$(\partial_x t)^2 \equiv 
\sum_{\mu =1}^D ( \partial_\mu t )^2
$,
and
$x \partial_x t \equiv 
\sum_{\mu =1}^D x^\mu  \partial_\mu t 
$.
$f, g, w, a, b$ are constants
that represent the contributions to the beta functions
generated from integrating out short distance modes
and rescaling the fundamental fields at every RG step.
The last term in  $\beta_1$ dilates the space  
because the coordinate in the $(l+1)$-th RG step 
is related to that in the previous step through
$x^{(l+1)} =x^{(l)} e^{-dz}$.
The rescaling makes sure that the UV cutoff remains  invariant under the RG flow, and the same coarse graining can be applied at all steps.
On the other hand, 
the rescaling of space reduces the size of the system in real space
by $e^{-dz}$ at every RG step.

\eq{eq:DDbeta}
fixes the bulk theory in \eq{eq:Lagrangian_evolution},
which in turn determines the fate of the field theory in the low-energy limit.
The wavefunction that fully determines the renormalized action at scale $z$ is given by  the path integration of the single-trace source and its conjugate variable,
\bqa
\Psi[ t, z ] 
&= \left.
\int_I  \mathcal{D} t (x,z') \mathcal{D} p(x,z') ~
\Psi_{J_1,J_2,...}[ t(0) ]
e^{- \int_0^{z} dz' L\left[ t, p, z' \right]}
\right|_{t(z) = t}.
\label{eq:zdependentstate}
\eqa
While the bulk Lagrangian is quadratic in the present case,
it depends on $x$ explicitly because of the dilatation term 
in $\beta_1$ of  \eq{eq:DDbeta}.
This gives rise to a mixing between different Fourier modes in the momentum space\footnote{
The mixing arises because 
the momentum in the $(l+1)$-th RG step is related to the momentum
in the previous step as  $k^{(l+1)} =k^{(l)} e^{dz}$. }.
The mixing makes it hard to compute the path integration directly. 
To bypass this problem, we follow the three steps described below.
\begin{enumerate}
\item
We introduce new variables in the bulk,
\bqa
{t}(x,z) =  -i \tilde{t}(x e^{z} ,z) e^{\frac{D}{2}z}, \quad 
{p}(x,z) = i \tilde{p}(x e^{z},z) e^{\frac{D}{2}z}.
\label{eq:change_variable}
\eqa
Besides the rescaling of spatial coordinate that undoes the dilatation,
the fields are also multiplied with a factor
$e^{\frac{D}{2}z}$ to compensate the $z$-dependent volume of the space.
$\pm i$ is multiplied so that $\tilde t$ and $\tilde p$ fluctuate
along the real axis.
In the new variables, 
the dilatation effect disappears and 
Fourier modes with different momenta do not mix
as will be shown later.
\item
The path integration in 
\eq{eq:zdependentstate}
is performed in $\tilde t$ and $\tilde p$.
This is done in the Hamiltonian picture.
\item
The scale transformation is reinstated
by expressing the $z$-dependent state in terms of 
\bqa
t'(x,z) =   \tilde{t}(x e^{z} ,z) e^{\frac{D}{2}z}, \quad 
p'(x,z) = \tilde{p}(x e^{z},z) e^{\frac{D}{2}z}.
\label{eq:change_variable2}
\eqa
\end{enumerate}
In the following sections,
we implement these steps 
to identify the IR fixed point 
and the spectrum of scaling operators
at the fixed point.

\subsubsection{The RG Hamiltonian}

In terms of the variables 
introduced in \eq{eq:change_variable},
the bulk Lagrangian is written as
\begin{equation}
\begin{aligned}
L \left[ t, p,z \right] 
&=\int^{1/\lambda} d^D X \Big(
  i\tilde{p} \partial_z \tilde{t}
  + i \frac{D}{2}\tilde{p}\tilde{t}
  - \sum_{n\geq0} \tilde \beta_{n}[ \tilde t;X]   \tilde{p}^n e^{\frac{(n-2)D}{2}z}
\Big),
\end{aligned}
\end{equation}
where
\bqa
\tilde \beta_1[ \tilde t; X] & = & \beta_1[ t; x] - i e^{\frac{D}{2} z}  X \partial_{X} \tilde t, \nn
\tilde \beta_n[ \tilde t; X] & = & \beta_n[ t; x] ~~~~ \mbox{for $n \neq 1$}.
\label{eq:beta_transf}
\eqa
with $X=xe^z$.
$\tilde{t}(X)$ and $\tilde{p}(X)$ obey the canonical commutation relation,
\begin{equation}
\left[ \tilde{t}(X),\tilde{p}(X') \right]= - i \delta(X-X').
\end{equation}
The RG Hamiltonian density is given by
\begin{equation}
\begin{aligned}
\mathcal{H} \left[ \tilde{t}, \tilde{p},z \right]&=
   i \frac{D}{2}\tilde{p}\tilde{t}
  - \sum_{n\geq0} \tilde \beta_{n}[ \tilde t;X]   \tilde{p}^n e^{\frac{(n-2)D}{2}z}.
\end{aligned}
\end{equation}
By shifting the Hamiltonian by a constant, 
we write the Hamiltonian density as
\begin{eqnarray}
\tilde{\mathcal{H}} \left[ \tilde{t}, \tilde{p},z \right] &=& \mathcal{H} \left[ \tilde{t}, \tilde{p},z \right]
+e^{-Dz}f 
-\frac{1}{2}(a-\frac{D}{2}) 
\delta(0)
\\
&=&  -g e^{2z}\left[\partial_{X} \tilde{t}(X)\right]^2-w \tilde{t}^2(X)  + i\frac{1}{2}(a+\frac{D}{2}) \left[\tilde{t}(X)\tilde{p}(X)+\tilde{p}(X)\tilde{t}(X)\right] + b\tilde{p}^2(X).\nonumber
\label{eq:DDH}
\end{eqnarray}
As expected,  the dilatation in \eq{eq:DDbeta}
cancels with that in \eq{eq:beta_transf}.
Instead, the RG Hamiltonian acquires explicit $z$ dependence.

In the Fourier basis,
\begin{equation}
\begin{aligned}
\tilde{t}(X) = \frac{1}{\sqrt{V}} 
\sum_{K} e^{i K \cdot X} \tilde{t}_{K} ,
\quad \tilde{p}(X) = \frac{1}{\sqrt{V}} 
 \sum_{K} e^{i K \cdot X}\tilde{p}_{K},
  \label{eq:fourier} 
\end{aligned}
\end{equation}
where $V = \lambda^{-D}$ is the volume of the system,
the RG Hamiltonian can be written as
\begin{equation}
\begin{aligned}
\tilde{\mathcal{H}}(z) &
&= \sum_{K} \tilde{h}_{K}, 
\end{aligned}
\end{equation}
where
\bqa
 \tilde{h}_K = b \Big\{ 
\left[ \tilde{p}_{K} +i\zeta \tilde{t}_{K} \right]\left[ \tilde{p}_{-K} +i\zeta\tilde{t}_{-K} \right] 
+\Omega_{K,z}^2   \tilde{t}_{K} \tilde{t}_{-K} \Big\}
\eqa
with
\begin{equation}
\begin{aligned}
\Omega_{K,z}^2 &= \sigma  + \alpha e^{2z} K^2.
\label{eq:Omega}
\end{aligned}
\end{equation}
Here, $\zeta=\frac{1}{2b}(a+\frac{D}{2})$, 
$\sigma= \frac{1}{4b^2}(a+\frac{D}{2})^2-\frac{w}{b}$ 
and 
$\alpha=-g/b$. 
Henceforth, we set $b=1/2$, resulting in $\zeta =a+\frac{D}{2}$, $\sigma=\zeta^2 -2w$ and 
$\alpha=-2g$.
$\tilde{t}_{K}$ and $\tilde{p}_{-K}$ are 
canonical conjugate variables that satisfy
$\left[  \tilde{t}_{K}, \tilde{p}_{K'} \right] = 
-i \delta_{K,-K'}$. 
While $\tilde{t}_{K= 0}$ and $\tilde{p}_{K= 0}$ are real,  
$\tilde{t}_{K\neq 0}$ and $\tilde{p}_{K\neq 0}$ are complex 
with $\tilde{t}_{K}=\tilde{t}_{-K}^\ast$ and $\tilde{p}_{K}=\tilde{p}_{-K}^\ast$.
The Hamiltonian can be decomposed into a sum of  time-dependent harmonic oscillators,
\begin{equation}
\tilde{\mathcal{H}}(z) =\tilde{h}_0 +\sum_{K>0}^{'} (\tilde h_{(R;K)} +\tilde h_{(I;K)} ),
\end{equation}
where $\sum_{K>0}^{'}$ runs over the half of non-zero momenta with $K$ identified with $-K$,
and 
\begin{equation}
\begin{aligned}
\tilde{h}_0 &= b \Big\{ 
\left[ \tilde{p}_{0} +i\zeta \tilde{t}_{0} \right]^2
+\Omega_{0,z}^2   \tilde{t}_{0}^2  \Big\}, \\
\tilde h_{(R;K>0)} &= b \Big\{ 
 \left[ \tilde{p}_{(R;K)} +i\zeta \tilde{t}_{(R;K)} \right]^2 
+\Omega_{K,z}^2   (\tilde{t}_{(R;K)})^2  \Big\}, \\
\tilde h_{(I;K>0)} &= b \Big\{ 
 \left[ \tilde{p}_{(I;K)} +i\zeta \tilde{t}_{(I;K)} \right]^2 
+\Omega_{K,z}^2   (\tilde{t}_{(I;K)} )^2  \Big\} \\
\end{aligned}
\end{equation}
with $\tilde{t}_{(R(I);K)}=\sqrt{2}{\rm Re (Im)~}\tilde{t}_{K}$ and $\tilde{p}_{(R(I);K)}=\sqrt{2}{\rm Re (Im)~}\tilde{p}_{K}$ that satisfy the commutation relation 
$\left[\tilde{t}_{(S;K)}, \tilde{p}_{(S';K')}\right]=-i\delta_{K,-K'}\delta_{SS'}$ with $S,S'=I,R$.

The RG flow is described by the imaginary time Schrodinger equation,
 \bqa
 \tilde{\mathcal{H}}(z) \Psi \left[\tilde{t},z\right] = -\frac{\partial}{\partial z} \Psi \left[\tilde{t},z\right]. \label{eq:schrodinger_DDim}
 \eqa
The three parameters $\zeta$, $\sigma$ and $\alpha$ fully determine the solution $\Psi \left[\tilde{t},z\right]$. 
The problem of the harmonic oscillator with time-dependent frequency 
has been studied extensively in Refs. ~\cite{burgan1979solution,cheng1988evolution,guasti2003solution},
which is reviewed in Appendix \ref{app:time_dependent_Schrodinger}. 
We consider a UV theory obtained by adding the single-trace and double-trace couplings 
to the reference theory $S_0$ in a translationally invariant way.
In this case, the initial wavefunction 
is a Gaussian product state in the $K$-space.
Because the Hamiltonian is non-interacting, 
$\Psi \left[\tilde{t},z\right]$ remains 
Gaussian at all $z$.
The solution is written as 
\bqa
\Psi \left[\tilde{t},z\right]= 
 \Psi_{0}\left[\tilde{t}_0,z\right]
\prod_{K > 0 }^{'} \Big\{ \Psi_{(R;K)}\left[\tilde{t}_{(R;K)},z\right] \Psi_{(I;K)}\left[\tilde{t}_{(I;K)},z\right] \Big\},
\label{eq:Productstate}
\eqa
where $\prod_{K > 0}^{'} $ runs over the half of the non-zero momenta.
The wavefunction for each mode satisfies 
$\tilde{h}_s \Psi_{s}\left[\tilde{t}_s,z\right]= -\frac{\partial}{\partial z} \Psi_{s}\left[\tilde{t}_s,z\right]$, 
where the subscript $s$ stands for $0$, $(R;K)$ or $(I;K)$.
The initial state can be written as 
\bqa
\Psi_{s}\left[\tilde{t}_s,0\right] &=& \sum_{m} c_{m} \Psi_{m,s}\left[\tilde{t}_s,0\right],
\eqa
where $\{  \Psi_{m,s}\left[\tilde{t}_s,0\right]  \}$
represents the eigenstates of the Hamiltonian $\tilde{h}_s$ at $z=0$
and $\{ c_{m}\}$ is a set of $z$-independent coefficients. 
Under the RG flow, the state evolves to
\bqa
\Psi_{s}\left[\tilde{t}_s,z\right] &=& \sum_{m} c_{m} \Psi_{m,s}\left[\tilde{t}_s,z\right], 
 \label{eq:superposition_modes}
\eqa
where 
\begin{equation}
\begin{aligned}
\Psi_{m,s}\left[\tilde t_s ,z\right] 
&=
\frac{1}{ \pi^{1/4} \sqrt{2^{m} m!}} 
e^{-\frac{1}{2}\Delta_{s,z}}
 \exp \left[-\frac{1}{2\Lambda_{s,z} }\tilde{t}_{s}^2  \right]  \times\\
& \exp\left[\frac{\omega_{s,z}}{2A_{s,z}^2} \tilde t_s^2 \right]
H_{m} \left[   -\frac{A_{s,z}}{\sqrt{\Omega_{s,0}}} \frac{\delta}{\delta  \tilde t_s } \right]  
\exp\left[-\frac{\omega_{s,z}}{2A_{s,z}^2} \tilde t_s^2 \right].
\label{eq:general_modes}
\end{aligned}
\end{equation}
Here 
$\omega_{s,z}=\left[\int_0^z \frac{dz'}{A^2_{s,z'}} +\frac{1}{\Omega_{s,0}} \right]^{-1}$.
$A_{s,z}$ is a function that satisfies $\ddot{A}_{s,z} -  A_{s,z} \Omega_{s,z}^2=0$ 
with $A_{s,0}=1$ and $\dot{A}_{s,0}=0$
 ( $\dot{A} \equiv \partial_z A$).
$e^{-\Delta_{s,z}}=\frac{\omega_{s,z}}{A_{s,z} \sqrt{\Omega_{s,0}}}$.
$\frac{1}{\Lambda_{s,z}} = \zeta + \frac{\dot{A}_{s,z}}{A_{s,z}}+\frac{ \omega_{s,z}}{A_{s,z}^2}$.
At $z=0$,
$\frac{1}{\Lambda_{s,z}}$ is reduced to $\Omega_{s,0}+\zeta$,
and  $e^{-\Delta_{s,z}}$ becomes $\sqrt{\Omega_{s,0}}$.

Finally, the $z$-dependent state is written in terms of 
the variables in \eq{eq:change_variable2},
\bqa
\Psi'[ t', z ] 
&=& 
 \Psi_{0}\left[ t_0',z\right]
\prod_{K>0}^{'} \Psi_{(R;K)}\left[  t_{(R;K e^{z}) }', z\right] \Psi_{(I;K)}\left[  t_{(I;K e^{z}) }', z\right] \nn
&=& 
 \Psi_{0}\left[t_0',z\right]
\prod_{ k>0}^{'} \Psi_{(R; k e^{-z}) }\left[   t_{(R; k)}', z\right] \Psi_{(I; k e^{-z}) }\left[   t_{ (I;k)}', z\right].
\eqa
Here we use $\tilde{t}_K
=   t'_{ k}$ for the Fourier modes,
where $ x = e^{-z} X$ and $ k = e^z K$~\footnote{
This follows from
\bqa
\tilde{t}_K
 & =& \int^{1/\lambda}  \frac{d^D X}{\sqrt{V}} e^{-iK X} \tilde t(X,z)  
  = \int^{1/\lambda}  \frac{d^D X}{\sqrt{V}} e^{-iK X}   e^{-\frac{D}{2}z} t'(X e^{-z},z)  \nn
 & =& \int^{e^{-z}/\lambda}  \frac{d^D  x}{\sqrt{V e^{-Dz}}} e^{-i   k  x}    t'(  x,z)  
  =  t'_{ k}. \nonumber
 \label{eq:inversefourier} 
\eqa
}.
For a finite system size, $k$ and $K$ are discrete,
\bqa
K  =  \frac{2 \pi}{L} ( n_1, n_2, .., n_D),  &~~~~&
k  =  \frac{2 \pi e^z }{L} ( n_1, n_2, .., n_D), 
\eqa 
where 
$L= V^{1/D}$ is the linear system size
and 
$n_i$'s are integers.

Our next goal is to extract the fixed point of the full Wilsonian RG and local scaling operators with their scaling dimensions from the scale dependent state obtained from the quantum RG.
As discussed in the previous sections,
the asymptotic ground state that emerges in the large $z$ limit corresponds to the stable fixed point, 
and eigenstates with local excitations and eigenvalues give scaling operators and scaling dimensions, respectively.
However, it is not easy to extract the asymptotic state in the large $z$ limit because the RG Hamiltonian is $z$-dependent.
Even if one prepares an initial state to be an eigenstate of the instantaneous RG Hamiltonian at $z=0$, the state does not remain the same under the RG evolution as is shown in \eq{eq:general_modes}.
Therefore, we use the following strategy.
Given that the RG Hamiltonian is invariant under the $\mathbbm{Z}_2$ symmetry, we consider a generic initial state in each of 
the  $\mathbbm{Z}_2$ even sector and 
the $\mathbbm{Z}_2$ odd sector. 
Under the quantum RG flow, those initial states evolve within each sector as
\bqa
\lim_{z \rightarrow \infty}
| \Psi^+ (z) \rangle &=& \sum_{n} 
e^{-\mathcal{E}^+_n z}
| n; + \rangle, \nn
\lim_{z \rightarrow \infty}
| \Psi^- (z) \rangle &=& \sum_{n} 
e^{-\mathcal{E}^-_n z}
| n; - \rangle, 
\eqa
where $|n, \pm \rangle$ corresponds to the
eigenstates of the RG Hamiltonian that emerges in the large $z$ limit in each parity sector, and 
$\mathcal{E}^\pm_n$ is the corresponding eigenvalue.
From this, we identify the eigenstate with the lowest eigenvalue in the even sector as the ground state that represents the stable IR fixed point.
The excited states in each parity sector  correspond to the states obtained by deforming the ground state with scaling operators with the corresponding $\mathbbm{Z}_2$ parity and scaling dimension, $\mathcal{E}^{\pm}_n - \mathcal{E}^+_0$.

\subsubsection{Fixed point
}

In this section, we identify the IR fixed point of the theory from quantum RG.
As an initial state, 
we choose the ground state 
of the instantaneous RG Hamiltonian at $z=0$, 
which has the translational invariance 
and even $\mathbbm{Z}_2$ parity,
\bqa
\Psi\left[\tilde{t},0\right]
=
 \Psi_{0,K=0}\left[\tilde{t}_0,0\right]
\prod_{K>0}^{'}
\Psi_{0,(R;K)}[\tilde t_{(R;K)}, 0] \Psi_{0,(I;K)}[\tilde t_{(I;K)}, 0].
\eqa
In the large $z$ limit, 
the $z$-dependent wavefunction for each $s$-mode becomes
\begin{equation}
\begin{aligned}
\Psi_{0,s}\left[\tilde{t}_s,z\right] &= 
\pi^{-1/4}  
e^{-\frac{1}{2}\Delta_{s,z}}   \exp \left[-\frac{1}{2\Lambda_{s,z} }\tilde{t}_{s}^2  \right].
 \label{eq:00mode}
\end{aligned}
\end{equation}
The asymptotic many-body wavefunction is written as 
\begin{equation}
\begin{aligned}
\Psi \left[\tilde{t},z\right] 
&= \mathcal{N}'(z)
\exp \left[-\sum_{K}\frac{\tilde{t}_{K}\tilde{t}_{-K}}{2\Lambda_{K,z} } \right],
\end{aligned}
\end{equation}
where  
$\mathcal{N}'(z)= \left[\prod_{K }  \pi^{-1/4}  e^{-\frac{1}{2}\Delta_{K,z}} \right]$ and $\Lambda_{K,z}$
are expressed in terms of $\alpha$, $\zeta$ and $\sigma$ 
as
\begin{equation}
\begin{aligned}
\Lambda_{K,z}
=\left[\mathbbm{G}_{K e^z} (\alpha,\sigma)+\zeta\right]^{-1},
\label{eq:Lambda_general}
\end{aligned}
\end{equation}
where 
\begin{equation}
\mathbbm{G}_k(\alpha,\sigma)=\frac{1}{2}\sqrt{\alpha} \frac{|k|}{I_{\sigma} \left[\sqrt{\alpha}|k|\right]}  (I_{-1+\sqrt{\sigma}}\left[ \sqrt{\alpha} |k| \right] + I_{1+\sqrt{\sigma}}\left[ \sqrt{\alpha} |k| \right])
\end{equation}
in the large $z$ limit with fixed $k = K e^z$ 
(see Appendix \ref{app:QD} for the details).
This shows that $\Lambda_{K,z}$ converges to a $z$-independent function when viewed as a function of $k$.
Physically, this is due to the fact that the scale invariance becomes
manifest if one zooms in toward the $K=0$ point 
progressively as $z$ increases.
The overall normalization of the wavefunction decreases with increasing $z$ 
due to the damping associated with the imaginary time evolution
\footnote{
The normalization factor $\mathcal{N}'(z)$ is determined by 
\begin{equation}
e^{-\Delta_{K,z}}  \approx \frac{\sigma^{1/4}}{2\mathbbm{A}_{Ke^z}(\alpha,\sigma)}e^{-\sqrt{\sigma}z} \nonumber
\label{eq:Delta_large_z_limit0}
\end{equation}
in the large $z$ limit, where $\mathbbm{A}_{k}$ is a function of $k=Ke^z$(\eq{eq:AAk_app} in Appendix \ref{app:QD}),
\begin{equation}
\mathbbm{A}_{k}(\alpha,\sigma)=- \frac{ 2^{-1+\sqrt{\sigma}}\pi \alpha^{-\frac{1}{2}\sqrt{\sigma}} }{ \sin (\pi \sqrt{\sigma})}  \frac{I_{ \sqrt{\sigma}} \left[ \sqrt{\alpha} |k|\right]  }{\Gamma(-\sqrt{\sigma})}  |k|^{-\sqrt{\sigma}} \nonumber
\label{eq:AA}
\end{equation}
}.

To show that the wavefunction approaches a scale invariant asymptotic in the large $z$ limit,
we need to go back to the scaled variable in 
\eq{eq:change_variable2}.
The wavefunction for $t'_{ k}, p'_{ k}$ is written as
\begin{equation}
\begin{aligned}
\Psi'[t',z] =
\mathcal{N}'(z) \exp \Big(-\sum_{k}\frac{1}{2 \tilde{\Lambda}_{k}} {t'}_{k} {t'}_{-k}\Big), 
\label{eq:largezground}
\end{aligned}
\end{equation}
where $\tilde \Lambda_{ k} = 
\Lambda_{k e^{-z},z}$.
In the large $z$ limit for a fixed $k$, $\tilde \Lambda_{ k}$  
takes the following forms 
(see Appendix \ref{app:QD}),
\bqa
\tilde{\Lambda}_{k} & = &
\left\{
\begin{array}{cc}
\frac{1}{(\sqrt{\sigma}+\zeta)}   & ~~ \mbox{for  $|k| \ll 1$}, \\ 
\frac{1}{  \left[\sqrt{\alpha}|k| +\zeta\right]} & ~~ \mbox{for  $|k| \gg 1$}
\end{array}
\right..
\eqa 
This confirms that in the large $z$ limit 
$\Psi'[t',z] $ evolves to a $z$-independent state
up to the $z$-dependent normalization factor.

Similar to what we studied in Sec.~\ref{sec:gs_0d_example}, 
the state in the large $z$ limit encodes the information on the IR fixed point. 
Defining $J_{2,k}^\ast \equiv \tilde \Lambda_k$,
we rewrite the asymptotic state in the large $z$ limit as
\bqa
\lim_{z\rightarrow \infty} |\Psi(z) \rangle  
&=& \mathcal{N}'(z) 
\int \mathcal{D} \phi~
 e^{- S_0}  \int_R \mathcal{D} t'
e^{ - \Big(
\frac{1}{2}
\sum_k   (J_{2,  k}^\ast)^{-1}  {t'}_k {t'}_{-k}
-i \sum_k t'_{k} \mathcal{O}_{-k}
 \Big)
 } | \phi \rangle \nn 
 &=& 
 \mathcal{N}(z) 
\int \mathcal{D} \phi~
 e^{- S^*}  | \phi \rangle,
 \label{eq:stable_fp}
\eqa
where 
$\mathcal{N}(z) = \mathcal{N}'(z) \det{[ 2 \pi J_2^* ]}^{1/2} $
and the fixed point action $S^*$ 
is given by
\begin{equation}
S^*= S_0+ \frac{1}{2} 
\sum_k
J_{2,k}^\ast 
\mathcal{O}_{k}  
\mathcal{O}_{-k} = S_0 + \frac{1}{2}\int d^Dx d^Dx' J_{2,x-x'}^\ast \mathcal{O}_{x}  
\mathcal{O}_{x'},
\end{equation}
where
\bqa
 J_{2,x-x'}^\ast = \frac{1}{ V e^{-Dz}} \sum_k J_{2,k}^\ast e^{-ik(x-x')} = \int \frac{d^D  k}{(2 \pi)^D} \tilde{\Lambda}_{k} e^{-i k(x'-x)}. 
  \label{eq:J2xx}
\eqa
Here we use 
$\mathcal{O}_k =\frac{1}{\sqrt{Ve^{-Dz}}} \int d^D x \mathcal{O}_x e^{-ikx}$.

\begin{figure}[h]
\includegraphics[width=.98\textwidth]{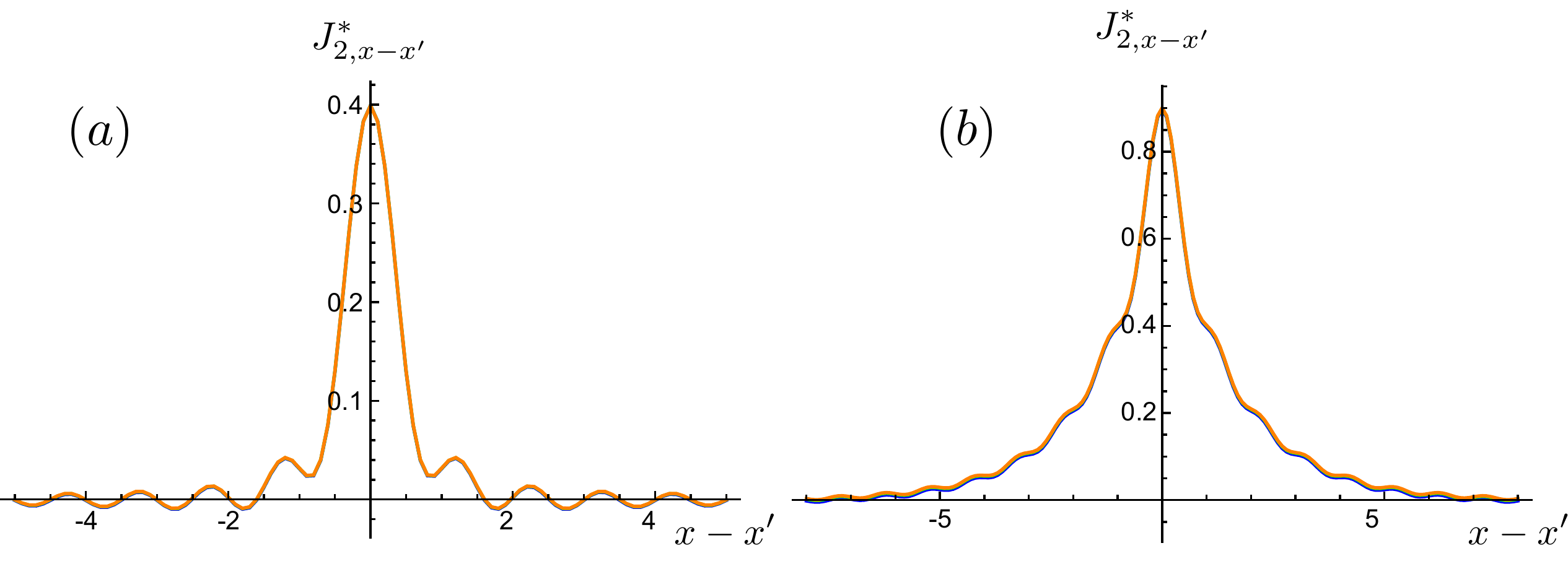}
\caption{
(a) $J_{2,x-x'}^\ast$ 
as a function of $x-x'$ at $\sigma=2.01$ and $\zeta=-0.1$ for $D=1$ at $z=22$ (orange), $z=23$ (green), $z=24$ (blue). 
For the computation, the lattice regularization is used with the total number of sites $L/a=e^{30}$. 
(b) $J_{2,x-x'}^\ast$  at $\sigma=0.01$ and $\zeta=0.1$ with the corresponding value of $z$ for each color as in (a). $\alpha=1$.
\label{fig:non_Hermitian_j2x} 
}
\end{figure}

As is shown in 
\fig{fig:non_Hermitian_j2x} (see Appendix \ref{app:fixed_J2xx} for  the details),
$\lim_{z \rightarrow \infty} J_{2,x-x'}$ converges to a universal profile in the thermodynamic limit.
$J_{2,x-x'}^\ast$ is peaked at $x-x'=0$ with a finite width that is order of  the short distance cutoff.
It decays exponentially at large $|x'-x|$. 
We emphasize that the IR fixed point that exists away from the subspace of the single-trace couplings
has been extracted solely from the \bfnc that are defined 
in the subspace.

\subsubsection{Scaling operators
}

In this  subsection, we extract  scaling operators from excited states of the RG Hamiltonian.

We first consider the $\mathbbm{Z}_2$ 
odd sector.
In the  $\mathbbm{Z}_2$  odd sector,
we consider an initial state in which
one of the Fourier modes is excited.
Suppose that the mode with momentum $P$ is in the first excited state with respect to the RG Hamiltonian at $z=0$,
where the momentum is measured in the coordinate system defined at $z=0$.
In the large $z$ limit, the state evolves to 
(see Appendix \ref{app:one_excited_modes} for derivation)
\begin{equation}
\begin{aligned}
|\Psi_{1,P}(z)\rangle 
  =(i\sqrt{2})\mathcal{N}(z) \int \mathcal{D}\phi ~ 
  \Big(  \frac{\sigma^{1/4}}{2\mathbbm{A}_{P e^z }}  \tilde \Lambda_{Pe^z } e^{-\sqrt{\sigma}z} \mathcal{O}_{P e^z} \Big)e^{-S^\ast}
  | \phi \rangle.
\label{eq:Psi1pz}
\end{aligned}
\end{equation}
$\mathcal{N}(z)$ is the  normalization of the ground state defined in \eq{eq:stable_fp}. 
Compared to the ground state,  the weight of the first excited state  with a definite momentum decays as $e^{-\sqrt{\sigma}z}$ in the large $z$ limit.
The state that supports an excitation at 
$ P \neq 0$ at $z=0$
can not be invariant under the RG flow
because a non-zero $P$ is pushed toward larger momenta 
in the large $z$ limit due to the rescaling.
Namely, a source that is added periodically in space at UV 
flows to a periodic source with a shorter wavelength at larger $z$
when measured in the rescaled coordinate system.  

The excited state with $P=0$ is an exception.
In the presence of a uniform source, the excited state flows to
 a scale invariant state in the large $z$ limit.
Using 
$\mathcal{O}_{p} = 
\frac{1}{\sqrt{Ve^{-Dz}}} \int d^Dx ~ e^{-i px } \mathcal{O}_{x}$
for the Fourier transformation at $z$, where $p=Pe^z$ and $x=Xe^{-z}$ ,
we rewrite \eq{eq:Psi1pz} for $P=0$ as
\bqa
|\Psi_{1,0}(z)\rangle =
e^{-(\sqrt{\sigma}-D/2)z}
\frac{ i\sqrt{2} \mathcal{N}(z)  \sigma^{1/4} \tilde \Lambda_{0} }{2\sqrt{V} \mathbbm{A}_{0}} 
\int \mathcal{D}\phi ~ 
\left( \int d^D x   ~ \mathcal{O}_{x} \right) e^{-S^\ast}
  | \phi \rangle.
\eqa
The first excited state with the uniform source flows to
$\int \mathcal{D}\phi ~ 
\left( \int d^Dx   ~ \mathcal{O}_{x} \right) e^{-S^\ast}
  | \phi \rangle$ in the large $z$ limit
  with the $z$-dependent amplitude, 
$e^{-(\sqrt{\sigma}-D/2)z}$ 
relative to the ground state.
This implies that the spatially uniform deformation of the 
$\mathbbm{Z}_2$ odd
single-trace operator
is relevant (irrelevant) if  $\sqrt{\sigma}<D/2$ ($\sqrt{\sigma}>D/2$).

\begin{figure}[h]
\includegraphics[width=1.0\textwidth]{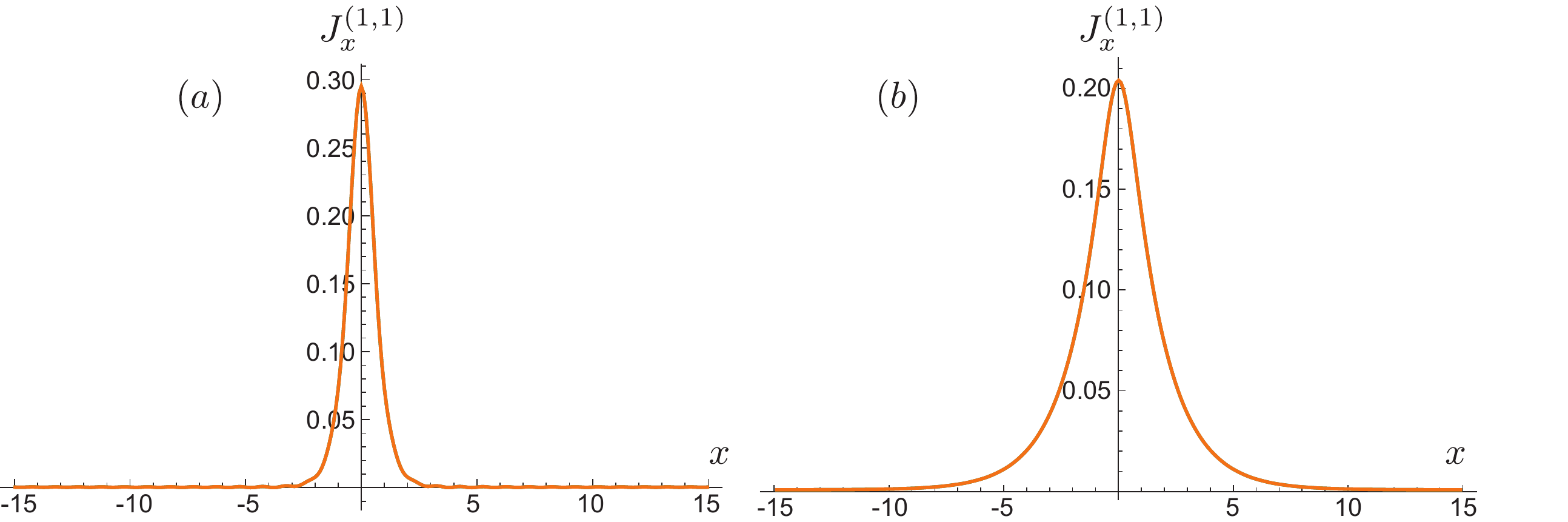}
\caption{ $J_{x}^{(1,1)}$ at $z=22$ (orange), $z=23$ (green), $z=24$ (blue) with $L/a= e^{30}$ in $D=1$. We set parameters to be $\alpha=1$ and (a) $\sigma=2.01$ and $\zeta=-0.1$; 
(b) $\sigma=0.01$ and $\zeta=0.1$.
\label{fig:jx_evolve_n_fixed} }
\end{figure}

The other type of eigenstates that are invariant under the RG evolution is the ones that support excitations localized in space.  
In order to find local scaling operators associated with states with local excitations,
we consider an initial state in which the single-trace operator is inserted at $X$. 
For $z > 0$, the state becomes
\bqa
| \Psi_{1,X} (z) \rangle &=& 
\frac{1}{\sqrt{V}}
\sum_{P} e^{iPX} | \Psi_{1,P}(z) \rangle.
\label{eq:real_space_one_excited_mode}
\eqa
In the large $z$ limit, the state evolves to 
\footnote{
\begin{equation}
\begin{aligned}
\lim_{z \rightarrow \infty} | \Psi_{1,X}(z)  \rangle
&=
\frac{(i\sqrt{2}) \mathcal{N}(z)}{\sqrt{V }}
\int \mathcal{D} \phi~
 \Big(   \sum_{P} \frac{ \sigma^{1/4}}{2\mathbbm{A}_{P e^z} } \tilde{\Lambda}_{Pe^z} e^{- \sqrt{\sigma} z} 
 {\mathcal{O}}_{Pe^z} 
 e^{iPX}
 \Big)
   e^{-S^\ast} | \phi \rangle \\
&=
\frac{(i\sqrt{2}) \mathcal{N}(z) e^{- \sqrt{\sigma} z} }{\sqrt{V }}
\int \mathcal{D} \phi~
 \Big(   \sum_{p} \frac{ \sigma^{1/4}}{2\mathbbm{A}_{ p} } \tilde{\Lambda}_{ p} 
 {\mathcal{O}}_{ p} 
 e^{i  p (X e^{-z})}
 \Big)
   e^{-S^\ast} | \phi \rangle \\
&= \frac{(i\sqrt{2}) \mathcal{N}(z) e^{-\sqrt{\sigma}z}}{V e^{-Dz/2} }
\int \mathcal{D} \phi~
 \Big(  \int d^D x' \sum_{ p} \frac{ \sigma^{1/4}}{2\mathbbm{A}_{ p} } \tilde{\Lambda}_{ p} 
 e^{i p(e^{-z} X-x')}{\mathcal{O}}_{x'} 
 \Big)
   e^{-S^\ast} | \phi \rangle \\
 &=(i\sqrt{2}) \mathcal{N}(z) 
 e^{-(\sqrt{\sigma}+D/2)z}
\int \mathcal{D} \phi~
\hat{\mathcal{A}}_{1}( Xe^{-z})
   e^{-S^\ast} | \phi \rangle, \nonumber
\end{aligned}
\end{equation}
where we used $\frac{1}{V e^{-Dz}} \sum_{ p} = \int \frac{ d^D p}{(2 \pi)^D}$ at $z$.
}
\begin{equation}
\begin{aligned}
\lim_{z \rightarrow \infty} | \Psi_{1,X}(z)  \rangle
 &=(i\sqrt{2}) \mathcal{N}(z) 
 e^{-(\sqrt{\sigma}+D/2)z}
\int \mathcal{D} \phi~
\hat{\mathcal{A}}_{1}( Xe^{-z})
   e^{-S^\ast} | \phi \rangle,
\label{eq:Psi1x}
\end{aligned}
\end{equation}
where
\bqa
\hat{\mathcal{A}}_{1}(x) =
  \int d^D x'
  J_{ x-x' }^{(1,1)}
  {\mathcal{O}}_{x'} 
\label{eq:scalingop_1_x}
\eqa
with
\bqa
J_{x}^{(1,1)}
=   \int \frac{d^D  p}{(2\pi)^D} 
\frac{ \sigma^{1/4}}{2\mathbbm{A}_{ p} } 
\tilde{\Lambda}_{ p}  
e^{ i  p x}. 
 \label{eq:coupling_single_trace_z}
\eqa
$\hat{\mathcal{A}}_{1}(x)$ inserts a single-trace operator around $x$ with distribution  given by $J_{x-x'}^{(1,1)}$.
Henceforth, we use $J^{(n,m)}$ to denote the contribution of the $m$-trace operator to the $n$-th scaling operator.
The local operator inserted at $X$ at the UV boundary evolves to a distribution of local operators centered at $X e^{-z}$ at $z>0$.
The shift of the central position is due to the rescaling of the space.
In the large $z$ limit, the local operator evolves to  $\hat{\mathcal{A}}_{1}(0)$,
which we identify
as the local scaling operator inserted at the origin.
The broadening of the distribution in $J_{x}^{(1,1)}$ is due to the correlation in the fluctuations of the single-trace coupling at the fixed point.
Irrespective of the initial profile of the local operator at the UV,
it converges to the universal profile $J_{x}^{(1,1)}$  at large $z$  in the thermodynamic limit. 
In \fig{fig:jx_evolve_n_fixed}, we numerically plot $J_{x}^{(1,1)}(z)$ which converges to a $z$-independent profile in the large $z$ limit.
Compared to the ground state,
the overall weight of the first excited state with the local excitation decays 
as $e^{-(\sqrt{\sigma}+D/2)z}$ in the large $z$ limit. 
This implies that the operator $\mathcal{O}_{x}$ has scaling dimension 
$\Delta_1 = \sqrt{\sigma}+D/2$. 
This is consistent with the fact that the operator is relevant (irrelevant) if 
$\sqrt{\sigma} < D/2$ ($\sqrt{\sigma} > D/2$).

If we impose the 
$\mathbbm{Z}_2$ 
symmetry,
the  
$\mathbbm{Z}_2$ 
odd operator is not allowed.
To see if the low-energy fixed point is stable in the presence of the $\mathbbm{Z}_2$ symmetry, 
we need to consider local scaling operators in the even parity sector.
The operator with the smallest scaling dimension in the even sector is the identity operator. 
In the following section, we obtain the next lowest scaling operator in the even sector.

Now, let us consider the $\mathbbm{Z}_2$ even sector.
Excited states in the  $\mathbbm{Z}_2$ even sector should include even number of excited modes.
Let us consider
an initial state with two excited modes labelled by momenta $P$ and $P'$.
Under the RG evolution, the state in general evolves into a linear superposition of the ground state (for $P+P'=0$)
and excited states.
Since we are interested in the excited state above the ground state,
we discard the slowest decaying state (the ground state).
The state with the next slowest decaying amplitude
in the large $z$ limit is given by
(see Appendix \ref{app:two_excited_modes})
\begin{equation}
\begin{aligned}
| \Psi_{2, P,  P' }(z) \rangle &=-\sqrt{2}
\mathcal{N}(z) 
e^{-2\sqrt{\sigma}z} 
\int \mathcal{D}\phi ~  \Big( \frac{\sigma^{1/2}}{4\mathbbm{A}_{Pe^z} \mathbbm{A}_{P'e^z}} \tilde{\Lambda}_{Pe^z} \tilde{\Lambda}_{P'e^z} 
\mathcal{O}_{Pe^z} \mathcal{O}_{ P'e^z} \\
&-
\delta_{P+P',0} \left[
\frac{\sigma^{1/2}}{4\mathbbm{A}_{Pe^z}^2} \tilde{\Lambda}_{Pe^z} +\frac{1}{2}\frac{\sqrt{\sigma}}{\sqrt{\alpha P^2+\sigma}}\mathbbm{W}_{Pe^z}\right]
\Big)e^{-S^\ast} |\phi\rangle,
\label{eq:two_excited_modes}
\end{aligned}
\end{equation}
where
$\mathcal{N}(z) $ encodes the rate at which the ground state decays
and 
$e^{-2\sqrt{\sigma}z} $ is the additional decay for the next slowest decaying state.
Again, the state with non-zero momenta can not be invariant under the RG evolution
due to the rescaling that shifts momenta to larger values with increasing $z$. 
To find a local scaling operator, we consider the state that evolves from an initial state that supports local excitations at positions $X$ and $X'$,
\bqa
|\Psi_{2,X,X'}(z)\rangle 
&=\frac{1}{V}
 \sum_{P,P'}e^{iPX+iP'X' }| \Psi_{2,P, P' }(z) \rangle.
 \eqa
 In the large $z$ limit, the state evolves to
\footnote{
\begin{equation}
\begin{aligned}
& |\Psi_{2,X,X'}(z)\rangle 
=\frac{1}{V}
 \sum_{P,P'}e^{iPX+iP'X' }| \Psi_{2,P, P' }(z) \rangle \\
&=-\sqrt{2}\mathcal{N}(z) e^{-2\sqrt{\sigma}z} \int \mathcal{D}\phi ~   
\Big( \frac{1}{V}\sum_{p,p'}e^{ip(Xe^{-z})+ip'(X'e^{-z}) } J_{p,p'}^{(2,2)}
\mathcal{O}_{p} \mathcal{O}_{ p'} 
-\frac{1}{V}\sum_{p}e^{ip(X-X')e^{-z}} J_{p}^{(2,0)} \mathbbm{1}\Big)e^{-S^\ast} |\phi\rangle \\
&=-\sqrt{2} \mathcal{N}(z)  e^{-(2\sqrt{\sigma}+D)z} \int \mathcal{D}\phi ~   \Big(\int d^Dyd^Dy' 
\left[
 \frac{1}{V^2  e^{-2Dz} }
 \sum_{p,p'}e^{ip(Xe^{-z}-y)+ip'(X'e^{-z}-y') } J_{p,p'}^{(2,2)} 
\right]
\mathcal{O}_{y} \mathcal{O}_{ y'}\Big)e^{-S^\ast} |\phi\rangle \\
&+ \sqrt{2} \mathcal{N}(z) e^{-(2\sqrt{\sigma}+D)z} \int \mathcal{D}\phi ~  
\Big(  
 \frac{1}{V  e^{-Dz} }
 \sum_{p} e^{ip(X-X') e^{-z}} 
 J_{p}^{(2,0)} 
\mathbbm{1}\Big)e^{-S^\ast} |\phi\rangle,  \nonumber
\label{eq:Psi2xx2}
\end{aligned}
\end{equation}
where $p=P e^z$ and
\begin{equation}
\begin{aligned}
J_{p,p'}^{(2,2)} = 
\frac{\sigma^{1/2}}{4\mathbbm{A}_p \mathbbm{A}_{p'}} \tilde{\Lambda}_{p} \tilde{\Lambda}_{p'}, ~~~
J_{p}^{(2,0)} =\frac{\sigma^{1/2}}{4\mathbbm{A}_p^2} \tilde{\Lambda}_p+\frac{\sqrt{\sigma}}{2\sqrt{\alpha p^2e^{-2z}+\sigma}}\mathbbm{W}_{p} \approx \frac{\sigma^{1/2}}{4\mathbbm{A}_p^2} \tilde{\Lambda}_p+\frac{1}{2}\mathbbm{W}_{p}.  \nonumber
\label{eq:J2pp2} 
\end{aligned}
\end{equation}
in the large $z$ limit.
}
\begin{equation}
\begin{aligned}
|\Psi_{2,X,X'}(z)\rangle 
&= -\sqrt{2} \mathcal{N}(z) e^{-(2\sqrt{\sigma}+D)z}  \int \mathcal{D}\phi ~   \hat{\mathcal{A}}_{2}( X e^{-z}, X'e^{-z}) e^{-S^\ast} |\phi\rangle,
\label{eq:Psi2xx}
\end{aligned}
\end{equation}
where
\begin{eqnarray}
\hat{\mathcal{A}}_2(x,x') =\int d^D y d^D y' J_{x-y,x'-y'}^{(2,2)}\mathcal{O}_y \mathcal{O}_{y'}- J_{x-x'}^{(2,0)} \mathbbm{1} \label{eq:scalingop_2_x}
\end{eqnarray}
with
\begin{eqnarray}
J_{x,x'}^{(2,2)}
&=& \int \frac{d^Dp d^Dp'}{(2\pi)^{2D}} ~
\frac{\sigma^{1/2}}{4\mathbbm{A}_p \mathbbm{A}_{p'}} \tilde{\Lambda}_{p} \tilde{\Lambda}_{p'}
e^{ipx+ip'x'}, \nonumber \\
J_{x}^{(2,0)} &=&
\int \frac{d^Dp}{(2\pi)^D} ~ e^{ipx} 
\left[\frac{\sigma^{1/2}}{4\mathbbm{A}_p^2} \tilde{\Lambda}_p +\frac{1}{2}\mathbbm{W}_p\right].
\end{eqnarray}

\begin{figure}[h]
\includegraphics[width=.8\textwidth]{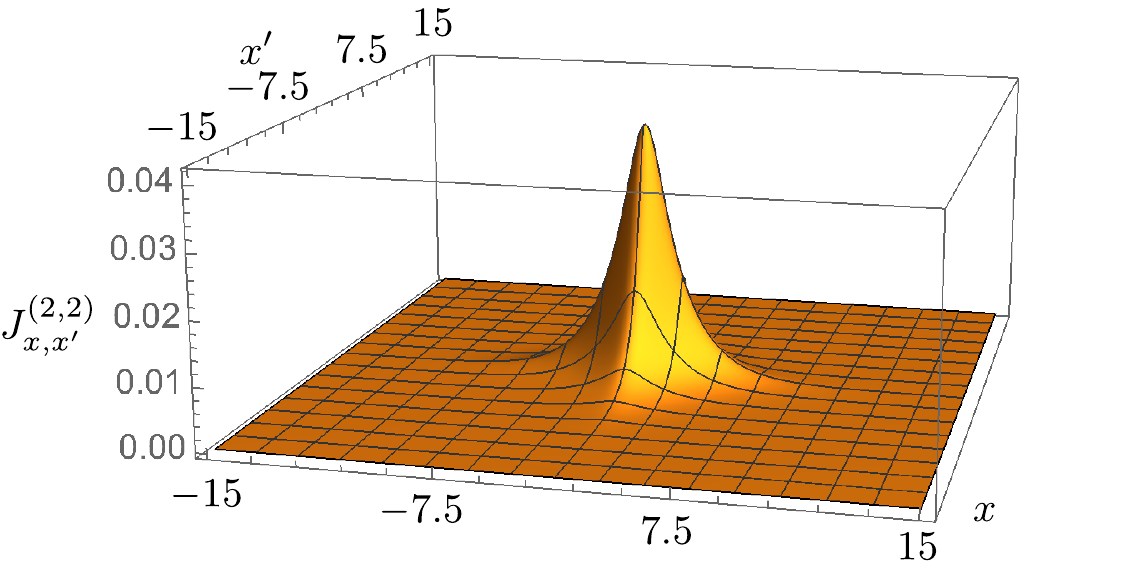}
\caption{
The profile of $J_{x,x'}^{(2,2)}$ as a function of $x$ and $x'$ at $z=25$ for a system with $L/a=e^{30}$ in $D=1$. 
We use $\alpha=1$, $\zeta=0.1$ and $\sigma=0.01$ for the plot. 
\label{fig:J2xxz}}
\end{figure}

$\hat{\mathcal{A}}_{2}(x,y)$
is a composite operator that supports
two single-trace operators centered at position $x$ and $y$ respectively.
In the large $z$ limit, the initial state flows to the state obtained by applying
$\hat{\mathcal{A}}_{2}(0,0)$ to the ground state.
Therefore,  we identify
$\hat{\mathcal{A}}_{2}(0,0)$
as the lowest scaling operator  above the identity operator in the even sector. 
It is noted that 
$\hat{\mathcal{A}}_{2}(0,0)$
is a linear superposition of a double-trace operator and an identity operator. 
This is because the double-trace operator and the identity operator mix under the RG flow.
$J_{x,x'}^{(2,2)}$, that describes the distribution of the two single-trace operators, can be written as 
$J_{x,x'}^{(2,2)} = J_{x}^{(1,1)} J_{x'}^{(1,1)}$,
and its profile in the real space is determined by that of $J_{x}^{(1,1)}$. 
In \fig{fig:J2xxz}, $J_{x,x'}^{(2,2)}$ is shown as a function of $x$ and $x'$ at a fixed $z$. 
It has a peak at the origin and decays exponentially away from the peak with the width that is comparable to the short distance cutoff scale. 
According to \eq{eq:Psi2xx}, the local deformation induced by this scaling operator decays with rate $2\sqrt{\sigma}+D$ relative 
to the ground state. 
Thus, its scaling dimension of the local operator is $2\sqrt{\sigma}+D$ which is twice of the single-trace operator. 
It is an irrelevant operator due to $\sqrt{\sigma}>0$.

It can be shown that general scaling operators have scaling dimensions given by
$n\left( \sqrt{\sigma}+\frac{D}{2} \right)$ 
for $n=1,2,3,..$.
See Appendix  \ref{appendix:OCO}  for the details.
All operators in the even sector are irrelevant for $\sqrt{\sigma}>0$.
This shows that the fixed point in \eq{eq:stable_fp} is stable  in the presence of the $\mathbbm{Z}_2$ symmetry.
The fact that the scaling dimensions are additive is a feature of the generalized free theory for which the bulk RG Hamiltonian is quadratic. 
For general theories whose RG Hamiltonian is not quadratic,  this is no longer the case.
It will be of great interest to consider large $N$ theories whose RG Hamiltonian includes interactions that are suppressed by $1/N$, and compute $1/N$ corrections to the scaling dimensions from the quantum RG.

\subsubsection{Operator Product Expansion}

Now we consider the OPE between two parity-odd operators with the lowest scaling dimension.
For this, we insert $\hat{\mathcal{A}}_1$ at $X$ and $-X$ in the stable fixed point theory.
The deformed theory corresponds to the initial state with two local excitations,
\begin{eqnarray}
|\Psi_{1\times 1;X,-X}(0)\rangle &=&-\sqrt{2}\mathcal{N}(0) \int \mathcal{D}\phi \hat{\mathcal{A}}_1(X)\hat{\mathcal{A}}_1(-X)e^{-S^\ast} |\phi \rangle.
\label{eq:A1_times_A1}
\end{eqnarray}
Then, following the steps explained in Sec.~\ref{sec:ope}, we evolve the state with the RG Hamiltonian  for $z= \ln |\frac{X}{a}|$
to obtain
\begin{eqnarray}
|\Psi_{1\times 1;X,-X}(z)\rangle  &=&-\sqrt{2}\mathcal{N}(z) e^{-(2\sqrt{\sigma}+D)z} \int \mathcal{D}\phi \hat{\mathcal{A}}_1(a )\hat{\mathcal{A}}_1(-a )e^{-S^\ast} |\phi \rangle  \nn
&=&  -\sqrt{2}\mathcal{N}(z)e^{-(2\sqrt{\sigma}+D)z}\int \mathcal{D}\phi \int d^D y d^D y' J_{a -y}^{(1,1)} J_{-a -y'}^{(1,1)} \mathcal{O}_y\mathcal{O}_{y'} e^{-S^\ast} |\phi \rangle, \nn
\label{eq:11XX}
\end{eqnarray}
where $a$ is the short distance cutoff length scale.
In the second equality, we use the expression for  $\hat{\mathcal{A}}_1$ in \eq{eq:scalingop_1_x}. 
By using \eq{eq:scalingop_2_x} for $\hat{\mathcal{A}}_2$, we rewrite 
\eq{eq:11XX} as
\footnote{ $F_{n,n',l}(x)$ introduced in \eq{eq:merging} is independent of $x$ because the RG Hamiltonian is quadratic in this case.}
\begin{eqnarray}
|\Psi_{1\times 1;X,-X}(z)\rangle &=& -\sqrt{2}\mathcal{N}(z)e^{-(2\sqrt{\sigma}+D)z} \int \mathcal{D}\phi \left[ 
 J_{2a}^{(2,0)} \mathbbm{1}
 +
\hat{\mathcal{A}}_2(a,-a) 
\right] e^{-S^\ast} |\phi \rangle.
\end{eqnarray}
We expand $\hat{\mathcal{A}}_2(a,-a)$ in $a$  to obtain
\begin{eqnarray}
|\Psi_{1\times 1;X,-X}(z)\rangle&=& -\sqrt{2}\mathcal{N}(z)e^{-(2\sqrt{\sigma}+D)z} \int \mathcal{D}\phi \Big[ 
 J_{2a}^{(2,0)} \mathbbm{1}
+ 
\hat{\mathcal{A}}_2(0,0) 
+...
 \Big] e^{-S^\ast} |\phi \rangle,
\end{eqnarray}
where the ellipsis includes derivative terms such as 
$a \partial_x\hat{\mathcal{A}}_2(x,y)|_{x=y=0}$ and $-a \partial_y\hat{\mathcal{A}}_2(x,y)|_{x=y=0}$.
Finally, the backward evolution for RG time $-z$ restores the initial state,
\begin{eqnarray}
|\Psi_{1\times 1;X,-X}(0)\rangle =  -\sqrt{2}\mathcal{N}(0)\int \mathcal{D}\phi \Big[ 
e^{-(2\sqrt{\sigma}+D)z} J_{2a}^{(2,0)}\mathbbm{1}
+ 
\hat{\mathcal{A}}_2(0,0) 
+...
\Big] e^{-S^\ast} |\phi \rangle.
\end{eqnarray}
Comparing this with \eq{eq:A1_times_A1}, we obtain the OPE for two $\hat{\mathcal{A}}_1$ operators inserting at $X$ and $-X$ as
\begin{eqnarray}
\hat{\mathcal{A}}_1(X)\times \hat{\mathcal{A}}_1(-X) &=&
J_{2a}^{(2,0)}
\Big|\frac{a}{X} \Big|^{2 \Delta_1}  
\hat{\mathcal{A}}_0 
+ 
\hat{\mathcal{A}}_2(0,0) 
+ ...,
\label{eq:A1A1OPE}
\end{eqnarray}
where $\hat{\mathcal{A}}_0=\mathbbm{1}$
and $\Delta_1 = \sqrt{\sigma}+D/2$.
\eq{eq:A1A1OPE} shows the channels in which two single-trace operators fuse into a local double-trace operator with spin $0$ and the identity operator.
The ellipsis includes double-trace operators with larger spins and descendants.
Following the procedure explained in Sec.~\ref{sec:beta_func}, one can compute the $\beta$-functions for general multi-trace couplings.

\section{Conclusion and discussion \label{sec:conclusion}}

In this paper, we show that the full $\beta$-function of the  exact Wilsonian RG is completely fixed by the $\beta$-function  defined in the subspace of single-trace couplings.
We establish this general constraints on $\beta$-functions using the quantum RG, 
which is an exact reformulation of the Wilsonian RG.
In quantum RG, the conventional RG flow in the space of couplings is replaced with a quantum evolution of a wavefunction defined in the subspace of single-trace couplings, where fluctuations of the dynamical single-trace couplings encode the information about all multi-trace couplings.
Since the quantum evolution of the RG flow is completely determined from the $\beta$-functions defined in the subspace of single-trace couplings, the full Wilsonian $\beta$-functions can be extracted from the  $\beta$-functions defined in the subspace.
This is used to compute the full \bfnc of two concrete models :
the $O(N)$ vector model
and the $O_L(N) \times O_R(N)$ matrix model.

We also provide the general algorithm for extracting other field theory data such as scaling operators and OPE.
The general procedure consists of two steps. 
First, we construct the RG Hamiltonian that generates the quantum RG flow 
from the $\beta$-functions defined in the subspace of the single-trace couplings.
Second, we establish the correspondence between the ground state of the RG Hamiltonian with the stable IR fixed point.
Similarly, excited states with local excitations are mapped to the IR fixed point deformed with corresponding local operators.
The energies of the excited states determine the scaling dimensions of the local operators.
From the completeness of the eigenstates of the RG Hamiltonian, one can also extract the OPE coefficients among general operators and reconstruct the full $\beta$-functions.

We conclude with open  questions and future directions.
First, QRG can be used to compute the exact quantum effective action.
The scale dependence of the  quantum effective action obeys the exact RG equation\cite{POLCHINSKI1984269,WETTERICH199390}.
In general solving the exact RG equation is challenging because the exact effective action includes operators made of arbitrarily many fields and derivatives. 
As a result, exact effective actions remain unknown even for relatively simple theories. 
In QRG, the exact RG equation
is mapped to a quantum evolution of a wavefunction for single-trace couplings.
Since the set of single-trace operators is far smaller than the set of all possible operators, 
QRG can be potentially more tractable.
For general quantum field theories, it is still difficult to solve the   corresponding quantum evolution problem in QRG. 
However, in the large $N$ limit, quantum fluctuations of the single-trace couplings become weak,
and the theory that describes QRG evolution becomes classical.
In the large $N$ limit, 
the solution to the exact RG equation can be obtained 
from the saddle-point solution.
Recently, the exact effective action for the $O(N)$ vector model has been computed from QRG in the large $N$ limit\cite{future}.
It would be of great interest  to compute exact effective actions for matrix models in the large $N$ limit.
Second, QRG provides a concrete prescription for constructing the holographic duals for general quantum field theories\cite{lee2014quantum}.
The construction gives a well-defined bulk theory 
that includes dynamical gravity
as far as the boundary theory is regularized\cite{lee2012background,lee2016horizon,LEE2012781}. 
However, the continuum limit of the bulk theory obtained from regularized boundary theories such as lattice models is not fully understood.
It is of great interest to understand how the regularized bulk theory obtained from QRG is related to 
continuum theories conjectured as holographic duals of known field theories in the semi-classical limit.
Third, the $\beta$-functions in the subspace of single-trace couplings include the information on all fixed points that exist away from the subspace as is discussed 
in Sec. \ref{sec:0d_example}.
As multiple fixed points collide with a parameter of the theory tuned, the stable fixed point can disappear 
in the space of real couplings.
It will be of interest to understand how the loss of conformality or an appearance of non-unitary fixed points \cite{kaplan2009conformality,wang2017deconfined, gorbenko2018walking1,gorbenko2018walking2}
manifests itself in the bulk.
Finally, it would be interesting to consider cases in which the bulk RG Hamiltonian supports multiple ground states.
One can consider a few scenarios in which degenerate ground states for the RG Hamiltonian arise.
Degenerate ground states can be related to each other through symmetry, in which case the emergence of degenerate ground state is a sign of a spontaneous symmetry breaking. 
A degeneracy can also arise due to a suppression of tunneling between topologically distinct RG paths\cite{LEE2012781}.
An exactly marginal deformation can also give rise to degenerate ground states.

\section*{Acknowledgement}

Research at Perimeter Institute is supported in part by the Government of Canada through the Department of Innovation, Science and Economic Development Canada and by the Province of Ontario through the Ministry of Colleges and Universities.
SL acknowledges the support of the Natural Sciences and Engineering Research Council of Canada.

\bibliography{ref}

\appendix

\section{
Wilsonian RG with Gaussian action as reference action
\label{app:Gaussian_reference_action}
}

In this appendix, 
we show the form of 
an RG Hamiltonian 
that generates the exact RG flow for the 
$\phi^4$- theory starting from the deformed Gaussian action\cite{POLCHINSKI1984269,lee2016horizon}. 
Let us consider a $D$-dimensional scalar field theory whose Euclidean action is written as
\bqa
S= S_0 + S_1.
\eqa
Here $S_0$ is a quadratic reference action,
\bqa
S_0 = \frac{1}{2} \int d^D k~ G^{-1}_\Lambda(k) \phi_k \phi_{-k},
\eqa
where  $k$ is momentum,
and
$G^{-1}_\Lambda(k) = e^{ \frac{k^2}{\Lambda^2} } k^2$
is a regularized kinetic term
that suppresses fluctuations at momenta larger than UV cutoff $\Lambda$.
$S_1$ is a deformation that includes interactions and higher derivative terms.
The standard exact RG flow equation can be obtained by lowering the UV cut-off as $\Lambda \rightarrow \Lambda e^{-dz}$
followed by a rescaling of field and momentum,
$\phi_k \rightarrow e^{ \frac{D+2}{2} dz} \phi_{ e^{dz} k}$\cite{POLCHINSKI1984269}.
The correction to the effective action generated from this coarse graining is obtained by applying an RG Hamiltonian to the wavefunction $e^{-S}$ as
\bqa
e^{-H dz} e^{-S}
= 
e^{-S - \delta S},
\eqa
where $\delta S$ is the correction to the effective action,
and
\bqa
\hat H = 
e^{-S_0}
\int d^Dk \left[
  \frac{\tilde G(k)}{2}  
\pi_k \pi_{-k}
-
i \left( \frac{D+2}{2}  \phi_k + k \partial_k  \phi_k  \right) \pi_{-k}
+ C
\right]
e^{S_0}
\eqa
is the RG Hamiltonian. 
Here $\pi_k = -i \frac{\delta}{\delta \phi_{-k}}$ is
the conjugate momentum of $\phi_k$.
$\tilde G(k) = \frac{\partial G_\Lambda(k)}{\partial \ln \Lambda}$
is the propagator of the high-energy modes that are integrated out in the coarse graining scheme.
$C = - \int d^Dk \delta^D(0)
\left[
\frac{\tilde G}{2} G^{-1}_\Lambda + 1
\right]
$ is a constant.
One can check that the RG Hamiltonian leaves the trivial state invariant,
$\langle \mathbbm{1} | H = 0$,
and the partition function is invariant under the RG evolution,
\bqa
 \langle \mathbbm{1} | S \rangle 
=
\langle \mathbbm{1} | e^{-dz H} | S \rangle.
\eqa

\section{Solving Non-Hermitian Harmonic oscillator 
 \label{app:non_Hermitian}}

The RG Hamiltonian considered in the paper takes the following form,
\begin{equation}
H=\frac{1}{2m} \bm{\pi}_{x}^2 +\frac{1}{2} m \omega^2 (\bm{x}+i\gamma \bm{\pi}_x)^2,
\end{equation}
where  $\bm{x}$ ($\bm{\pi}_x$)
corresponds to  $p$ ($j$)
in the RG Hamiltonian in the main text.
The conjugate variables satisfy the commutation relation 
$\left[ \bm{x},\bm{\pi}_x\right]=i$. 
The RG Hamiltonian is invariant under
the $\mathcal{P}$ and $\mathcal{T}$ symmetries,
\begin{equation}
\begin{aligned}
\mathcal{P} \bm{x}\mathcal{P} = -\bm{x},\quad \mathcal{P} \bm{\pi}_x\mathcal{P} = -\bm{\pi}_x, \quad \mathcal{P} i\mathcal{P} = i, \\
\mathcal{T} \bm{x}\mathcal{T} = \bm{x},\quad \mathcal{T} \bm{\pi}_x\mathcal{T} = -\bm{\pi}_x, \quad \mathcal{T} i\mathcal{T} = -i.
\end{aligned}
\end{equation}

Under the similarity transformation $S = e^{\frac{\gamma}{2} \bm{\pi}_x^2}$,
the non-Hermitian RG Hamiltonian is transformed to an Hermitian RG Hamiltonian as
\begin{equation}
H_0 \equiv S H S^{-1}= 
\left[ \frac{1}{2m}\bm{\pi}_x^2 +\frac{1}{2} m \omega^2 \bm{x}^2 \right].
\end{equation}
In terms of the $n$-th eigenstate
$| \psi_n \rangle $ of $H_0$,
the right and left eigenstates of $H$
are given by
\bqa
|\psi_n^R \rangle =  
S^{-1} |\psi_n \rangle, & ~~~~~&
\langle \psi_n^L  |  =  
\langle \psi_n  | S
\eqa
with eigenvalue $(n+\frac{1}{2})\omega$. 
Their wavefunctions in the $\bm{\pi}_x$ basis are given by
\begin{equation}
\begin{aligned}
\psi^{R}_n(\bm{\pi}_x)=\langle \bm{\pi}_x | \psi_n^R \rangle &=
\mathcal{N}^R
e^{-\frac{1}{2}(\frac{1}{m\omega}  +\gamma)\bm{\pi}_x^2}H_n (\frac{\bm{\pi}_x}{\sqrt{m\omega}}), \\
\psi^{L}_n(\bm{\pi}_x)=\langle \bm{\pi}_x | \psi_n^L \rangle &=
\mathcal{N}^L e^{-\frac{1}{2}(\frac{1}{m\omega}  -\gamma)\bm{\pi}_x^2}H_n (\frac{\bm{\pi}_x}{\sqrt{m\omega}}),
\end{aligned}
\end{equation}
where 
$\mathcal{N}^{R(L)}$ are the normalization constants.
The right (left) eigenstates are normalizable if 
$\gamma > -\frac{1}{m\omega}$ 
($\gamma < \frac{1}{m \omega}$).

From the standard ladder operators of the Hermitian Hamiltonian,
\begin{equation}
a^\dag = \sqrt{\frac{m\omega}{2}} (\bm{x}-\frac{i}{m\omega}\bm{\pi}_x), \quad
a = \sqrt{\frac{m\omega}{2}} (\bm{x}+\frac{i}{m\omega}\bm{\pi}_x),
\end{equation}
the raising and lowering operators for the non-Hermitian Hamiltonian can be obtained via the similarity transformation,
\bqa
a_1 &=&
S^{-1} a^\dag S
=\sqrt{\frac{m\omega}{2}} \left[ \bm{x}+i(\gamma-\frac{1}{m\omega}) \bm{\pi}_x \right], \nn
a_2 &=&
S^{-1} a  S
=\sqrt{\frac{m\omega}{2}} \left[\bm{x}+i (\gamma+\frac{1}{m\omega})\bm{\pi}_x \right].
\eqa
The non-Hermitian RG Hamiltonian can be written in terms of $a_1$ and $a_2$ as
\begin{equation}
H= \omega a_1 a_2 +\frac{\omega}{2}.
\end{equation}

\section{\bfnc for the $0$-dimensional model 
in limiting cases}
\label{app:limitingcases}

In this appendix, we analyze the RG flow of the $0$-dimensional example
in various limits.

\begin{itemize}
\item $w \rightarrow 0^{-}$ limit with $a<0$ and $b>0$ : 

In the $w \rightarrow 0^{-}$ limit, $\xi_R$ approaches $0^+$. 
In this limit, the two fixed points are pushed to the region with large $J_2$ as is shown in 
\fig{fig:fp_move}(a).
As $\xi_R$ vanishes in the small $w$ limit, the eigenstates  of the RG Hamiltonian become non-normalizable.

\item $w \rightarrow 0^{-}$ limit with $a>0$ and $b>0$ : 

In this case, 
$\xi_R=\frac{a}{2b}$ and $\varepsilon=\frac{a}{b}$ are finite and positive,
and the right wave function is normalizable.
Because $\varepsilon-2\xi_R=0$, the $\beta$-functions become
\begin{eqnarray}
\frac{d}{dz} J_{1}  &=& - \sqrt{\eta} J_{1} 
  + O\Big((J-J^\ast)^3 \Big), \nonumber \\
\frac{d}{dz} J_{2}  & = &
 \frac{\sqrt{\eta}}{2\xi_R}  
 - 2\sqrt{\eta} J_{2} 
 + O\Big((J-J^\ast)^3 \Big) .\nonumber
\end{eqnarray}
The unstable fixed point is pushed to positive infinity, 
and only the stable fixed point   located at
$(J_1^\ast,J_2^\ast)_-=\Big(0,\frac{1}{4\xi_R}\Big)$ survives.

\item 
$b \rightarrow 0^{+}$ limit with $w<0$  and $a<0$ : 

In the small $b$ limit, $\xi_R = \frac{1}{4b}(|a|+a-\frac{2bw}{|a|})$ and $\varepsilon -2\xi_R=\frac{1}{4b}(|a|-a-\frac{2bw}{|a|})$. 
For $a<0$, $\xi_R =\frac{w}{2a}$ and $\varepsilon -2\xi_R \approx -\frac{a}{2b} \rightarrow \infty^+$.  
The $\beta$-functions in \eq{eq:0DJ1} and \eq{eq:0DJ2} become  
\begin{eqnarray}
 \frac{d}{dz} J_{1}  &=& - \sqrt{\eta}\frac{\sqrt{2}(\varepsilon-2\xi_R)}{\xi_R}\left[\frac{2\sqrt{2}\xi_R^2 +(\varepsilon-2\xi_R)}{4(\varepsilon-2\xi_R)\xi_R}-J_2 \right] J_{1} 
 + O\Big((J-J^\ast)^3 \Big), 
 \nonumber\\
\frac{d}{dz} J_{2}  & = &
 \sqrt{2 \eta}\frac{\varepsilon-2\xi_R}{\xi_R} \Big(J_{2}-\frac{1}{4\xi_R}\Big)^2- 2\sqrt{ \eta}\Big(J_{2}-\frac{1}{4\xi_R}\Big)     + O\Big((J-J^\ast)^3 \Big).\nonumber
\end{eqnarray}
As $\varepsilon-2\xi_R$ approaches infinity in the small $b$ limit, the second term in $\frac{d}{dz}J_2$ is negligible. 
Thus, the two fixed points are getting closer to each other until  they collide at $(J_1^\ast,J_2^\ast)_\pm = \Big(0,\frac{1}{4\xi_R}\Big)$.
This is shown in \fig{fig:fp_move}($\rm c_{1,2}$). 
Before collision, the stable (unstable) fixed point has two irrelevant (relevant) directions.
As the fixed points collide, both directions become marginal because the negative and positive scaling dimensions can meet only at $0$.
It turns out that the perturbations are marginally irrelevant from one side and marginally relevant from the 
 other side.
If $b$ becomes negative, the normalizable ground state disappears, which suggests a loss of stable fixed point in the real space of the couplings~\cite{kaplan2009conformality,wang2017deconfined,gorbenko2018walking1,gorbenko2018walking2}.
It will be of interest to understand constraints on the range of conformality from unitarity.

\item $b \rightarrow 0^{+}$ limit with $w<0$  and $a>0$ :

If $a>0$, $\xi_R=\frac{a}{2b} \rightarrow \infty^+$ and $\varepsilon -2\xi_R =-\frac{w}{2a}$ is finite. 
In the $b\rightarrow 0^+$ limit, 
the unstable fixed point moves towards the positive infinity, while the stable fixed point moves to the origin.
The $\beta$-functions in this limit are given by
\begin{eqnarray}
\frac{d}{dz} J_{1}  &\approx &  -\sqrt{\eta} J_{1}+ O\Big((J-J^\ast)^3 \Big), \nonumber \\
\frac{d}{dz} J_{2}  &\approx &
 -2\sqrt{\eta}J_2+ O\Big((J-J^\ast)^3 \Big), \nonumber
\end{eqnarray}
where $\eta \approx a^2$.

\end{itemize}

\section{
Solution to the 
Schrodinger equation 
for the time-dependent 
harmonic oscillator \label{app:time_dependent_Schrodinger} 
}

We first
review the evolution of a time-dependent harmonic oscillator with Hamiltonian
\begin{equation}
\begin{aligned}
H &=\frac{1}{2} \bm{x}^2 + \frac{1}{2} \Omega_{z}^2 \bm{\pi}_x^2,
\label{eq:H1}
\end{aligned}
\end{equation}
where $\Omega_z$ is a function of imaginary time $z$. 
An initial state can be written as a superposition of the eigenstates of the instantaneous Hamiltonian at $z=0$ as $\Psi\left[\bm{\pi}_x,0\right] = \sum_m c_m \Psi_m \left[\bm{\pi}_x,0\right]$, where
\begin{equation}
\begin{aligned}
\Psi_{m}\left[ \bm{\pi}_x,0\right]&=\frac{1}{\sqrt{2^{n} n!}} \Big( \frac{\Omega_{0}}{\pi} \Big)^{1/4} H_{n} \left[ \sqrt{\Omega_{0}} \bm{\pi}_x \right]  \exp \left[ -\frac{1}{2}\Omega_{0}\bm{\pi}_x^2 \right].
\label{eq:app_eigenwf}
\end{aligned}
\end{equation}
The time dependent state satisfies the Schrodinger equation,
\begin{equation}
- \frac{1}{2}\frac{\partial^2}{\partial \bm{\pi}_x^2 }\Psi[ \bm{\pi}_x, z]+\frac{1}{2} \Omega_{z}^2  \bm{\pi}_x^2 \Psi[ \bm{\pi}_x, z]= -\frac{\partial \Psi\left[\bm{\pi}_x,z\right]}{\partial z}.
\end{equation}
We introduce a new variable,  $\xi= \frac{\bm{\pi}_x}{A_{z}}$ 
and write the time-dependent solution as
\begin{equation}
\Psi \left[\bm{\pi}_x,z\right] = C_{z} \exp \left[- A^2_{z}  K_{z} \xi^2\right] 
\tilde{\Psi} \left[\xi, \theta \right],
\label{eq:Ansatz}
\end{equation}
where $K_z$, $C_z$ and $\theta_z$ are function of $z$
which are related to $A_z$ through
$K_{z} = \frac{\dot{A}_{z}}{2 A_{z}}$,
$\dot{C}_{z}=-C_{z}K_{z}$
and 
$\dot{\theta}(z)=1/A^2_{z}$ 
with $\dot{A}_{z} = dA_{z}/dz$. 
$\tilde{\Psi} \left[\xi, \theta \right]$ satisfies 
\begin{equation}
\begin{aligned}
\frac{1}{2} A^3_{z} \left[   \ddot{A}_{z} -  A_{z} \Omega^2_{z}\right]\xi^2 \tilde{\Psi} \left[\xi, \theta \right]-  \frac{\delta \tilde{\Psi} \left[\xi, \theta \right]}{\delta \theta }+ \frac{1}{2}    \frac{\delta^2}{\delta \xi^2}\tilde{\Psi} \left[\xi, \theta \right]   =0   
\end{aligned}
\end{equation}
We choose $A_z$ so that   $\ddot{A}_{z} -  A_{z} \Omega^2_{z}=0$,
and  $\tilde{\Psi}$ satisfies
\begin{equation}
-  \frac{\delta \tilde{\Psi} \left[\xi, \theta \right]}{\delta \theta } +  \frac{1}{2}   \frac{\delta^2}{\delta \xi^2} \tilde{\Psi} \left[\xi, \theta \right]   =0.
\end{equation}
This is the Schrodinger equation of a particle in the free space, which has 
a one-parameter family of solutions,
\begin{equation}
\tilde{\Psi}_\ell \left[\xi, \theta  \right]= \exp \left[ - \frac{\ell^2}{2} \theta(z) \right] \exp \left[ -i \ell \xi(z) \right],
\end{equation}
where $\ell$ is a real parameter
that corresponds 
to the momentum conjugae to $\xi$
in the particle analogy.
The general solution is given by a linear superposition of $\Psi_\ell$ as
\begin{equation}
\begin{aligned}
\Psi \left[ \bm{\pi}_x,z\right]&=\frac{1}{\sqrt{A_{z}}}\exp \Big( - \frac{\dot{A}_{z}}{2A_{z}} \bm{\pi}_x^2 \Big) \int_{-\infty}^\infty d \ell  \Big\{ \phi \left[\ell\right] \exp \left[ - \frac{\ell^2}{2}  \int_0^z \frac{dz'}{A^2_{z'}} \right] \exp \left[ -i  \frac{ \ell \bm{\pi}_x}{A_{z}}\right]\Big\},
\label{eq:solution_time_dependent}
\end{aligned}
\end{equation}
where 
$\phi \left[\ell \right]$ is the weight for the mode labelled by $\ell$,
and $\theta(z) = \int_0^z \frac{dz'}{A^2_{z'}} $ is used.
At $z=0$, the initial conditions $\dot{A}_{0}=0$ and $A_{0}=1$ lead to
\begin{equation}
\begin{aligned}
\Psi \left[\bm{\pi}_x ,0\right]
&= \int d\ell  \Big\{ \phi \left[\ell \right]  \exp ( -i  \ell \bm{\pi}_x )\Big\}.
\end{aligned}
\end{equation}
Thus, $\phi \left[\ell \right]$ is the Fourier transformation of $\Psi \left[\bm{\pi}_x,0\right]$,
\begin{equation}
\phi \left[\ell \right] = 
\frac{1}{2\pi}
\int \Psi\left[ \bm{\pi}_x',0\right] \exp ( i  \ell \bm{\pi}_x' ) d \bm{\pi}_x'
\end{equation}

The $n$-th eigenstate of the $H$ at $z=0$ has the Fourier components given by
\begin{equation}
\begin{aligned}
\phi_{n}\left[\ell \right] 
&=\frac{1}{2 \pi \sqrt{2^{n} n!}} \Big( \frac{\Omega_{0}}{\pi} \Big)^{1/4} \int   H_{n} \left[ \sqrt{\Omega_{0}} \bm{\pi}_x \right] \exp \Big(-\frac{1}{2}\Omega_{0} \bm{\pi}_x^2 \Big)  \exp \left[ i  \ell \bm{\pi}_x \right] d \bm{\pi}_x \\
&=\frac{\sqrt{2}}{2 \pi \sqrt{2^{n} n}} \Big( \frac{\pi}{\Omega_{0}} \Big)^{1/4} H_{n} \Big(  -i\sqrt{\Omega_{0}} \frac{\delta}{\delta \ell } \Big)   \exp \left[-\frac{\ell^2 }{2\Omega_{0}} \right].   \\
\end{aligned}
\end{equation}
Because the Hermite polynomial is complete, 
$\phi_{n}\left[\ell \right] $ can be decomposed as a linear superposition of 
\begin{equation}
\begin{aligned}
\tilde \phi_{n'}\left[\ell \right]
&=\frac{\sqrt{2}}{2 \pi  \sqrt{2^{n'} n'!}}  \Big( \frac{\pi}{\Omega_{0}} \Big)^{1/4}  H_{n'} \left[  i \frac{\ell}{\sqrt{\Omega_{0}}} \right]  \exp \left[-\frac{\ell^2}{2\Omega_{0}} \right]. 
\label{eq:alternativephil}
\end{aligned}
\end{equation}
Inserting \eq{eq:alternativephil} 
into \eq{eq:solution_time_dependent},
we obtain a solution,
\begin{equation}
\begin{aligned}
\Psi_{n'}\left[\bm{\pi}_x,z\right] &=\frac{\sqrt{4\pi }}{2 \pi \sqrt{2^{n'} n'!}} \Big( \frac{\pi}{\Omega_{0}} \Big)^{1/4} \sqrt{\frac{\omega_z}{A_{z}}}   \exp \Big( - \frac{\dot{A}_{z}}{2A_{z}} \bm{\pi}_x^2  \Big)   H_{n'} \left[   -\frac{A_{z}}{\sqrt{\Omega_{0}}} \frac{\delta}{\delta \bm{\pi}_x} \right]  \exp\left[-\frac{\omega_z}{2A_z^2}\bm{\pi}_x^2 \right],
\label{eq:Psinprime}
\end{aligned}
\end{equation}
where we use $\omega_{z}=\left[\int_0^z \frac{dz'}{A^2_{z'}} +\frac{1}{\Omega_{0}} \right]^{-1}$ as in the main context. 
First three states are given by
\bqa
\Psi_{0}\left[\bm{\pi}_x,z\right] &=&
\frac{1}{ \sqrt{\pi}} 
\Big( \frac{\pi}{\Omega_{0}} \Big)^{1/4} \sqrt{\frac{\omega_z}{A_{z}}}   
\exp \left[  -\frac{1}{2}\Big(\frac{\dot{A}_{z}}{A_{z}}+ \frac{\omega_{z}}{A_{z}^2} \Big) \bm{\pi}_x^2 \right], \nn
\Psi_{1}\left[\bm{\pi}_x,z\right] &=&
\frac{1}{ \sqrt{2 \pi}} 
\Big( \frac{\pi}{\Omega_{0}} \Big)^{1/4} \sqrt{\frac{\omega_z}{A_{z}}}   
\exp \left[  -\frac{1}{2}\Big(\frac{\dot{A}_{z}}{A_{z}}+ \frac{\omega_{z}}{A_{z}^2} \Big) \bm{\pi}_x^2 \right]
\left[  2 \frac{\omega_{z}}{A_{z}\sqrt{\Omega_{0}}} \bm{\pi}_x \right],       \nn
\Psi_{2}\left[\bm{\pi}_x,z\right] &=&
\frac{1}{ 2 \sqrt{2 \pi}} 
\Big( \frac{\pi}{\Omega_{0}} \Big)^{1/4} \sqrt{\frac{\omega_z}{A_{z}}}   
\exp \left[  -\frac{1}{2}\Big(\frac{\dot{A}_{z}}{A_{z}}+ \frac{\omega_{z}}{A_{z}^2} \Big) \bm{\pi}_x^2 \right]
\left[  4 
\left( \frac{\omega_{z}}{A_{z}\sqrt{\Omega_{0}}} \bm{\pi}_x \right)^2
-2
- \frac{4 \omega_{z}}{\Omega_{0}} 
\right]. \nn
\eqa

Now we consider the non-Hermitian RG Hamiltonian which is of our interest :
\begin{equation}
H'=\frac{1}{2} (\bm{x}+i\gamma \bm{\pi}_x)^2+\frac{1}{2} \Omega_z^2 \bm{\pi}_x^2.
\end{equation}
This is related to \eq{eq:H1} through the similarity transformation,
$H' = 
e^{-\frac{\gamma}{2} \bm{\pi}_x^2 }
H
e^{\frac{\gamma}{2} \bm{\pi}_x^2 }
$.
Accordingly, its solution is related to 
\eq{eq:Psinprime}
through
$\Psi_n' = 
e^{-\frac{\gamma}{2} \bm{\pi}_x^2 }
\Psi_n$,
\begin{equation}
\begin{aligned}
\Psi'_{n}\left[\bm{\pi}_x,z\right] &=\frac{\sqrt{4\pi }}{2 \pi \sqrt{2^{n} n!}} \Big( \frac{\pi}{\Omega_{0}} \Big)^{1/4} \sqrt{\frac{\omega_z}{A_{z}}}   
\exp \left[-\frac{1}{2}\Big(\gamma +\frac{\dot{A}_{z}}{A_{z}} \Big) \bm{\pi}_x^2 \right] 
H_{n} \left[   -\frac{A_{z}}{\sqrt{\Omega_{0}}} \frac{\delta}{\delta \bm{\pi}_x} \right]  \exp\left[-\frac{\omega_z}{2A_z^2}\bm{\pi}_x^2 \right],
\label{eq:Psinprime5}
\end{aligned}
\end{equation}
where we use $\omega_{z}=\left[\int_0^z \frac{dz'}{A^2_{z'}} +\frac{1}{\Omega_{0}} \right]^{-1}$ as in the main context.

\section{ 
Computation of 
$ A_{s,z} $, $\omega_{s,z}$, 
$ \Lambda_{s,z} $ and
$ \Delta_{s,z}$ 
}
\label{app:QD}

In this appendix, 
we provide the expressions for 
$ A_{s,z} $, $\omega_{s,z}$, 
$ \Lambda_{s,z} $ and
$ \Delta_{s,z}$ 
that appear in the solution 
for the RG Hamiltonian 
in Sec.~\ref{sec:dd_example}. 
Since the expressions of
$A_{s,z}$, $\omega_{s,z}$, 
$ \Lambda_{s,z} $ and
$ \Delta_{s,z}$ 
are the same for
$s=(R;K)$ and $s=(I;K)$, 
we will just denote them as 
$A_{K,z}$, $\omega_{K,z}$
$ \Lambda_{K,z} $ and
$ \Delta_{K,z}$ 
in this appendix.
For $b=1/2$, 
we express the solution for the Schrodinger equation \eq{eq:schrodinger_DDim} in terms of
$\alpha=-2g$,
$\zeta=a+\frac{D}{2}$ and $\sigma= \zeta^2-2w $. 

\subsection{$A_{s,z}$ \label{app:bigA}}

We start with $A_{s,z}$ 
that satisfies
$\ddot{A}_{s,z}-A_{s,z}\Omega^2_{s,z}=0$
with the initial conditions $\dot{A}_{0,0}=0$ and $A_{0,0}=1$.
For $K=0$, 
$A_{0,z}= \cosh (\sqrt{\sigma} z)$
is the solution. 
For general $K$, $A_{K,z}$ is given by
\begin{equation}
\begin{aligned}
A_{K,z } =\frac{ \pi \sqrt{\alpha} }{4 \sin (\pi \sqrt{\sigma})} |K| &\Big\{ I_{- \sqrt{\sigma}} \left[ \sqrt{\alpha} |K| e^{z}\right] \Big( I_{-1+\sqrt{\sigma}} \left[  \sqrt{\alpha}|K| \right] + I_{1+\sqrt{\sigma}}\left[ \sqrt{\alpha}|K| \right] \Big) \\&-I_{ \sqrt{\sigma}} \left[ \sqrt{\alpha} |K| e^{z}\right]  \Big( I_{-1-\sqrt{\sigma}} \left[  \sqrt{\alpha}|K|\right] + I_{1-\sqrt{\sigma}}\left[ \sqrt{\alpha} |K|\right] \Big)\Big\}.  \label{eq:A_solution}
\end{aligned}
\end{equation}
In the large $z$ limit with fixed $k\equiv K e^z$,
$A_{K,z}$ can be written as
\begin{eqnarray}
\tilde{A}_{k,z}
\equiv
\left. \lim_{z \rightarrow \infty} 
A_{K,z}
\right|_{k}
=\mathbbm{A}_{k}(\alpha,\sigma) e^{\sqrt{\sigma}z}, \label{eq:A_k}
\end{eqnarray}
where
\begin{equation}
\begin{aligned}
\mathbbm{A}_{k} (\alpha,\sigma) &\approx - \frac{ 2^{-1+\sqrt{\sigma}}\pi \alpha^{-\frac{1}{2}\sqrt{\sigma}} }{ \sin (\pi \sqrt{\sigma})}  \frac{I_{ \sqrt{\sigma}} \left[ \sqrt{\alpha} |k|\right]  }{\Gamma(-\sqrt{\sigma})}  |k|^{-\sqrt{\sigma}}.
\label{eq:AAk_app}
\end{aligned}
\end{equation}
This has the following limiting behaviour in  
$k = K e^z$.

\paragraph{$ |K| e^{z} \ll 1$ : }

For $\sqrt{\alpha} |K| e^{z} \ll \sqrt{1-\sqrt{\sigma}}<\sqrt{1+\sqrt{\sigma}}$, 
one can approximate $A_{K,z}$ to 
\begin{equation}
\begin{aligned}
A_{K,z} &\sim  \mathcal{A}_{K,1} e^{-\sqrt{\sigma}z} + \mathcal{A}_{K,2} e^{\sqrt{\sigma}z}, \label{eq:kez_small}
\end{aligned}
\end{equation}
where 
\begin{equation}
\begin{aligned}
\mathcal{A}_{K,1} &= \frac{ \pi \sqrt{\alpha}^{1-\sqrt{\sigma}} }{2^{2-\sqrt{\sigma}} \sin (\pi \sqrt{\sigma})} \frac{1}{\Gamma(1- \sqrt{\sigma})}  |K|^{1-\sqrt{\sigma}}
 \Big( I_{-1+\sqrt{\sigma}} \left[  \sqrt{\alpha}|K|\right] + I_{1+\sqrt{\sigma}}\left[  \sqrt{\alpha}|K|\right] \Big), \\
\mathcal{A}_{K,2} &=- \frac{ \pi \sqrt{\alpha}^{1+\sqrt{\sigma}} }{2^{2+\sqrt{\sigma}} \sin (\pi \sqrt{\sigma})} \frac{1}{\Gamma(1+ \sqrt{\sigma})}  |K|^{1+\sqrt{\sigma}}
 \Big( I_{-1-\sqrt{\sigma}} \left[  \sqrt{\alpha}|K|\right] + I_{1-\sqrt{\sigma}}\left[  \sqrt{\alpha}|K|\right] \Big).
 \label{eq:appD4}
\end{aligned}
\end{equation}
In the large $z$ limit with fixed $k$, 
\eq{eq:appD4} becomes
\begin{eqnarray}
\tilde{\mathcal{A}}_{k,1} &\equiv& 
\left. \lim_{z \rightarrow \infty} 
\mathcal{A}_{K,1}
\right|_{k}
= \frac{1}{2} +\frac{ \alpha k^2 e^{-2z} }{8} \frac{1}{\sqrt{\sigma}(1+\sqrt{\sigma})} + O\Big((ke^{-z})^4\Big), \nonumber \\
\tilde{\mathcal{A}}_{k,2} &\equiv& 
\left. \lim_{z \rightarrow \infty} 
\mathcal{A}_{K,2}
\right|_{k}
=\frac{1}{2}+\frac{ \alpha k^2 e^{-2z} }{8} \frac{1}{\sqrt{\sigma}(\sqrt{\sigma}-1)} + O\Big((ke^{-z})^4\Big). \label{eq:curlyA12k}
\end{eqnarray}

\paragraph{$ |K| e^z \gg 1$ : }

For $|K|e^{z} \gg 1$, 
$A_{K,z}$ can be expanded in 
powers of $1/|K|e^{z}$,
\begin{equation}
\begin{aligned}
A_{K,z} &
=
\mathcal{A}_{K,3} 
\frac{e^{\sqrt{\alpha}|K| e^z}}{e^{z/2}}
\left[
1
-\frac{1}{8}\frac{(4 \sigma-1)}{\sqrt{\alpha}
|K|e^{z}
} 
+ O\left(
\frac{1}{(K e^{z})^{2}}
\right)
\right],
\label{eq:large_kz}
\end{aligned}
\end{equation}
where
\begin{equation}
\mathcal{A}_{K,3} =\frac{ \pi^{1/2} \alpha^{1/4}  g_K }{4 \sqrt{2} \sin (\pi \sqrt{\sigma})} |K|^{1/2}
\end{equation}
and
\begin{equation}
g_K =   I_{-1+\sqrt{\sigma}} \left[  \sqrt{\alpha}|K| \right] + I_{1+\sqrt{\sigma}}\left[  \sqrt{\alpha}|K| \right] - I_{-1-\sqrt{\sigma}} \left[  \sqrt{\alpha}|K|\right] - I_{1-\sqrt{\sigma}}\left[ \sqrt{\alpha}|K| \right].
\end{equation}
In the large $z$ limit with fixed $k$, 
we obtain
\begin{eqnarray}
\tilde{\mathcal{A}}_{k,3} 
&\equiv& 
\left. \lim_{z \rightarrow \infty} 
\mathcal{A}_{K,3} 
\right|_{k}
= V_3 k^{-\frac{1}{2}-\sqrt{\sigma}} e^{(\frac{1}{2}+\sqrt{\sigma})z}+ O\Big((ke^{-z})^{\frac{3}{2}-\sqrt{\sigma}}\Big), \label{eq:curlyA3k}
\end{eqnarray}
where 
\begin{eqnarray}
V_3 &=& -\frac{ \pi^{1/2} \alpha^{-1/4-\frac{\sqrt{\sigma}}{2}}   }{2^{-\sqrt{\sigma}+\frac{3}{2}} \sin (\pi \sqrt{\sigma})}  \frac{1}{\Gamma(-\sqrt{\sigma})}.
\end{eqnarray}

\subsection{$\omega_{s,z}$ \label{app:omega}}

Here, we show that
\begin{eqnarray}
\tilde{\omega}_{k,z} &\equiv&
\left. \lim_{z \rightarrow \infty}
\omega_{K,z} 
\right|_k
\approx \frac{1}{2}\sqrt{\sigma} \left[1
-\mathbbm{W}_k e^{-2\sqrt{\sigma}z}\right]^{-1}, \label{eq:omega_k}
\end{eqnarray}
where 
\begin{eqnarray}
\mathbbm{W}_k \approx \Big\{ \begin{array}{cc}
    1 &  k \ll 1 \\
    k^{2\sqrt{\sigma}}\left[ 1- \frac{\sqrt{\sigma}}{4V_3^2\sqrt{\alpha} } (e^{-2\sqrt{\alpha} }-e^{-2\sqrt{\alpha} k} )\right] &  k \gg 1
\end{array}
\end{eqnarray}
in the large $z$ limit with fixed $k = K e^z$. 

\paragraph{$|K|e^z \ll 1$ : }

Using \eq{eq:kez_small}, 
we obtain 
\begin{equation}
\begin{aligned}
\omega_{K,z} &=  \left[\frac{1}{\Omega_{K,0} }+\int_0^z \frac{dz'}{A^2_{K,z'}}\right]^{-1} 
\approx \left[\frac{1}{\Omega_{K,0} }+ \frac{e^{2\sqrt{\sigma}z}-1}{2\sqrt{\sigma}(\mathcal{A}_{K,1}+\mathcal{A}_{K,2}) ( \mathcal{A}_{K,1} +\mathcal{A}_{K,2} e^{2\sqrt{\sigma}z})} \right]^{-1} .
\label{eq:omega_ksmall}
\end{aligned}
\end{equation}
According to \eq{eq:curlyA12k},
we find
\begin{eqnarray}
\tilde{\omega}_{k,z} &\equiv&
\left. \lim_{z \rightarrow \infty}
\omega_{K,z} 
\right|_k
\approx \frac{1}{2}\sqrt{\sigma} \left[1
- e^{-2\sqrt{\sigma}z}\right]^{-1},
\label{eq:omega_k_small}
\end{eqnarray}
where we used 
$\left. \lim_{z\rightarrow \infty}\Omega_{K,0} \right|_k 
=\sqrt{\sigma}$

\paragraph{$|K|e^z \gg 1$ :}

According to \eq{eq:large_kz}, 
we have
\begin{equation}
\begin{aligned}
 \omega_{K ,z} & \approx \Big[ \frac{1}{\Omega_{K,0}} + \frac{ K^{-2\sqrt{\sigma}}-1}{2\sqrt{\sigma}(\mathcal{A}_{K,1}+\mathcal{A}_{K,2}) ( \mathcal{A}_{K,1} +\mathcal{A}_{K,2} K^{-2\sqrt{\sigma}})}
+  \frac{1}{\mathcal{A}_{K,3}^2} \Big(-\frac{e^{-2\sqrt{\alpha} |K|e^z}}{2\sqrt{\alpha}|K|} +\frac{e^{-2\sqrt{\alpha} }}{2\sqrt{\alpha}|K|}\Big)\Big]^{-1}.
\label{eq:large_kz_omega}
\end{aligned}
\end{equation}
In the large $z$ limit, based on \eq{eq:curlyA3k}
we find
\begin{eqnarray}
\tilde{\omega}_{k,z} &\equiv&
\left. \lim_{z \rightarrow \infty}
\omega_{K,z} 
\right|_k
\approx \frac{1}{2}\sqrt{\sigma} \left[1
-k^{2\sqrt{\sigma}}\Big( 1- \frac{\sqrt{\sigma}}{4V_3^2\sqrt{\alpha} } (e^{-2\sqrt{\alpha} }-e^{-2\sqrt{\alpha} k} )\Big) e^{-2\sqrt{\sigma}z}\right]^{-1},
\label{eq:omega_k_large}
\end{eqnarray}
for $k\gg 1$.

\subsection{$\Lambda_{s,z}$ \label{app:Lambda}}

\begin{figure}[h]
\includegraphics[width=.98\textwidth]{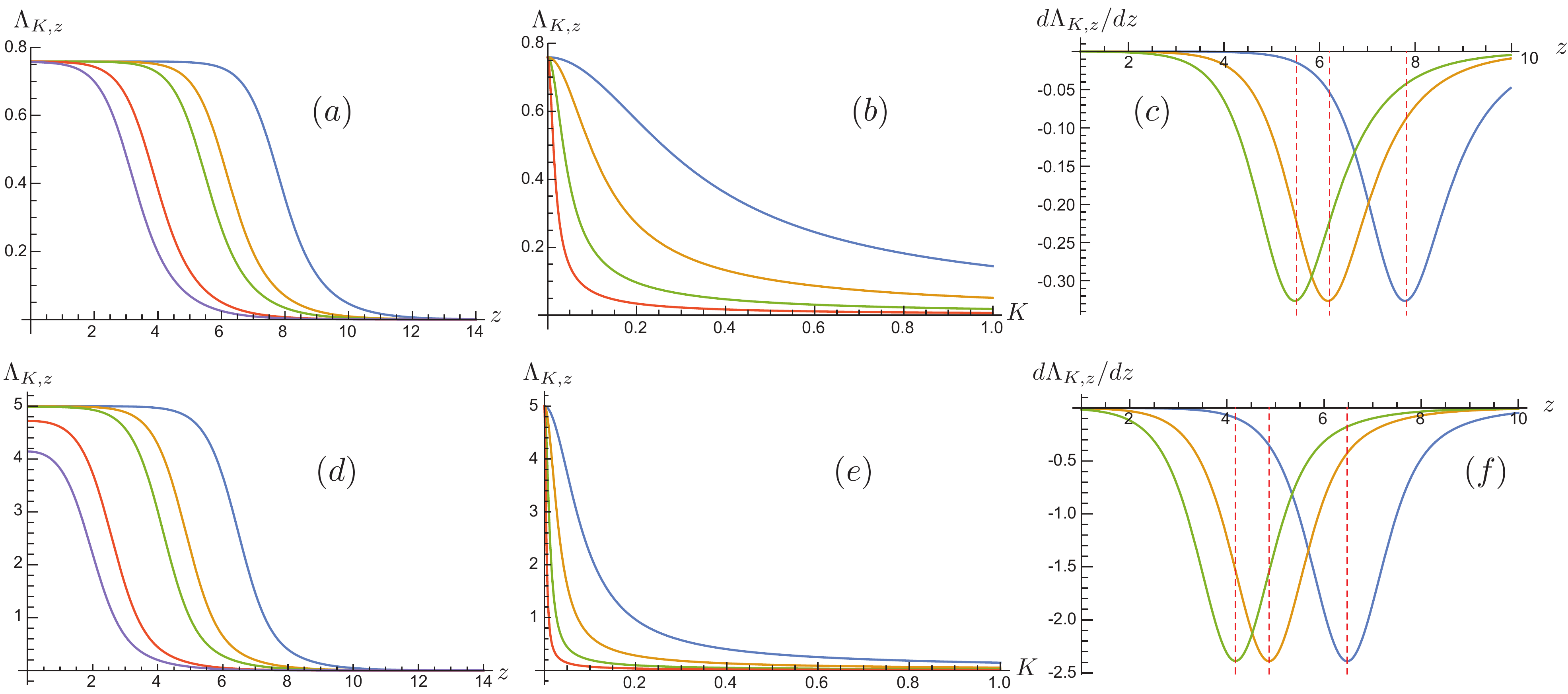}
\caption{
$\Lambda_{K,z}$ plotted as a functions of $K$ or $z$
with $\alpha=1$ in $D=1$.
For (a), (b), (c), we choose 
$\zeta=-0.1$ and $\sigma=2.01$.
For (d), (e), (f), we choose 
$\zeta=0.1$ and $\sigma=0.01$.
(a) $\Lambda_{K,z}$ vs $z$ at $K=0.001$ (blue), $0.005$ (orange), $0.01$ (green), $0.05$ (red) and $0.1$ (purple). 
(b) $\Lambda_{K,z}$ vs $K$ at $z=2$ (blue), $3$ (orange), $4$ (green), $5$ (red). 
(c) $\frac{d}{dz}\Lambda_{K,z}$ vs $z$ for $K=0.001$, $K=0.005$ and $K=0.01$. 
For each value of $K$, 
the minimum occurs at $z=7.82184$, $z=6.2124$ and $z=5.51925$, respectively. 
At all minima, $Ke^z$ takes  the same value, $2.49448$. 
(d) $\Lambda_{K,z}$ vs $z$ at $K=0.001$ (blue), $0.005$ (orange), $0.01$ (green), $0.05$ (red) and $0.1$ (purple). 
(e) $\Lambda_{K,z}$ vs $K$ at $z=2$ (blue), $3$ (orange), $4$ (green), $5$ (red). 
(f) $\frac{d}{dz}\Lambda_{K,z}$ vs $z$ for $K=0.001$, $K=0.005$ and $K=0.01$. 
For each value of $K$, the minimum  is located at $z=6.47645$, $z=4.86718$ and $z=4.17465$. 
At all minima, $Ke^z$ takes  the same value, $0.649$. 
\label{fig:non_Hermitian_Lambda_k_z}}
\end{figure}

\begin{figure}[h]
\includegraphics[width=.98\textwidth]{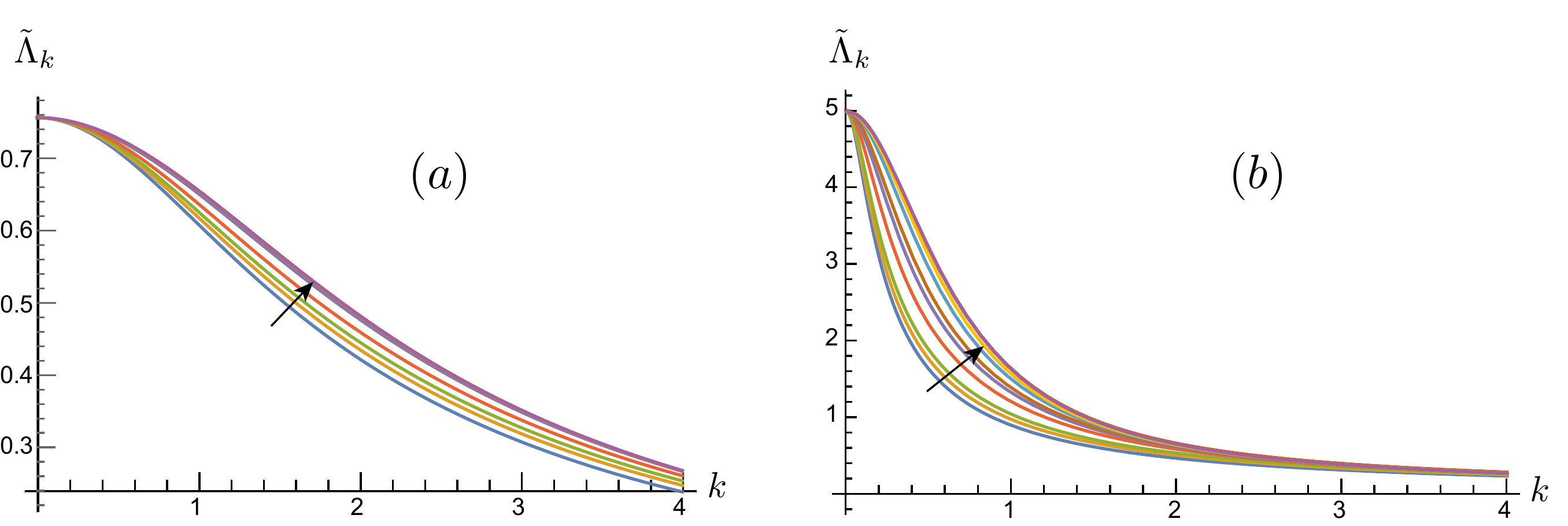}
\caption{
$\Lambda_{K,z}=\tilde{\Lambda}_k$ vs $k$ 
for various  values of $z$ between $0$ and $20$ in $D=1$. 
The arrows point towards the  direction of increasing $z$.
(a)  $\alpha=1$, $\zeta=-0.1$ and $\sigma=2.01$ 
(b) 
$\alpha=1$, $\zeta=0.1$ and $\sigma=0.01$.
The curves converge to a universal one in the large $z$ limit.
\label{fig:non_Hermitian_Lambda_ktilde}}
\end{figure}

In this section, we compute
$\Lambda_{s,z}$ 
defined by
\begin{equation}
\frac{1}{\Lambda_{s,z}}
=\left[\zeta +\frac{\dot{A}_{s,z}}{A_{s,z}}+\frac{\omega_{s,z}}{A_{s,z}^2} \right],
\label{eq:appD11}
\end{equation}
where $\omega_{s,z} = \left[\int_0^z \frac{dz'}{A^2_{s,z'}} +\frac{1}{\Omega_{s,0}} \right]^{-1}$. 
In \fig{fig:non_Hermitian_Lambda_k_z}, we show the profile of  $\Lambda_{K,z}$ 
for two sets of parameters.
$\Lambda_{K,z}$ smoothly interpolates the two limiting behaviours 
of the 
$K e^z \gg 1$ 
and
$K e^z \ll 1$ 
limits.
If one takes the large $z$ limit with fixed $k=Ke^z$,
$\Lambda_{K,z}$ approaches a universal function
as is  shown in \fig{fig:non_Hermitian_Lambda_ktilde}.
Now, let us find the analytic expression for 
\bqa
\tilde \Lambda_k = 
\left. \lim_{z \rightarrow \infty} \Lambda_{K,z} \right|_{k},
\eqa
where the limit is taken with $k=K e^z$ fixed.

In the large $z$ limit, as we shown in \eq{eq:A_k} and \eq{eq:omega_k}, $\tilde{\omega}_{k,z}$ approaches $\frac{1}{2}\sqrt{\sigma}$, and
$\tilde{\omega}_{k,z} \ll \tilde{A}_{k,z}$. 
So the dominant contribution to $\Lambda_{K,z}$ 
in \eq{eq:appD11}
is from
$\left[ \frac{\dot{A}_{K,z}}{A_{K,z}} +\zeta\right]^{-1}$. 
Therefore, 
$\Lambda_{K,z} $ approaches a universal form as a function of  $k=|K|e^z$,
\begin{equation}
\tilde{\Lambda}_{k}
=\left[\mathbbm{G}_{k} (\alpha,\sigma)+\zeta\right]^{-1} 
\label{eq:Lambda_app},
\end{equation}
as the large $z$ limit is taken with fixed $k$, where
\begin{equation}
\begin{aligned}
\mathbbm{G}_{k}(\alpha,\sigma) 
&=\frac{\dot{A}_{K,z}}{A_{K,z}}
&= \frac{1}{2}\sqrt{\alpha} \frac{|k|}{I_{\sigma} \left[\sqrt{\alpha}|k|\right]}  (I_{-1+\sqrt{\sigma}}\left[ \sqrt{\alpha} |k| \right] + I_{1+\sqrt{\sigma}}\left[ \sqrt{\alpha} |k| \right]).
\label{eq:Lambda_G}
\end{aligned}
\end{equation}
$\mathbbm{G}_{k}(\alpha,\sigma) $
becomes
\begin{equation}
\mathbbm{G}_{k}(\alpha,\sigma) \approx  \sqrt{\sigma}
\end{equation}
for $k \ll 1$, 
and 
\begin{equation}
\mathbbm{G}_{k}(\alpha,\sigma) \approx  \sqrt{\alpha}|k|
\end{equation}
for $k \gg 1$.
In order for the wavefunction to be normalizable, 
the width of the Gaussian wavefunction in \eq{eq:general_modes}  should be finite.
This requires
$\Lambda_{K,z}>0$ for all $K$ and $z$.
This, in turn, implies that
$\Omega_{K,0} > -\zeta$ for all $K$, equivalently $\sqrt{\sigma}>-\zeta$.

\subsection{$\Delta_{s,z}$ \label{app:Delta}}

\begin{figure}[h]
\includegraphics[width=.98\textwidth]{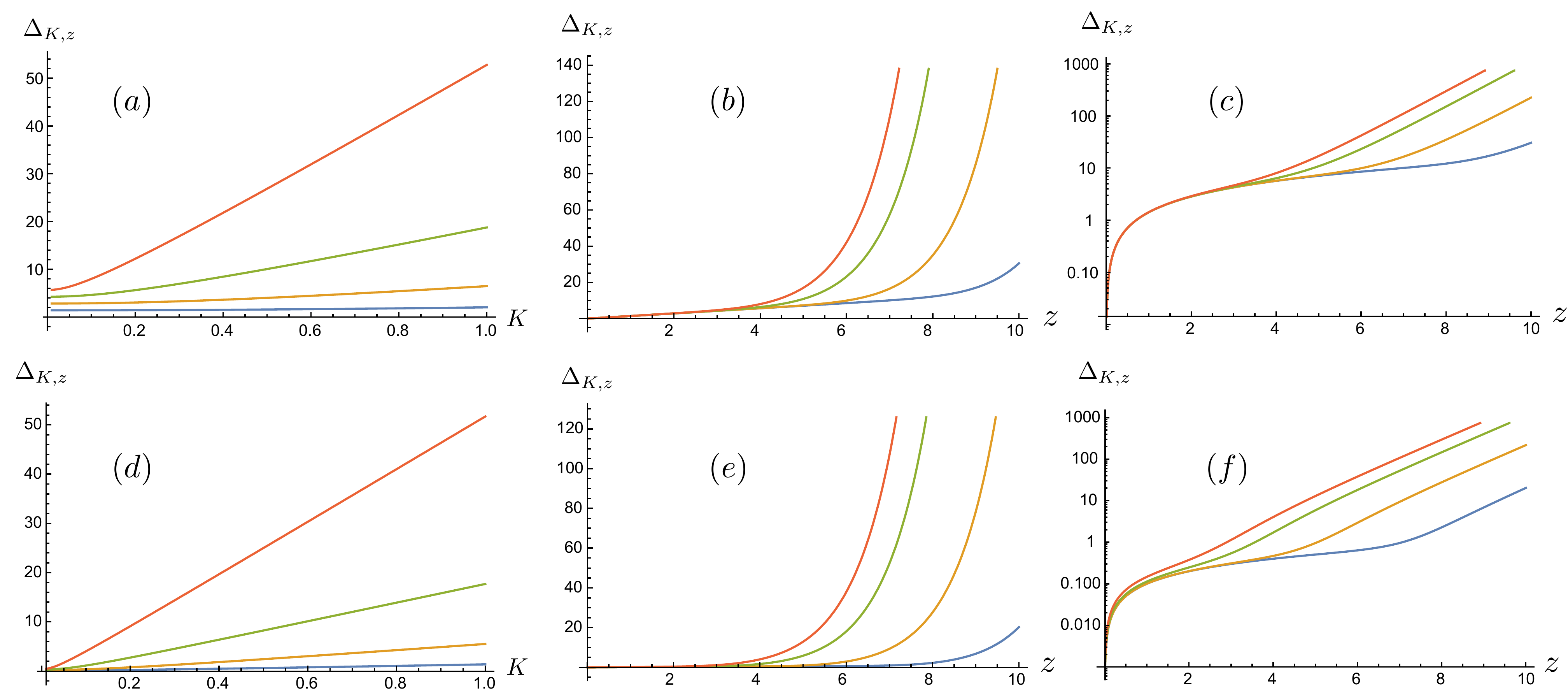}
\caption{
(a) and (d): $\Delta_{K,z}$ plotted as a function of $K$ at $z=1$ (blue), $z=2$ (orange), $z=3$ (green) and $z=4$ (red)
for (a) $\zeta=0.1$, $\sigma=0.01$,
and (d) $\zeta=-0.1$, $\sigma=2.01$. 
(b) and (e): 
$\Delta_{K,z}$ plotted as a function of $z$ at $K=0.001$ (blue), $K=0.01$ (orange), $K=0.05$ (green) and $K=0.1$ (red)
for (b) $\zeta=0.1$, $\sigma=0.01$,
and (e) $\zeta=-0.1$, $\sigma=2.01$. 
(c) and (f) are (b) and (e) shown in the logarithmic scale.
In all plots, we set $D=1$
and $\alpha=1$.
\label{fig:Delta_k}}
\end{figure}

\begin{figure}[h]
\includegraphics[width=0.98\textwidth]{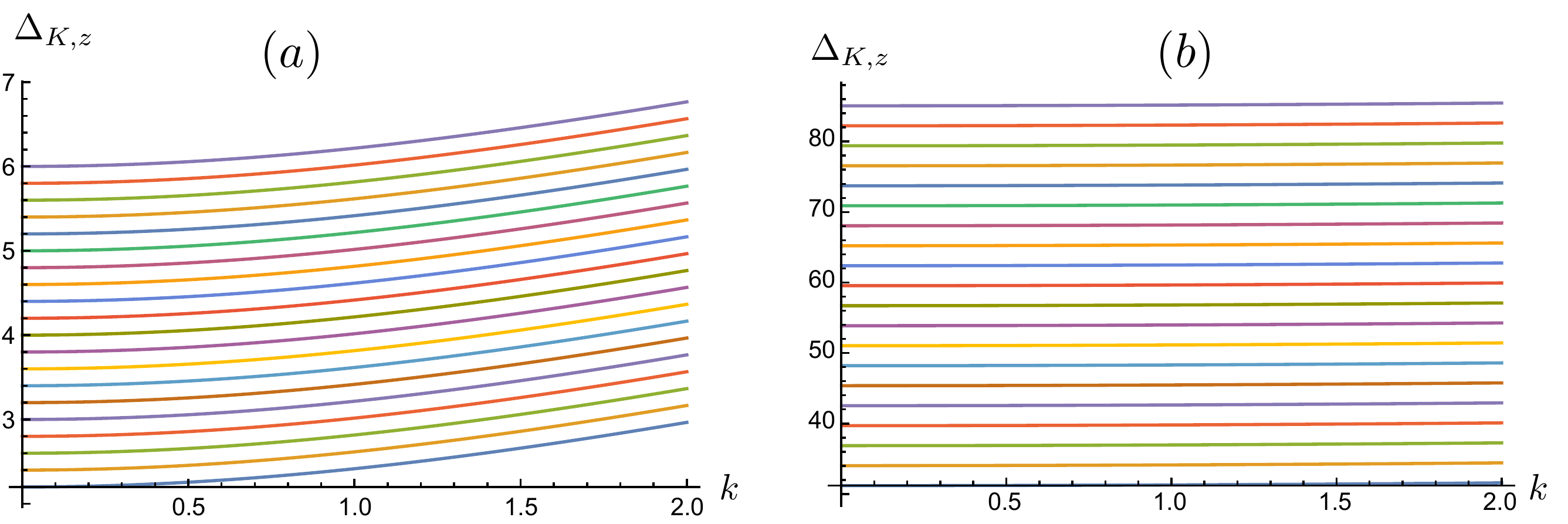}
\caption{
$\Delta_{K,z}$ as a function of $k=Ke^z$ at different values of $z$ ranging from $z=22$ (bottom) to $z=60$ (top) for
(a) $\zeta=0.1$, $\sigma=0.01$,
and 
(b) $\zeta=-0.1$, $\sigma=2.01$. 
$D=1$ and $\alpha=1$ are used for both plots.
\label{fig:Delta_k_k0}} 
\end{figure}

According to \eq{eq:A_k} and \eq{eq:omega_k}, we have expressed $\Omega_{K,0}$, $A_{K,z}$ and $\omega_{K,z}$ in terms of $k$ and $z$. In the large $z$ limit, we have
\begin{equation}
e^{-\Delta_{K,z}} = \frac{\omega_{K,z}}{A_{K,z} \sqrt{\Omega_{K,0}}} \approx \frac{1}{2} \frac{\sigma^{1/4}}{\mathbbm{A}_{Ke^z}(\alpha,\sigma)}e^{-\sqrt{\sigma}z}.
\label{eq:Delta_large_z_limit}
\end{equation}
This analytical expression is consistent with the numerical  plot  of $\Delta_{K,z}$ shown in \fig{fig:Delta_k}. 
As a function of $k$, $\Delta_{K,z}$ at different $z$ behave in the same way except for a vertical shift,
as is shown in \fig{fig:Delta_k_k0}. 
This agrees with our analytical expression, $\Delta_{K,z}=-\log 2 -\frac{1}{4}\log \sigma +\log \mathbbm{A}_k + \sqrt{\sigma}z $. 
Under the RG transformation from length scale $z$ to $z+dz$ followed by the rescaling of $K$ to $Ke^{-z}$, 
$\Delta_{K,z}$ transforms to 
$\Delta_{K,z}+ \sqrt{\sigma} dz$.

\section{Numerical calculation of $J_{2,x-x'}^\ast$ \label{app:fixed_J2xx}}

It is hard to obtain a full expression for $J_{2,x-x'}^\ast$ in a closed form. 
In this appendix, we compute it numerically for $D=1$. 
This requires UV and IR   regularizations.
Consider a one-dimensional lattice with $N$ sites. 
Before doing the scale transformation in \eq{eq:change_variable2}, 
if the lattice spacing is $a$, 
a function $f_{X=xe^{z},z}$ 
at $z=0$ 
can be expressed as
\begin{equation}
f_{X,0} =\frac{1}{N}\sum_{m=1}^{N} f_{\frac{2\pi}{Na} m} e^{i \frac{2\pi}{Na} m X} = a \int^{\Lambda_0\sim \frac{2\pi}{a}}_0 \frac{dK}{2\pi} f_{K,0} e^{iK X}.
\label{eq:fXz}
\end{equation}
For $z \neq 0$, 
the lattice spacing increases to $ae^z$ and the number of sites decreases to
$N(z)=N e^{-z}$. 
Then \eq{eq:fXz} becomes
\begin{eqnarray}
f_{xe^z,z} &=& 
\frac{1}{N(z)} \sum_{m=1}^{N(z)} f_{\frac{2\pi}{N(z)ae^z} m} e^{i \frac{2\pi}{N(z) a} m x}
= a \int^{\Lambda_0 e^{-z}}_0 \frac{e^zdK}{2\pi} f_{K,z} e^{iKX}
=
a \int^{\Lambda_0}_0 \frac{ d k}{2\pi} f_{ke^{-z},z} e^{i k x}, \nonumber \\
\end{eqnarray}
where  
$k= K e^z$ and $x=e^{-z}X$.
If $f_{ke^{-z},z} = \tilde{f}_{k}$ is independent of $z$ for a fixed $k$, $\tilde f_{x}=f_{xe^z,z}$ is scale invariant, i.e. $z$-independent. 
Since $\Lambda_{K,z}= \tilde{\Lambda}_{k}$, 
the profile of $J_{2,x-x'}$ would be invariant under RG transformation.

\begin{figure}[h]
\centering
\includegraphics[width=.8\textwidth]{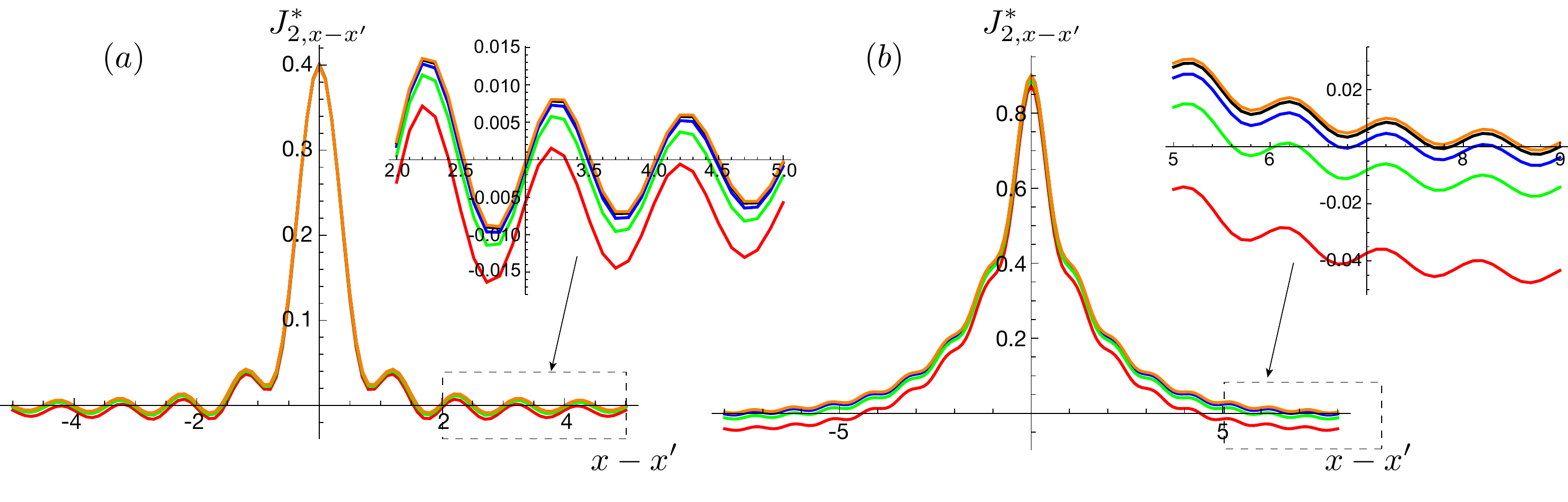}
\caption{
(a) $J_{2,x-x'}^\ast$ in $D=1$ plotted as a function of $x-x'$ for $\sigma=2.01$ and $\zeta=-0.1$ for $z=26$ at $N =e^{30}$ (red), $N = e^{31}$ (green), $N = e^{32}$ (blue), $N = e^{33}$ (black), $N = e^{34}$ (orange). 
(b) The profile at $\sigma = 0.01$ and $\zeta=0.1$. 
Curves with a same color are the ones with a same $N$. 
$\alpha$ is set to $1$. }
\label{fig:finite_size_j2x}
\end{figure}

Now, let us numerically compute $J_{2,x-x'}^\ast$, which is expressed as
\begin{equation}
\begin{aligned}
J_{2,x-x'}^\ast & =\frac{1}{N(z)}\sum_{m=1}^{N(z)}   \tilde{\Lambda}_{\frac{2\pi m}{N(z)}} \cos \left[ \frac{2\pi m}{N(z)} (x-x')\right] 
\end{aligned}
\end{equation}
with $a=1$. 
The profile is shown in \fig{fig:non_Hermitian_j2x}. 
For large enough system size $N$, the coupling in the real space reaches a $z$-independent profile at large $z$ provided $N e^{-z} \gg 1$. 
This profile is universal because it does not depend on $\Omega_{K,0}$ at UV.
We note that there are regions of negative coupling at large $|x-x'|$. We attribute this phenomenon as a finite size effect. In \fig{fig:finite_size_j2x}, as the system size $N$ increases, the coupling becomes more positive.

\section{Wavefunction with one excited mode in the D-dimensional example 
\label{app:one_excited_modes}}

Suppose that the mode $s$ is in its first excited state,
where $s$ can be either $P=0$, $(R;P)$ or $(I;P)$.
The wavefunction for mode $s$ is given by 
\begin{equation}
\begin{aligned}
\Psi_{1,s}\left[\tilde{t}_{s},z\right] &= 
  \pi^{-1/4} e^{-\frac{1}{2}\Delta_{s,z}}   \Big(\sqrt{2} e^{-\Delta_{s,z}} \tilde{t}_{s} \Big) \exp \left[-\frac{1}{2\Lambda_{s,z} }\tilde{t}_{s}^2  \right],
\end{aligned}
\end{equation}
where  $H_1(x)=2x$ is used. 
The excited state corresponds to the following state in the rescaled variables in the large $z$ limit,
\begin{equation}
\begin{aligned}
| \Psi_{1,0}(z) \rangle
&=\mathcal{N}(z)\int \mathcal{D}\phi ~ e^{-S_0} \Big(i \sqrt{2} \frac{\sigma^{1/4}}{2\mathbbm{A}_0}
e^{- \sqrt{\sigma} z} 
\tilde{\Lambda}_0 \mathcal{O}_0 \Big)e^{-S^\ast} |\phi\rangle , \\
| \Psi_{1,(R;P)}(z) \rangle
&=\mathcal{N}(z) \int \mathcal{D}\phi ~ 
  \Big( i \frac{\sigma^{1/4}}{2\mathbbm{A}_{Pe^z }}e^{- \sqrt{\sigma} z} 
 \tilde \Lambda_{Pe^z } (\mathcal{O}_{Pe^z }+ \mathcal{O}_{-Pe^z }) \Big)e^{-S^\ast}
  | \phi \rangle ,\\
 | \Psi_{1,(I;P)}(z) \rangle
&=\mathcal{N}(z) \int \mathcal{D}\phi ~ 
  \Big(  \frac{\sigma^{1/4}}{2\mathbbm{A}_{Pe^z }}e^{- \sqrt{\sigma} z} 
  \tilde \Lambda_{Pe^z } (\mathcal{O}_{Pe^z }- \mathcal{O}_{-P e^z }) \Big)e^{-S^\ast}
  | \phi \rangle,
 \label{eq:partition_function_single_excitation}
\end{aligned}
\end{equation}
where we used 
$
e^{-\Delta_{K,z}}  \approx \frac{\sigma^{1/4}}{2\mathbbm{A}_{Ke^z}(\alpha,\sigma)}e^{-\sqrt{\sigma}z}
$
and
 $ k = e^z K$. 
$\mathcal{N}(z)$  is the $z$-dependent normalization of the ground state in  \eq{eq:stable_fp}. 
 $P$ labels the initial momentum of the excited mode. 
 It is scaled to be $P e^z$ as $z$ increases.
 For $P \neq 0$, we can construct the excited state with 
 a momentum $\pm P$ by making linear superpositions of 
$  | \Psi_{1,(R;P)}(z) \rangle$ and
 $| \Psi_{1,(I;P)}(z) \rangle$ :
 $|\Psi_{1,\pm P}(z)\rangle=\frac{1}{\sqrt{2}}\Big(|\Psi_{1,(R:P)}(z)\rangle \pm i |\Psi_{1,(I:P)}(z)\rangle \Big)$. 
This leads to \eq{eq:Psi1pz}.

\section{Possible wavefunctions with two excited modes in the D-dimensional example \label{app:two_excited_modes}}

In order to derive \eq{eq:two_excited_modes}, 
we first list the wave functions for two excited modes as
\begin{eqnarray}
\Psi_{2,0,0} \left[\tilde{t}_0,z \right] &=& \pi^{-1/4} e^{-\frac{1}{2}\Delta_{0,z}} \frac{1}{\sqrt{2}}  \left[2 ( e^{-\Delta_{0,z}}\tilde{t}_0)^2-1 -\frac{2\omega_{0,z}}{\Omega_{0,0}}\right] \exp \left[ -\frac{1}{2\Lambda_{0,z}}\tilde{t}_0^2 \right]  \\&=& \frac{1}{\sqrt{2}} \Psi_{1,0}\left[\tilde{t}_0,z \right]\times \Psi_{1,0}\left[\tilde{t}_0,z \right] -\frac{1}{\sqrt{2}}(1+\frac{2\omega_{0,z}}{\Omega_{0,0}})\Psi_{0,0}\left[\tilde{t}_0,z \right], \nonumber \\
\Psi_{2,(S;P),(S;P)} \left[\tilde{t}_{S;P},z \right] &=& \pi^{-1/4} e^{-\frac{1}{2}\Delta_{P,z}} \frac{1}{\sqrt{ 2}} \left[ 2(e^{-\Delta_{P,z}}\tilde{t}_{S;P})^2-1-\frac{2\omega_{P,z}}{\Omega_{P,0}} \right] \exp \left[ -\frac{1}{2\Lambda_{P,z}}\tilde{t}_{S;P}^2 \right] \nonumber \\
&=&\frac{1}{\sqrt{2}} \Psi_{1,(S;P)}\left[\tilde{t}_{S;P},z \right]\times \Psi_{1,(S;P)}\left[\tilde{t}_{S;P},z \right] -\frac{1}{\sqrt{2}}(1+\frac{2\omega_{P,z}}{\Omega_{P,0}})\Psi_{0,(S;P)}\left[\tilde{t}_{S;P},z \right], \nonumber\\
\Psi_{2,(S;P),(S';P') \neq (S;P)} &=&\Psi_{1,S;P}\left[\tilde{t}_{S;P},z \right] \times \Psi_{1,S';P'}\left[\tilde{t}_{S';P'},z \right]. \nonumber
\end{eqnarray}
Here $S$ and $S'$ can be $R$ or $I$. 
Using \eq{eq:omega_k} and \eq{eq:Delta_large_z_limit},
we rewrite the excited states 
in terms of rescaled variables $p$ 
in the large $z$ limit as
\begin{equation}
\begin{aligned}
&| \bar{\Psi}_{2,0,0}(z) \rangle
 =
 \frac{\mathcal{N}_2(z)}{\sqrt{2}}
 \int \mathcal{D}\phi ~  \Big(- 2 \frac{\sigma^{1/2}}{4\mathbbm{A}_0^2}\left[\tilde{\Lambda}_0^2 \mathcal{O}_0^2\right] \Big)e^{-S^\ast} |\phi\rangle 
 \\
 & \hspace{2cm} 
 + \frac{1}{\sqrt{2}}\Big(2\frac{\sigma^{1/2}}{4\mathbbm{A}_0^2}e^{-2\sqrt{\sigma}z}\tilde{\Lambda}_0 -1-\left[1-\mathbbm{W}_0 e^{-2\sqrt{\sigma}z}\right]\Big)|\Psi_0(z)\rangle, \\
&| \Psi_{2,(R;P),(R;P)}(z) \rangle
=
 \frac{\mathcal{N}_2(z)}{\sqrt{2}}
\int \mathcal{D}\phi ~  \Big(-  \frac{\sigma^{1/2}}{4\mathbbm{A}_{Pe^z}^2} \tilde{\Lambda}_{Pe^z}^2  (\mathcal{O}_{Pe^z}+\mathcal{O}_{-Pe^z})^2 \Big)e^{-S^\ast} |\phi\rangle \\
& \hspace{2cm} + \frac{1}{\sqrt{2}} \Big(2\frac{\sigma^{1/2}}{4\mathbbm{A}_{Pe^z}^2}e^{-2\sqrt{\sigma}z} \tilde{\Lambda}_{Pe^z}-1-\frac{\sqrt{\sigma}}{\sqrt{\alpha P^2+\sigma}}\left[1-\mathbbm{W}_{Pe^z} e^{-2\sqrt{\sigma}z}\right]\Big)|\Psi_0(z)\rangle, \\
& | \Psi_{2,(I;P),(I;P)}(z) \rangle
=
 \frac{\mathcal{N}_2(z)}{\sqrt{2}}
\int \mathcal{D}\phi ~ \Big(  \frac{\sigma^{1/2}}{4\mathbbm{A}_{Pe^z}^2} \tilde{\Lambda}_{Pe^z}^2 (\mathcal{O}_{Pe^z}-\mathcal{O}_{-Pe^z})^2\Big)e^{-S^\ast} |\phi\rangle \\
& \hspace{2cm} + \frac{1}{\sqrt{2}} \Big(2\frac{\sigma^{1/2}}{4\mathbbm{A}_{Pe^z}^2}e^{-2\sqrt{\sigma}z} \tilde{\Lambda}_{Pe^z}-1-\frac{\sqrt{\sigma}}{\sqrt{\alpha P^2+\sigma}}\left[1-\mathbbm{W}_{Pe^z} e^{-2\sqrt{\sigma}z}\right]\Big)|\Psi_0(z)\rangle, \\
&| \Psi_{2,(R;P),(I;P)}(z) \rangle
=
 \mathcal{N}_2(z)
\int \mathcal{D}\phi ~  \Big( i \frac{\sigma^{1/2}}{4\mathbbm{A}_{Pe^z}^2} \tilde{\Lambda}_{Pe^z}^2\left[ \mathcal{O}_{Pe^z}^2-\mathcal{O}_{-Pe^z}^2\right] \Big)e^{-S^\ast} |\phi\rangle , 
\end{aligned}
\end{equation}
\begin{equation}
\begin{aligned}
&| \Psi_{2,0,(R;P)}(z) \rangle
=
 \mathcal{N}_2(z)
\int \mathcal{D}\phi ~  \Big( -\sqrt{2} \frac{\sigma^{1/2}}{4\mathbbm{A}_{Pe^z} \mathbbm{A}_0} \tilde{\Lambda}_0 \tilde{\Lambda}_{Pe^z} \mathcal{O}_0\left[ \mathcal{O}_{-Pe^z}+\mathcal{O}_{Pe^z}\right] \Big)e^{-S^\ast} |\phi\rangle , \\
&| \Psi_{2,0,(I;P)}(z) \rangle 
=
 \mathcal{N}_2(z)
\int \mathcal{D}\phi ~  \Big( i\sqrt{2} \frac{\sigma^{1/2}}{4\mathbbm{A}_{Pe^z} \mathbbm{A}_0} \tilde{\Lambda}_0 \tilde{\Lambda}_{Pe^z} \mathcal{O}_0\left[ \mathcal{O}_{Pe^z}-\mathcal{O}_{-Pe^z}\right] \Big)e^{-S^\ast} |\phi\rangle ,  \\
& | \Psi_{2,(R;P),(R;P')}(z) \rangle
=
 \mathcal{N}_2(z)
\int \mathcal{D}\phi ~  \Big(  \frac{-\sigma^{1/2}}{4\mathbbm{A}_{Pe^z} \mathbbm{A}_{P'e^z}} \tilde{\Lambda}_{P'e^z} \tilde{\Lambda}_{Pe^z} \left[ \mathcal{O}_{Pe^z}+\mathcal{O}_{-Pe^z}\right]\left[ \mathcal{O}_{P'e^z}+\mathcal{O}_{-P'e^z}\right] \Big)e^{-S^\ast} |\phi\rangle , \\
&| \Psi_{2,(R;P),(I;P')}(z) \rangle
=
 \mathcal{N}_2(z)
\int \mathcal{D}\phi ~  \Big(  \frac{i\sigma^{1/2}}{4\mathbbm{A}_{Pe^z} \mathbbm{A}_{P'e^z}} \tilde{\Lambda}_{P'e^z} \tilde{\Lambda}_{Pe^z} \left[ \mathcal{O}_{Pe^z}+\mathcal{O}_{-Pe^z}\right]\left[ \mathcal{O}_{P'e^z}-\mathcal{O}_{-P'e^z}\right] \Big)e^{-S^\ast} |\phi\rangle , \\
&| \Psi_{2,(I;P),(R;P')}(z) \rangle
=
 \mathcal{N}_2(z)
\int \mathcal{D}\phi ~ \Big(  \frac{i\sigma^{1/2}}{4\mathbbm{A}_{Pe^z} \mathbbm{A}_{P'e^z}} \tilde{\Lambda}_{P'e^z} \tilde{\Lambda}_{Pe^z} \left[ \mathcal{O}_{Pe^z}-\mathcal{O}_{-Pe^z}\right]\left[ \mathcal{O}_{P'e^z}+\mathcal{O}_{-P'e^z}\right] \Big)e^{-S^\ast} |\phi\rangle , \\
&| \Psi_{2,(I;P),(I;P')}(z) \rangle
=
 \mathcal{N}_2(z)
\int \mathcal{D}\phi ~  \Big(  \frac{\sigma^{1/2}}{4\mathbbm{A}_{Pe^z} \mathbbm{A}_{P'e^z}} \tilde{\Lambda}_{P'e^z} \tilde{\Lambda}_{Pe^z} \left[ \mathcal{O}_{Pe^z}-\mathcal{O}_{-Pe^z}\right]\left[ \mathcal{O}_{P'e^z}-\mathcal{O}_{-P'e^z}\right] \Big)e^{-S^\ast} |\phi\rangle , \\
 \label{eq:partition_function_double_excitations}
\end{aligned}
\end{equation}
where
$ \mathcal{N}_2(z)
=\mathcal{N}(z)e^{-2\sqrt{\sigma}z}$.
For $P\neq P'$, we can superpose the wavefunctions above to obtain
\begin{equation}
\begin{aligned}
| \Psi_{2,P, P' }(z) \rangle &= \frac{\sqrt{2}}{4}\Big( | \Psi_{2,(R;P),(R;P')}(z) \rangle+i| \Psi_{2,(R;P),(I;P')}(z) \rangle - | \Psi_{2,(I;P),(I;P')}(z) \rangle +i| \Psi_{2,(I;P),(R;P')}(z) \rangle\Big),  \\
| \Psi_{2,P, -P' }(z) \rangle &=\frac{\sqrt{2}}{4}\Big( | \Psi_{2,(R;P),(R;P')}(z) \rangle -i| \Psi_{2,(R;P),(I;P')}(z) \rangle + | \Psi_{2,(I;P),(I;P')}(z) \rangle +i| \Psi_{2,(I;P),(R;P')}(z) \rangle\Big), \\
| \Psi_{2,-P, P' }(z) \rangle &=\frac{\sqrt{2}}{4}\Big( | \Psi_{2,(R;P),(R;P')}(z) \rangle+i| \Psi_{2,(R;P),(I;P')}(z) \rangle + | \Psi_{2,(I;P),(I;P')}(z) \rangle -i| \Psi_{2,(I;P),(R;P')}(z) \rangle \Big),\\
| \Psi_{2,-P, -P' }(z) \rangle &=\frac{\sqrt{2}}{4}\Big( | \Psi_{2,(R;P),(R;P')}(z) \rangle -i| \Psi_{2,(R;P),(I;P')}(z) \rangle - | \Psi_{2,(I;P),(I;P')}(z) \rangle -i| \Psi_{2,(I;P),(R;P')}(z) \rangle \Big), \\
| \Psi_{2,0,  P}(z) \rangle 
&=  \frac{1}{2}(  | \Psi_{2,0,(R;P)}(z) \rangle +i| \Psi_{2,0,(I;P)}(z) \rangle), \\
| \Psi_{2,0,  -P}(z) \rangle 
&= \frac{1}{2} (  | \Psi_{2,0,(R;P)}(z) \rangle -i| \Psi_{2,0,(I;P)}(z) \rangle). 
\end{aligned}
\end{equation}
For non-zero $P$, we have
\begin{equation}
\begin{aligned}
| \Psi_{2,P, P }(z) \rangle &=\frac{1}{2}\Big(  | \Psi_{2,(R;P),(R;P)}(z) \rangle- | \Psi_{2,(I;P),(I;P)}(z) \rangle +\sqrt{2}i | \Psi_{2,(R;P),(I;P)}(z) \rangle \Big), \\
| \Psi_{2,-P, -P }(z) \rangle &=\frac{1}{2}\Big( | \Psi_{2,(R;P),(R;P)}(z) \rangle- | \Psi_{2,(I;P),(I;P)}(z) \rangle -\sqrt{2}i | \Psi_{2,(R;P),(I;P)}(z) \rangle \Big), \\
| \Psi_{2,P, -P }(z) \rangle &= \frac{1}{2}\Big( | \Psi_{2,(R;P),(R;P)}(z) \rangle+ | \Psi_{2,(I;P),(I;P)}(z) \rangle  \Big) +\frac{1}{\sqrt{2}}\left[1+\frac{\sqrt{\sigma}}{\sqrt{\alpha P^2+\sigma}}\right]|\Psi_0(z)\rangle.
\end{aligned}
\end{equation}
Finally, together with 
\begin{equation}
| \Psi_{2,0, 0 }(z) \rangle = | \bar{\Psi}_{2,0, 0 }(z) \rangle + \sqrt{2} |\Psi_0(z)\rangle ,
\end{equation}
$| \Psi_{2,P, P' }(z) \rangle$ for any $P$ and $P'$ 
can be written in the general form given in \eq{eq:two_excited_modes}.

\section{Other scaling operators}
\label{appendix:OCO}

In this section, 
we consider general excited states.
The wavefunction for $n$ excited modes is 
\begin{equation}
\begin{aligned}
| \Psi_{n, \{P\} }(z) \rangle &=\left[(-i)^n \frac{\sqrt{2^n}}{\sqrt{ n!}}\right] \mathcal{N}(z) \int \mathcal{D}\phi ~  \Big(\sigma^{n/4} \frac{\prod_i^n \tilde{\Lambda}_{P_ie^z}}{2^n \prod_i^n \mathbbm{A}_{P_ie^z} }   e^{-n\sqrt{\sigma}z} \left[\prod_i^n \mathcal{O}_{P_ie^z} \right]+\dots \Big)e^{S^\ast} |\phi\rangle,
\label{eq:n_excited_modes}
\end{aligned}
\end{equation}
where $\dots$ includes terms with less number of $\mathcal{O}_{Pe^z}$ operators.
This wave function leads to state of the system as
\begin{equation}
\begin{aligned}
|\Psi_{n,\{X\}}(z)\rangle 
&=\frac{1}{V^{n/2}}
 \sum_{\{P\}}e^{i\sum_i^n P_iX_i  }| \Psi_{n,\{P\} }(z) \rangle \\
&=\left[(-i)^n \frac{\sqrt{2^n}}{\sqrt{ n!}} \right] \mathcal{N}(z)  e^{-(n\sqrt{\sigma}+\frac{nD}{2})z}  \int \mathcal{D}\phi ~   \hat{\mathcal{A}}_n (\{x \equiv X e^{-z}\})e^{-S^\ast} |\phi\rangle+\dots,
\label{eq:Psinx}
\end{aligned}
\end{equation}
where the scaling operator is defined as
\begin{eqnarray}
\hat{\mathcal{A}}_n ( \{ x\} ) =  
\sum_{m=0}^{[n/2]}
\int \left[\prod_i^{n-2m} d^Dy_i\right] J_{\{x-y\}}^{(n,n-2m)} \left[\prod_i^{n-2m} \mathcal{O}_{y_i} \right]
\end{eqnarray}
with
\begin{equation}
J_{\{x-y\}}^{(n,n-2m)}  = \int \left[ \prod_i^{n-2m} \frac{d^D p_i}{(2\pi)^D}\right] J_{\{p\}}^{(n,n-2m)} e^{i \sum_i^{n-2m} p_i (x_i-y_i)} e^{i \sum_{i=n-2m+1}^{n}p_i x_i} \delta (\sum_{i=n-2m+1}^{n} p_i).
\end{equation}
Here $J^{(n,n-2m)}$ represents the weight for $(n-2m)$-trace operators to the $n$-th scaling operator.
For example, the contribution from the $n$-trace operator is given by
\begin{eqnarray}
J_{\{p\}}^{(n,n)} = \sigma^{n/4} \frac{\prod_i^n \tilde{\Lambda}_{p_i}}{2^n \prod_i^n \mathbbm{A}_{p_i} }.
\end{eqnarray}
The local operator $\hat{\mathcal{A}}_n$ has scaling dimension 
$n \left( \sqrt{\sigma}+\frac{D}{2} \right)$.

\end{document}